\documentclass[11pt]{article}

\usepackage{amssymb,amsmath, amsthm, color, natbib, commath,graphicx,amsfonts,amstext,amsthm,ifthen}
\usepackage{latexsym,indentfirst,verbatim,moreverb,alltt, bm, url, cancel, subfig}
\usepackage{setspace}
\usepackage{JASA_manu}
\usepackage[mathscr]{euscript}
\usepackage{algorithm}
\usepackage{algpseudocode}
\usepackage{pifont}
\usepackage{enumerate}
\usepackage{comment}
\usepackage{ulem}
\usepackage{soul, xcolor}
\usepackage{graphicx}
\usepackage{multirow}
\usepackage{tikz}
\usepackage{hhline}
\usepackage{enumitem}
\allowdisplaybreaks

\usepackage{xr}
\makeatletter
\newcommand*{\addFileDependency}[1]{
  \typeout{(#1)}
  \@addtofilelist{#1}
  \IfFileExists{#1}{}{\typeout{No file #1.}}
}
\makeatother

\theoremstyle{definition}

\newtheorem{theorem}{Theorem}

\newtheorem{prop}{Proposition}

\newtheorem{lem}{Lemma}

\newtheorem{defn}{Definition} 

\def\real{\mathop{\mathbb R}}      

\newcommand{\be}{\begin{eqnarray*}}
\newcommand{\ee}{\end{eqnarray*}}
\newcommand{\ff}{\infty}
\newcommand{\ra}{\rightarrow}

\newcommand{\la}{\lambda}

\newcommand{\ta}{\theta}

\newcommand{\beq}{\begin{eqnarray}}
\newcommand{\eeq}{\end{eqnarray}}

\newboolean{includefigs}
\setboolean{includefigs}{true} 
\newcommand{\condcomment}[2]{\ifthenelse{#1}{#2}{}}

\setstcolor{red}
\begin{document}
\baselineskip=22pt
\begin{titlepage}
\title{Grouping predictors via network-wide metrics}
\author{Brandon Woosuk Park$^1$, Anand N. Vidyashankar$^2$, and Tucker S. McElroy$^3$ \\
$^1$Department of Statistics, Volgeneau School of Engineering,
George Mason University \\
$^2$Department of Statistics, Volgeneau School of Engineering,
George Mason University\\
$^3$ U.S. Census Bureau }


\maketitle

\setlength{\parindent}{2em}
\setlength{\baselineskip}{20pt}
\setlength{\parskip}{1ex}
\setlength{\textheight}{8.7in}

\begin{abstract}
{When multitudes of features can plausibly be associated with a response, both privacy considerations and model parsimony suggest grouping them to increase the predictive power of a regression model. Specifically, the identification of groups of predictors significantly associated with the response variable eases further downstream analysis and decision-making. This paper proposes a new data analysis methodology that utilizes the high-dimensional predictor space to construct an implicit network with weighted edges 
to identify significant associations between the response and the predictors. Using a population model for groups of predictors defined via network-wide metrics, a new supervised grouping algorithm is proposed to determine the correct group, with probability tending to one as the sample size diverges to infinity. For this reason, we establish several theoretical properties of the estimates of network-wide metrics. A novel model-assisted bootstrap procedure that substantially decreases computational complexity is developed,
facilitating the assessment of uncertainty in the estimates of network-wide metrics. The proposed methods account for several challenges that arise in the high-dimensional data setting, including (i) a large number of predictors, (ii) uncertainty regarding the true statistical model, and (iii) model selection variability. The performance of the proposed methods is demonstrated through numerical experiments, data from sports analytics, and breast cancer data.
}
\end{abstract}

\vspace{0.2cm}

\thispagestyle{empty}
\end{titlepage}


\section{\textbf{Introduction}}

\noindent High-dimensional data are pervasive in several areas of modern science, and complexities in such data are increasing over time. While some data, such as those from GWAS (Genome-Wide Association Study), are generated from high throughput experiments, newly emerging data may also arise from integrating several ``smaller'' data sets. While the defining characteristics of variables in different data sets are usually hard to identify, it is, however, expected that they share few attributes or traits. As a general motivating example, in the study of emerging antimicrobial resistance, it is typical that several ``atomic level features'' are combined to define a macro feature, mainly using data from dry labs. These macro features from various labs are then related to responses from biological experiments from wet labs. Since the macro features tend to share few atomic features and the effect of a macro feature on the responses is unknown, it is an important scientific question to identify groups of macro features with a ``similar'' effect on the responses.

To resolve problems of this type in real applications,  methods such as incorporating an $L_2$-penalty or bridge penalty in a regression model tend to be used. However, 
these methods do not take into account the correlation structure explicitly. Additionally, even if the correlation structure could be integrated into the methodology, what is missing currently is an understanding of the effect of dependent covariates on the response variable after accounting for the remaining variables. The problem is more complicated in the setting when the number of covariates $p$ is much larger than $n$, the sample size.  Thus, again focusing on the widely used regression models, the following question becomes pertinent: should one first group (cluster) the variables and then perform a regression using a reasonable representative of the cluster, or use the regression model to cluster and perform analyses simultaneously? 

In this manuscript, we illustrate that the latter approach can be valuable in applications such as sports and health analytics. Additionally, we show that it is feasible to incorporate a \emph{learning} component into the methodology and use \emph{learned} clusters for additional downstream data analyses. This, in effect, allows one to account for variable selection uncertainty in the data analyses and estimation of clusters. Specifically, the proposed approach facilitates a supervised clustering of the feature set, that is, a grouping of features that accounts for information about the responses. The grouped feature sets can then be used to perform data analyses to investigate the effect of summary information from a cluster on the response variable.

The proposed methodology involves the construction of a weighted implicit network and analyzing the network-wide metrics (NWM), which account for the structure of the underlying statistical model. A sequential hypothesis testing procedure is then employed to identify the cluster. The proposed clustering algorithm adopts statistical hypothesis testing to compare the NWMs between two covariates based on their statistical distribution. This generates a statistically associated group of covariates with respect to the response variable. The end product is a pre-determined number of groups (clusters) where the members of a specified cluster have ``similar'' NWM values. The similarity (based on the NWM values) is evaluated using the association between the response and the covariates; this association is a metric, so that ``similar" NWM values of two covariates indicate their proximity in the weighted implicit network. Furthermore, the weights of the implicit network can be chosen to generate other clusters (such as correlation-based, canonical correlation-based, or information theory-based) investigated in the literature. Section \ref{Clustering} details the proposed clustering algorithm and its statistical properties.

A key issue in our methodological development is that the statistical model is unknown. One approach to identification,
assuming a linear relationship between the response variable and the feature set is to invoke
data-splitting followed by regularization methods.
However, this yields a \emph{random predictor set}. Hence, to reduce the effect of such randomness, we perform the splitting multiple times and refine the recently developed techniques in \cite{KV18} to estimate a statistical model. We then construct a weighted implicit network using the random model and investigate the properties of clusters obtained from the proposed supervised algorithm.
	
Our main contributions in this manuscript are as follows: First, we develop a rigorous approach to constructing a weighted implicit network that explicitly takes into account the structure of the statistical model and analyzes the statistical properties of several NWMs. Second, we use the NWM to describe a new supervised clustering algorithm. The algorithm yields feature sets that can then be utilized to study the effect of the response variable on the summary information from the cluster. As a consequence, the clustering of the feature set yields a dimension-reduction mechanism. Third, under a posited population model, we provide theoretical guarantees for the consistency of the supervised clustering algorithm. We illustrate our contributions with extensive simulations and data analyses.

The rest of the paper is organized as follows: Section \ref{sec:Sec2} introduces basic concepts of networks, including types of networks and definitions of network-wide metrics. Section \ref{sec:Sec3} describes our methodology to construct the implicit network. Section \ref{Post-selection} presents asymptotic distributions of sample network-wide metrics that can be computed from our implicit network. Section \ref{Clustering} describes our method to detect clusters in the network. Section \ref{NS} is devoted to Numerical Studies. Section \ref{Sec:SA} and Section \ref{BC} contain the data analyses on major league baseball (MLB) data and breast cancer data. Section \ref{Conclusion} contains a few concluding remarks. Appendix \ref{App:B} includes proofs of the theoretical results, while regularity conditions and technical assumptions are provided in Appendix \ref{App:A}.

\section{\textbf{Network and network-wide Metrics}}{\label{sec:Sec2}}

This section provides a brief introduction to networks and network-wide metrics. 
We begin with the following definition \citep{C18}:

\begin{defn}
A {\bf network} is a graph $\mathcal{G}$ with vertices and edges. We denote the vertex set by $\mathcal{V}$ and the edge set by $\mathcal{E}$.
\end{defn}

A critical role in the analysis of network topology is played by the \emph{adjacency matrix}, which 
contains information about the edges and vertices of a graph.


  

\begin{defn}{\label{adjmat}}
An {\bf adjacency matrix} for a network with $p$ vertices, namely $\bm{A} \equiv ((a_{ij}))$, is defined as follows: for $i \ne j$, let
\begin{align*}
a_{ij} =
\begin{cases}
1 & \mbox{ \ \ \ if there exists an edge between vertex $i$ and vertex $j$} \\
0 & \mbox{ \ \ \ otherwise.}
\end{cases}
\end{align*}
\end{defn}
We do not allow for self-loops (i.e., edges between a vertex and itself), so $a_{ii}=0$ for all $ 1\le i \le p$. We note that a graph with $a_{ij}=1$ for all $i \neq j$ is ``fully-connected,'' and symmetric $\bm{A}$ corresponds to an ``undirected graph.''
Alternatively, one could consider a weighted adjacency matrix (WAM), where $\bm{W} = ((w_{ij}))$ and $w_{ij}$ represents the weight on the edge between vertex $i$ and vertex $j$. Evidently, when $w_{ij}$ equals 0 or 1, the weighted adjacency matrix reduces to the adjacency matrix given in Definition \ref{adjmat}. 
In this manuscript we consider  NWMs based on an \emph{undirected weighted fully-connected network with no self-loops}.



Network-wide metrics summarize important connectedness properties of a graph.  These metrics can be used to characterize a single node, a subset of nodes, or an entire graph. Metrics pertaining to a single vertex or a subset of vertices are called local metrics, whereas those pertaining to the entire graph are called global metrics. We provide the definitions of weighted network-wide metrics, following \cite{OAS10}, \cite{LRG04}, and \cite{AT08}.

\subsubsection{\textbf{Degree Centrality}}~\\
Degree centrality of a vertex $v \in \mathcal{V}$, denoted by $D_v$, is the sum of the weights on the edges of nodes linked to $v$; that is, 
\[
D_v = \sum\limits_{u \in \mathcal{V}(v)} w_{vu},
\]
where $\mathcal{V}(v) = \{ u| w_{vu}>0\}$. In the above expression for the sum one could replace $\mathcal{V}(v)$ by $\mathcal{V}$, since the network is fully-connected and there are no self-loops. However, in some applications, it is common to use the standardized degree centrality defined by $|\mathcal{V}(v)|^{-1} D_v$, where  $|\mathcal{V}|$ denotes the cardinality of a set $\mathcal{V}$. Note that $|\mathcal{V}(v)| \ne |\mathcal{V}|$ in general. 


\subsubsection{\textbf{Clustering Coefficient}}~\\
Clustering coefficient measures the tendency of a vertex being in a cluster. It is based on the triangle corresponding to a triplet of vertices.  The formal definition of a clustering coefficient of node $v \in \mathcal{V}$ is as follows:
\begin{defn}
The clustering coefficient of node $v \in \mathcal{V}$ in a weighted graph $\mathcal{G}$ is defined as
\[
C_{v} = \sum\limits_{(u,k) \in \mathcal{N}(v)}\frac{w_{uk}}{r_v (r_v-1)},
\]
where $\mathcal{N}(v) = \{ (u,k) \in \mathcal{V} \times \mathcal{V}|u \neq k \neq v$\} and $r_v = |\mathcal{V}(v)|$.
\end{defn}
\noindent For the metrics described above, the value of a NWM in a weighted graph that is not fully-connected is the same as that in a fully-connected weighted graph, where the weight is set to 0 on the potential edges between unconnected vertices.  Hence, there is no loss of generality
in our restriction to  fully-connected networks.

\noindent {\bf{Interpretation of NWM:}} We now provide an interpretation of the above metrics when the weights represent correlations between the random variables. Specifically, given $p$ random variables $X_1, \dots, X_p$ with $\mbox{Var} (X_i)=1$ for all $1 \le i \le p$, let the correlations between the variables be denoted by $\rho_{ij}$, that is,
\[
\rho_{ij}= \mbox{Corr} (X_i, X_j),~~ i, j=1, 2, \dots, p ~ \text{and}~ i \neq j.
\]
Then if $v_1, \dots, v_p$ denote the vertices corresponding to $X_1, \dots, X_p$ and $w_{ij} = \rho_{ij}$, then 
\[
D_{v_i}= \mbox{Corr} (X_i, \sum_{j \ne i=1}^p X_j)=\sum_{j\ne i=1}^p  \mbox{Corr} (X_i, X_j).
\]
Informally, the degree centrality of vertex $i$   measures the sum of all correlations between the $i^{th}$ vertex and all other vertices in the graph. Next, the clustering coefficient of $v_i$ can be expressed in terms of the degree centrality as follows:
\[
C_{v_i}=\frac{1}{r_{v_i}(r_{v_i}-1)}(\sum_{u \neq k}\rho_{uk}-2D_{v_i}).
\]
Using any of these centralities for clustering is then  analogous to using correlation as a \emph{similarity measure}. 


\section{\textbf{Model based weights for the Implicit Network}}{\label{sec:Sec3}}

This section describes the essential ingredients of the manuscript's sparse modeling framework. Consider a collection of $n$ independent and identically distributed random vectors $\left(Y_i, X_{i1}, \dots, X_{ip}\right)$ (for $1 \le i \le n$), where the response $Y$ is associated with $q$ of the $p$  ($q < p$) explanatory variables. 
We begin by constructing a weighted implicit network with vertices corresponding to these explanatory variables.  Toward that end,  we assume that the association between the response variable and the covariates $X_1, X_2, \dots, X_p$ follows a linear model:
\begin{align}\label{Model}
\bm{Y} = \bm{X\beta} + \bm{\epsilon},
\end{align}
where $\bm{Y}$ is a $n \times 1$ vector of response variables, $\bm{X}$ is a $n \times p$ design matrix, $\bm{\beta}$ is a $p \times 1$ vector of regression coefficients, and $\bm{\epsilon}$ is a $n \times 1$ vector of errors. We denote the column vector of $\bm{X}$ by $\bm{C}_i = (X_{1i},X_{2i},\dots,X_{ni})'$ for $i = 1,\dots,p$ and the row vector of $\bm{X}$ by $R_j = (X_{j1},X_{j2},\dots,X_{jp})$ for $j = 1,\dots, n$. We assume throughout the manuscript that (i) $\mbox{Var} (\bm{\epsilon}) = \sigma^2 \bm{I}$,
where $\bm{I}$ is the identity matrix; (ii) the rows of $\bm{X}$ are i.i.d random vectors; (iii) $\mbox{Cov} (X_i, \epsilon)= 0$ for $i=1,\dots,p$; and (iv) $\Sigma_{\bm{X}}$, which denote the covariance matrix of $\bm{X}$, is positive definite. In the sequel, we will refer to the structure of the data set by $(Y, {\bm{R}})$.


\subsection{\textbf{Implicit Weighted Fully-Connected Network}}
An implicit weighted fully-connected network for a data set $(Y, {\bm{R}})$ is a weighted graph, where  the $p$ covariates $X_1, X_2, \dots, X_p$ in ${\bm{R}}$ correspond to  vertices;
accordingly the vertex set will be $\{1,2, \dots, p\}$, where vertex $i$ represents the covariate $X_i$.  
The edge set $\mathcal{E}$ is the set of all links between vertices (recall that the network is 
fully connected).  The edge weights are functions of the regression coefficients. 
These concepts are formally parsed below. 


\begin{defn}\label{Def:Network}
An  {\bf implicit weighted fully-connected network} for a data set $(Y, \bm{R})$ is a weighted graph $\mathcal{G}$ with vertex set $\mathcal{V}$ composed of the indices of the covariates, and with a given adjacency matrix $\bm{W}$.
\end{defn}
In the rest of the manuscript, we will use the terminology \emph{implicit network} to mean the \emph{implicit weighted fully connected network}. To complete the description of the implicit network, we need to describe the weights on the edges of the network. While there are several choices based on the model (\ref{Model}), we focus on (i) weights that are functions of regression coefficients and (ii) partial correlation weights.

\subsection{\textbf{Regression Weights}}
Let $f: \real^2 \rightarrow [0, \ff)$ be a twice continuously differentiable function (on both the coordinates) and set  $w_{ij}=f(\beta_i, \beta_j)$. Using the definitions in Section \ref{sec:Sec2}, the weighted degree centrality of node $i \in \mathcal{V}$ is given by
\begin{align}\label{D}
D_i = \sum\limits_{j \in \mathcal{V}(i)} w_{ij} = \sum\limits_{j \in \mathcal{V}(i)} f(\beta_i, \beta_j),
\end{align}
where $\mathcal{V}(i) = \{j | w_{ij}>0\}$. The weighted clustering coefficient of node $i \in \mathcal{V}$ is given by
\begin{align}\label{C}
C_i = \sum\limits_{(j,k) \in \mathcal{N}(i)} \frac{w_{jk}}{r_i(r_i-1)} = \sum\limits_{(j,k) \in \mathcal{N}(i)} \frac{f(\beta_j, \beta_k)}{(p-1)(p-2)} \quad\text{,}
\end{align}
where $\mathcal{N}(i)= \{ (j,k) \in \mathcal{V} \times \mathcal{V} | j \neq k \neq i\}$ and $r_i = |\mathcal{V}(i)|$. Notice that since the implicit network is fully-connected, $r_i = p-1$ for $i \in \mathcal{V}$.
The graph below represents an example of a weighted implicit network with three vertices.
\begin{center}
\begin{tikzpicture}[auto, node distance=3cm, every loop/.style={},
                    thick,main node/.style={circle,draw,font=\sffamily\Large\bfseries}]

  \node[main node, fill=red!10] (1) {$Y$};
  \node[main node, fill=blue!40] (2) [below left of=1] {$X_1$};
  \node[main node, fill=blue!40] (3) [below right of=2] {$X_3$};
  \node[main node, fill=blue!40] (4) [below right of=1] {$X_2$};
  \node[main node, fill=red!10] (5) [below left of=2] {$Y$};
  \node[main node, fill=red!10] (6) [below right of=4] {$Y$};

  \path[every node/.style={font=\sffamily\small}]
    (1) edge node [left] {$\beta_2$} (4)
        edge [right] node[left] {$\beta_1$} (2)
    (2) edge node {$f(\beta_1, \beta_2)$} (4)
        edge [below] node[left] {$\beta_1$} (5)
    (3) edge node [right] {$f(\beta_1, \beta_3)$} (2)
    	edge node [below] {$\beta_3$} (6)
        edge [left] node[below] {$\beta_3$} (5)
    (4) edge node [right] {$f(\beta_2 , \beta_3)$} (3)
        edge [right] node[right] {$\beta_2$} (6);
\end{tikzpicture}
\end{center}

\subsection{\textbf{Partial Correlation Weights}}\label{ParCor}
Another choice of $f $ that has received attention in the literature is a function of $\bm{\beta}$ that yields the partial correlation coefficient. A partial correlation is the strength of relationship between two variables, while controlling the effect of other variables. Here, the partial correlation network represents the association between the covariate and the response variable while accounting for other covariates. Partial correlation networks have been widely used in biology and genetic studies to construct biological networks on genes and proteins. \cite{RC08} introduced a method to reconstruct the gene co-expression networks using the partial correlation coefficient with information theory to identify gene associations. \cite{DBH04} constructed approximate undirected dependency graphs of large-scale biochemical data based on the hypothesis testing of partial correlations between two biochemical compounds. In addition, \cite{AWQ20} applied Gaussian Graphical models -- whose network defines a weight between two nodes as their partial correlation -- to genomics data, in order to study the relationship among genes. Hence, the use of partial correlation weights on the network is consistent with other network analysis methods that have been used.

We refer to this network as a \emph{weighted partial correlation network}. It is well-known that the partial correlation can be computed using the components of a precision matrix \citep{MRJ09}. Specifically, let $\mathcal{Y} = (Y, X_1, \dots, X_p)$, and let $\bm{\sigma}_{Y,\bm{X}} = (\mbox{Cov} (Y, X_1),  \mbox{Cov}(Y, X_2), \dots, \mbox{Cov}(Y, X_p))$ denote a $p \times 1$ vector of covariances between $Y$ and the covariates. Then the covariance matrix of $\mathcal{Y}$, denoted by $\Sigma$, can be expressed as follows:
\begin{align*}
\Sigma =\left(
\begin{array}{c|c}
\sigma_Y^2 & \bm{\sigma}_{Y, \bm{X}} \\ \hline
\bm{\sigma}_{Y, \bm{X}}' & \Sigma_{\bm{X}}
\end{array}
\right)
=
\begin{pmatrix}
\sigma_Y^2 & \sigma_{YX_1} & \cdots & \sigma_{YX_p} \\
\sigma_{YX_1} & \ddots & \cdots & \vdots \\
\vdots & \cdots & \ddots & \vdots \\
\sigma_{YX_p} & \cdots & \cdots & \sigma_{X_p}^2
\end{pmatrix}.
\end{align*}
Now using (\ref{Model}), $\sigma_Y^2 = \bm{\beta}'\Sigma_{\bm{X}}\bm{\beta} + \sigma^2$ and $\sigma_{YX_i} = \bm{\beta}'\Sigma_{\bm{X},i}$, where $\Sigma_{\bm{X},i}$ is the $i^{th}$ column of $\Sigma_{\bm{X}}$, and the elements of $\Gamma = \Sigma^{-1}$ are function of $\bm{\beta}$. Let $\gamma_{ij}$ denote the $(i,j)^{th}$ element of $\Gamma$. Then, $\rho_{i|-i}$, which denotes the partial correlation coefficient between $Y$ and $X_i$, is given by
\[
\rho_{i|-i} = - \frac{\gamma_{1,i+1}}{\sqrt{\gamma_{11}}\sqrt{\gamma_{i+1, i+1}}}.
\]
Let ${\mathfrak{M}}$ denote the space of the class of all precision matrices, and $h: \mathfrak{M} \ra [-1, 1]^p$ be such that $h(A)={\bm{\ta}}$, where
$\bm{\theta}=(\ta_1, \ta_2, \dots, \ta_p)$ and
\[
\ta_i= \frac{a_{1,i+1}}{\sqrt{a_{11}a_{i+1,i+1}}}.
\]
We study two functions of $\rho_{i|-i}$, viz.  $\rho_i = (1 + \rho_{i|-i} )/2$, which yields a number  between 0 and 1, and  $f_1 (\rho_i, \rho_j)$
where 
\begin{eqnarray}
\label{Choice:f}
f_1(x, y)=\sqrt{2(1-x^2)}+\sqrt{2(1-y^2)}.
\end{eqnarray}
It is clear that both $\rho_i$ and $f_1(\rho_i, \rho_j)$ are functions of $\bm{\beta}$, since partial correlation coefficients can be expressed as functions of the precision matrix.

We notice that the description of the implicit network assumes that the regression model, and hence the set of covariates and its dimension $p$, is known. However, in real applications, especially those involving clustering (grouping), this assumption is not valid.  Additionally, if the the model is unknown a priori,
then one needs to ``estimate'' the network, which is equivalent to identifying the set of features associated with the response. These issues occur more frequently in high-dimensional settings, and ``regularization'' methods are typically used to address these challenges. 

\subsection{\textbf{The Case of Categorical Covariates}}

A data set with categorical variables requires an alternative approach to construct an implicit network, since partial correlations do not exist for categorical variables. 
Instead of a multiple linear regression model, we assume a fixed effect model with $p$ covariates and their interactions. The model is given by
\[
y_{ijkl} = \mu + \alpha_i + \beta_j + \cdots + \gamma_k + \alpha\beta_{ij} + \alpha\gamma_{ik} + \cdots + \beta\gamma_{jk} + \epsilon_{ijkl},
\]
where $i = 1,\dots,r_1$, $j=1,\dots,r_2$, $k=1,\dots,r_p$, and $l=1,\dots,n$. Here $r_m$ is the number of levels in $X_m$ for $m=1,\dots,p$. To take account for the effect of covariates $X_i$ and $X_j$ on $Y$, we compute $\mbox{SS}_{ij}/ \mbox{TSS}$, where $\mbox{SS}$ and $\mbox{TSS}$ denote the sum of squares of the interaction between variable $i$ and $j$ and the total sum of squares, respectively, from model (\ref{Model}). We use this quantity as the weight on the edge between $i$ and $j$.

\subsubsection{\textbf{Implicit Network Construction}}

Let $\hat{Y}$ denote the estimated $y$ from model (\ref{Model}). Then, $\mbox{SS}_{ij}$ and $\mbox{TSS}$ are given by
\begin{align*}
\mbox{SS}_{ij} &= n\sum\limits_{i=1}^{r_1}\sum\limits_{j=1}^{r_2} (\bar{Y}_{ij\cdot}-\bar{Y}_{i\cdot\cdots} - \bar{Y}_{\cdot j \cdot} + \bar{Y}_{\cdots})^2 \\
\mbox{TSS} &= \sum\limits_{l=1}^n (Y_l -\bar{Y}_{\cdots})^2.
\end{align*}
We will use these measures to define the weight 
$w_{ij}$ on the edge between $X_i$ and $X_j$:
\[
w_{ij} =\frac{\mbox{SS}_{ij}}{\mbox{TSS}}.
\]
Since $0 \le \mbox{SS}_{ij} \le \mbox{TSS}$, the weights are bounded by one.   A graph below illustrates this construction of the implicit network.
\begin{center}
\begin{tikzpicture}[auto, node distance=3cm, every loop/.style={},
                    thick,main node/.style={circle,draw,font=\sffamily\Large\bfseries}]

  \node[main node, fill=red!10] (1) {$Y$};
  \node[main node, fill=blue!40] (2) [below left of=1] {$X_1$};
  \node[main node, fill=blue!40] (3) [below right of=2] {$X_3$};
  \node[main node, fill=blue!40] (4) [below right of=1] {$X_2$};
  \node[main node, fill=red!10] (5) [below left of=2] {$Y$};
  \node[main node, fill=red!10] (6) [below right of=4] {$Y$};

  \path[every node/.style={font=\sffamily\small}]
    (1) edge node [left] {} (4)
        edge [right] node[left] {} (2)
    (2) edge node {$\frac{SS_{12}}{TSS}$} (4)
        edge [below] node[left] {} (5)
    (3) edge node [left] {$\frac{SS_{13}}{TSS}$} (2)
    	edge node [below] {} (6)
        edge [left] node[below] {} (5)
    (4) edge node [right] {$\frac{SS_{23}}{TSS}$} (3)
        edge [right] node[right] {} (6);
\end{tikzpicture}
\end{center}

\noindent {\bf{Interpretation of NWM:}} We now provide an interpretation of the above metrics when the weights represent sums of squares of an interaction of two covariates in a fixed effect model. Then for $p$ vertices $X_1, \dots, X_p$, the degree centrality of $X_i$ is given by 
\[
D_i=\frac{1}{\mbox{TSS}}\sum_{j\ne i=1}^p \mbox{SS}_{ij}.
\]
Thus, informally, the degree centrality of vertex $i$ can be described as a quantity that measures the sum of all sums of squares of interaction terms between the $i^{th}$ vertex and all other vertices in the graph. Next, the clustering coefficient of $v_i$ can be expressed in terms of the degree centrality as follows:
\[
C_i=\frac{1}{(p-1)(p-2)}\frac{1}{\mbox{TSS}}(\sum_{u \neq k}\mbox{SS}_{uk}-2D_i).
\]
In a fixed effects model, the proportion of variation of the response variable explained by each covariate will work as a \emph{similarity measure}. Hence, using any of these centralities for clustering assigns covariates with similar effects on the response variable to the same cluster. 

\subsection{\textbf{Model Estimation via Regularization}}

We assume a sparse modeling framework, i.e.,   in (\ref{Model}) only $q$ ($< p$) of the regression coefficients are non-zero. We denote by
\[
\mathcal{S} = \{j \in \{1,2, \dots, p\} | \beta_j \neq 0 \}
\] 
the true \emph{active predictor set}. Let $S^c$ denote the complement of $S$, and note that $q=|\mathcal{S}|$ (which is unknown). 
Simultaneous consistent variable selection and parameter vector estimation  is possible using regularization methods, as discussed in \cite{FL01}, \cite{T96}, \cite{Z10}, and \cite{YL06}.
Specifically, let $\hat{\bm{\beta}}_n$ denote the solution to the optimization problem
\beq\label{PenalizedQ}
Q_n({\bm{\beta}})=\left(\bm{Y}-\bm{X\beta}\right)^{\prime}\left(\bm{Y}-\bm{X\beta}\right)+p_{\la}(|\bm{\beta}|),
\eeq
where $p_{\la}(\cdot)$ is a non-concave penalty \citep{FL11}. It is known, under appropriate regularity conditions (see Appendix \ref{App:A}), that the estimated active predictor set
\[
\hat{\mathcal{S}}_n = \{j \in \{1,2,\dots,p\} | \hat{\beta}_j \neq 0\}
\]
converges to $\mathcal{S}$ in probability; this in turn implies that $\hat{q}_n \equiv |\hat{\mathcal{S}}_n|$ converges in probability to $q$ under certain regularity conditions. Let $\bm{\beta}_{\mathcal{S}} = \{\beta_j | j \in \mathcal{S}\}$, $\bm{\beta}_{\mathcal{S}^c} = \{\beta_j | j \notin \mathcal{S}\}$, and $\bm{X}_{\mathcal{S}} = \{ \bm{X}_{j_1}, \bm{X}_{j_2},\dots,\bm{X}_{j_q}| j_i \in \mathcal{S} \mbox{\ for $i = 1,\dots,q$}\}$. Let $\bm{V}_{\mathcal{S}}= E[\bm{X}_{\mathcal{S}}'\bm{X}_{\mathcal{S}}]$  and $\Sigma = \mbox{diag} \{ p_{\lambda}''(|\beta_1|), \dots,$ $p_{\lambda}''(|\beta_q|)\}$, where $\mbox{diag} \{\ldots \}$ represents the diagonal matrix.
Also, let $\hat{\bm{\beta}}_{\mathcal{S}}=\{\hat{\beta}_j | j \in \mathcal{S}\}$ and $\hat{\bm{\beta}}_{\mathcal{S}^c} = \{\hat{\beta}_j | j \notin \mathcal{S}\}$.  The lemma below describes the properties of regularized estimators, and is due to \cite{FL01}.
\begin{lem}{\label{lem:FL}} Assume that the assumptions \ref{A1}-\ref{A3} and \ref{P1}-\ref{P6} hold. Then,
\begin{enumerate}[label=(\roman*)]
\item $\bm{P}(\hat{\bm{\beta}}_{\mathcal{S}^C} = \bm{0}) \rightarrow 1$ as $n \rightarrow \infty$, \quad \text{and}
\item \label{ROP} $\lim\limits_{n \rightarrow \infty} \bm{P}\left(\sqrt{n}(\hat{\bm{\beta}}_{\mathcal{S}} - \bm{\beta}_{\mathcal{S}} + (\bm{V}_{\mathcal{S}}+\Sigma)^{-1}\bm{b}) \le \bm{x}\right) = \bm{P}\left( \bm{Z} \le \bm{x}\right)$,
\end{enumerate}
\noindent where  
$\bm{Z} \sim 
N(\bm{0},\sigma^2(\bm{V}_{\mathcal{S}}+\Sigma)^{-1} \bm{V}_{\mathcal{S}}(\bm{V}_{\mathcal{S}}+\Sigma)^{-1})$ and
$\bm{b} = \left(p_{\lambda}'(|\beta_1|)sgn(\beta_1), \dots, p_{\lambda}'(|\beta_q|)sgn(\beta_q)\right)'.$
\end{lem}
\noindent We focus on a fully-connected weighted network on the \emph{true active predictor set}. Hence, by an abuse of notation, we will use $\mathcal{S}$ (instead of $\mathcal{V}$) to denote the vertex set as well the  true active predictor set. Furthermore, since we are mainly concerned with non-zero weights, we will assume in the rest of the manuscript that $\mathcal{R}(f)=\text{range of }f(\cdot, \cdot)=(0, \ff).$  Our first result is concerned with the joint asymptotic distribution of the estimated \emph{degree centralities} derived using regularized estimates of the regression coefficients. To this end, we need additional notations. Following (\ref{D}) and using the above assumption on $f(\cdot, \cdot)$, $D_i$ and $\hat{D}_{n, i}$ can be expressed as
\[
D_i = \sum\limits_{j \in \mathcal{S}} f(\beta_i,\beta_j), \ \ \
\hat{D}_{n,i} = \sum\limits_{j \in \mathcal{S}} f(\hat{\beta}_i,\hat{\beta}_j).
\]
Additionally, set $\bm{D} = (D_1,D_2,\dots,D_q)'$ and $ \hat{\bm{D}}_n = (\hat{D}_{n,1}, \hat{D}_{n,2},\dots, \hat{D}_{n,q})'$. Let $\bm{c}=\left(c_1, c_2, \cdots , c_q\right)\\=(\bm{V}_{\mathcal{S}}+\Sigma)^{-1}\bm{b}$, where $\bm{b}$ is as in Lemma \ref{lem:FL}. Let $\bm{L}_D(\mathcal{S},\bm{\beta})$ be a $q \times q$ matrix of gradients whose $(j, k)^{\text{th}}$ element is given by
\begin{eqnarray}\label{L:D:Prop1}
l_D(j,k; \mathcal{S}, \bm{\beta}) = \sum_{r \in {\mathcal{S}(i)}} \frac{\partial f(\beta_j, \beta_r)}{\partial \beta_k} = 
\begin{cases}
\sum\limits_{r \in \mathcal{S}} \frac{\partial f(\beta_j, \beta_r)}{\partial \beta_j} & j=k \\
\frac{\partial f(\beta_j,\beta_k)}{\partial \beta_k} & j \neq k.
\end{cases}
\end{eqnarray}
And $\hat{l}_D(j,k; \mathcal{S},\hat{\bm{\beta}})$ can be obtained by evaluating (\ref{L:D:Prop1}) at $\bm{\beta}^*$, which is between $\hat{\bm{\beta}}$ and $\bm{\beta}$. Let $\bm{H}_i^{(D)} (\mathcal{S},\bm{\beta}) \equiv ((H_{i,jk}^{(D)}(\mathcal{S},\bm{\beta})))$ and $\hat{\bm{H}}_{n,i}^{(D)}(\mathcal{S}, \hat{\bm{\beta}}) \equiv ((\hat{H}_{i,jk}^{(D)}(\mathcal{S}, \hat{\bm{\beta}})))$ denote the Hessian matrix (the matrix of second partials with respect to $\bm{\beta}$) associated with $D_i$. Here the notation $H_{i,jk}^{(D)}(\mathcal{S},\bm{\beta})$ indicates the $(j,k)^{th}$ element of $\bm{H}_i^{(D)} (\mathcal{S},\bm{\beta}) $, and the same format is used for the notation $\hat{\bm{H}}_{n,i}^{(D)}(\mathcal{S}, \hat{\bm{\beta}})$. That is, 
\begin{eqnarray}\label{H:D:Prop1}
{H}^{(D)}_{i,jk} (\mathcal{S},\bm{\beta}) = \frac{\partial^2 D_i}{\partial \beta_j \partial \beta_k} = 
\begin{cases}
\sum\limits_{r \in \mathcal{S}(i)} \frac{\partial^2 f(\beta_i, \beta_r)}{\partial \beta_i^2} & j=k=i \\
\frac{\partial^2 f(\beta_i,\beta_k)}{\partial \beta_i \partial \beta_k} & j=i \text{ and } k \neq i\\
\frac{\partial^2 f(\beta_i,\beta_j)}{\partial \beta_j \partial \beta_i} &  j \neq i \text{ and } k=i\\
\frac{\partial^2 f(\beta_i,\beta_j)}{\partial \beta_j^2} & j=k\neq i \\
0 & \mbox{otherwise},
\end{cases}
\end{eqnarray}
and $\hat{H}_{i,jk}^{(D)}(\mathcal{S},\hat{\bm{\beta}})$ is obtained by evaluating (\ref{H:D:Prop1}) at $\bm{\beta}^*$, which is between $\hat{\bm{\beta}}$ and $\bm{\beta}$. Finally, setting $\bm{H}^{(D)}(\mathcal{S},\bm{\beta}) = (\bm{H}_1^{(D)}(\mathcal{S},\bm{\beta}), \dots,\bm{H}_q^{(D)}(\mathcal{S},\bm{\beta}))$ and $ \hat{\bm{H}}_n^{(D)}(\mathcal{S},\hat{\bm{\beta}}) = (\hat{\bm{H}}_{n,1}^{(D)}(\mathcal{S},\hat{\bm{\beta}}), \dots, \hat{\bm{H}}_{n,q}^{(D)}(\mathcal{S},\hat{\bm{\beta}}))$, let $\bm{B} =- [\bm{c}'(\hat{\bm{L}}_D(\mathcal{S},\hat{\bm{\beta}})+\hat{\bm{H}}_n^{(D)} (\mathcal{S},\hat{\bm{\beta}})'\bm{c})]'$.  We remark that both $\bm{H}^{(D)}(\mathcal{S},\bm{\beta})$ and $\hat{\bm{H}}_n^{(D)}(\mathcal{S},\hat{\bm{\beta}})$ are $q \times q \times q$ tensors.
\begin{prop}\label{Prop-vector}
Assume that the assumptions \ref{A1}-\ref{A3} and \ref{P1}-\ref{P6} hold. Then,
\begin{eqnarray*}
\lim\limits_{n \rightarrow \infty} \bm{P} \left( \sqrt{n} (\hat{\bm{D}}_n - \bm{D} + \bm{B}) \le \bm{x} \right) = \bm{P}\left( \bm{Z} \le \bm{x}\right),  
\end{eqnarray*}
where $\bm{Z} \sim  N(0,\bm{\Sigma}_D)$ and
\[
 \bm{\Sigma}_D = \sigma^2 [\bm{L}_D(\mathcal{S},\bm{\beta}) + \bm{H}^{(D)'}(\mathcal{S},\bm{\beta})\bm{c}]' (\bm{V}_{\mathcal{S}} + \Sigma)^{-1} \bm{V}_{\mathcal{S}} (\bm{V}_{\mathcal{S}} + \Sigma)^{-1}[\bm{L}_D(\mathcal{S},\bm{\beta}) + \bm{H}^{(D)'}(\mathcal{S},\bm{\beta})\bm{c}]
\]
is a positive definite matrix.
\end{prop}

\noindent Our next result is concerned with the joint asymptotic distribution of the estimated \emph{clustering coefficients} derived using regularized estimates of the regression coefficients. To this end, we introduce additional notations. First recall that,
\begin{align*}
C_i = \sum\limits_{(j,k) \in \mathcal{N}(i)}\frac{f(\beta_j,\beta_k)}{(q-1)(q-2)} , \ \ \
\hat{C}_{n,i} = \sum\limits_{(j,k) \in \mathcal{N}(i)} \frac{f(\hat{\beta}_j,\hat{\beta}_k)}{(\hat{q}_n-1)(\hat{q}_n-2)},
\end{align*}
where $\mathcal{N}(i) = \{ (j,k) \in \mathcal{S} \times \mathcal{S} | j \neq k \neq i \}$. Set $\bm{C} = (C_1,C_2,\dots,C_q)'$ and $ \hat{\bm{C}}_n = (\hat{C}_{n,1}, \dots, \hat{C}_{n,\hat{q}_n})'$. Let $\bm{c}=\left(c_1, c_2, \cdots , c_q\right)=(\bm{V}_{\mathcal{S}}+\Sigma)^{-1}\bm{b}$, where $\bm{b}$ is as in Lemma \ref{lem:FL}. Let $\bm{L}_C(\mathcal{S},\bm{\beta})$ is a $q \times q$ matrix of gradients whose $(j,k)^{th}$ element is given by
\begin{align}\label{L:C:Prop2}
    l_C(j,k; \mathcal{S}, \bm{\beta}) = \frac{1}{(q-1)(q-2)} \sum\limits_{(u,v) \in \mathcal{N}(j)} \frac{\partial f(\beta_u,\beta_v)}{\partial \beta_k} = 
    \begin{cases}
    0 & j=k \\
    \frac{1}{(q-1)(q-2)}\sum\limits_{(k,u) \in \mathcal{N}(j)} \frac{\partial f(\beta_k,\beta_u)}{\partial \beta_k} & j \neq k,
    \end{cases}
\end{align}
and $\hat{\bm{l}}_C(j,k ;\mathcal{S},\hat{\bm{\beta}})$ is obtained by evaluating (\ref{L:C:Prop2}) at $\bm{\beta}^*$, which is between $\hat{\bm{\beta}}$ and $\bm{\beta}$. Let $\bm{H}_i^{(C)}(\mathcal{S},\bm{\beta}) \equiv ((H_{i,jk}^{(C)}(\mathcal{S},\bm{\beta})))$ and $\hat{\bm{H}}_{n,i}^{(C)}(\mathcal{S},\hat{\bm{\beta}}) \equiv ((\hat{H}_{i,jk}^{(C)}(\mathcal{S},\hat{\bm{\beta}})))$ denote the Hessian matrix associated with $C_i$. Here the notation $H_{i,jk}^{(C)}(\mathcal{S},\bm{\beta})$ indicate the $(j,k)^{th}$ element of $\bm{H}_i{(C)} (\mathcal{S},\bm{\beta})$, and the same format is applied to $\hat{\bm{H}}_{n,i}^{(C)}(\mathcal{S}, \hat{\bm{\beta}})$. That is, 
\begin{eqnarray}\label{H:C:Prop2}
{H}^{(C)}_{i,jk} (\mathcal{S},\bm{\beta}) = \frac{\partial^2 C_i}{\partial \beta_j \partial \beta_k} = \frac{1}{(q-1)(q-2)} 
\begin{cases}
\frac{\partial^2 f(\beta_j,\beta_k)}{\partial \beta_j \partial \beta_k} & j\neq k \neq i \\
 \sum\limits_{(k,u) \in \mathcal{N}(i)} \frac{\partial^2 f(\beta_k,\beta_u)}{\partial \beta_k^2} & j=k \text{ and } k \neq i \\
0 & \mbox{otherwise},
\end{cases}
\end{eqnarray}
and $\hat{H}_{i,jk}^{(C)}(\mathcal{S},\hat{\bm{\beta}})$ is obtained by evaluating (\ref{H:C:Prop2}) at $\bm{\beta}^*$, which is between $\hat{\bm{\beta}}$ and $\bm{\beta}$. Finally, setting $\bm{H}^{(C)}(\mathcal{S},\bm{\beta}) = (\bm{H}_1^{(C)}(\mathcal{S},\bm{\beta}), \dots, \bm{H}_q^{(C)} (\mathcal{S},\bm{\beta}))$ and $ \hat{\bm{H}}_n^{(C)}(\mathcal{S},\hat{\bm{\beta}}) = (\hat{\bm{H}}_{n,1}^{(C)}(\mathcal{S},\hat{\bm{\beta}}), \dots, \hat{\bm{H}}_{n,q}^{(C)}(\mathcal{S},\hat{\bm{\beta}}))$, let $\bm{B} = - [\bm{c}'(\hat{\bm{L}}_C(\mathcal{S},\hat{\bm{\beta}})+\bm{H}_n^{(C)}(\mathcal{S},\hat{\bm{\beta}})' \bm{c})]'$.  Again, both $\bm{H}^{(C)}(\mathcal{S},\bm{\beta})$ and $\hat{\bm{H}}_n^{(C)}(\mathcal{S},\hat{\bm{\beta}})$ are $q \times q \times q$ tensors.
\begin{prop}\label{Prop-vector2}
Assume that the assumptions \ref{A1}-\ref{A3} and \ref{P1}-\ref{P6} hold. Then,
\begin{eqnarray*}
\lim\limits_{n \rightarrow \infty} \bm{P} \left( \sqrt{n} (\hat{\bm{C}}_n - \bm{C} + \bm{B}) \le \bm{x} \right) = \bm{P}\left( \bm{Z} \le \bm{x}\right), 
\end{eqnarray*}
where $\bm{Z} \sim N(0,\bm{\Sigma}_C)$ and
\[
\bm{\Sigma}_C = \sigma^2 [\bm{L}_C(\mathcal{S},\bm{\beta}) + \bm{H}^{(C)'}(\mathcal{S},\bm{\beta})\bm{c}]' (\bm{V}_{\mathcal{S}} + \Sigma)^{-1} \bm{V}_{\mathcal{S}} (\bm{V}_{\mathcal{S}} + \Sigma)^{-1}[\bm{L}_C(\mathcal{S},\bm{\beta}) + \bm{H}^{(C)'}(\mathcal{S},\bm{\beta})\bm{c}]
\]
is a positive definite matrix.
\end{prop}
The proofs of Propositions \ref{Prop-vector} and   \ref{Prop-vector2} are provided in Appendix \ref{App:B}. We notice that the asymptotic properties depend on the unknown $\mathcal{S}$, which hinders the routine use of this methodology for inferential and clustering problems. In such situations, it is reasonable to replace $\mathcal{S}$ by  $\hat{\mathcal{S}}_n$. However, the dimension of $\hat{\mathcal{S}}_n$ and $\mathcal{S}$ can be different. This difference causes technical difficulties in providing theoretical guarantees concerning the properties of clusters that are derived in Sections \ref{PSNWM} and \ref{Clustering}.  To address this, we refine the \emph{data-splitting and dimension-matching technique} described in \cite{KV18};
the resulting \emph{post-selection} estimator of the degree centrality $\bm{D}$ and the clustering coefficient $\bm{C}$ do not involve the asymptotic bias  $\bm{B}$ as in Proposition \ref{Prop-vector}.

\section{\textbf{POST-SELECTION ESTIMATORS OF NETWORK-WIDE METRICS}}\label{Post-selection}

In this section we provide post-selection estimators of NWMs and describe their asymptotic properties. We begin with a brief description of the data-splitting and dimension-matching methods. 

\subsection{\textbf{Data-splitting and dimension-matching methods}}\label{Data-Split}
 
We first divide the data set into two parts of approximately equal size ($\lfloor n/2 \rfloor$), viz. $\mathcal{D}_{1n}$ and $\mathcal{D}_{2n}$, respectively. Using $\mathcal{D}_{1n}$, we estimate the active predictor set as follows: first, estimate the true active predictor set $\mathcal{S}$ using a consistent regularization method and then perform a hypothesis test to check if the selected regression coefficients are zero, \emph{using the same data} $\mathcal{D}_{1n}$. Let $\hat{\mathcal{S}}_n^{(1)}$ denote the estimated active predictor set obtained minimizing (\ref{PenalizedQ}), and let $\hat{\mathcal{S}}_n^{(2)}$ denote the estimated active predictor set obtained from the hypothesis test using $\mathcal{D}_{1n}$, i.e.,
\[
 \hat{\mathcal{S}}_n^{(2)} = \{ j \in \hat{\mathcal{S}}_n^{(1)} | |t_{n,j} > c_{\alpha_n} \}.
\]
 Here $t_{n,j}$ is the t-statistic for testing $H : \beta_j =0$ vs. $K : \beta_j \neq 0$, i.e.,
\[
  t_{n,j} = \frac{\hat{\beta}_j}{se(\hat{\beta}_j)} \quad \text{for} \; j \in \hat{\mathcal{S}}_n^{(1)},
\]
where $\hat{\beta}_j$ is the least square estimate of $\beta_j$ and $se(\hat{\beta}_j)$ is its standard error. Let $\hat{\mathcal{S}}_n(\mathcal{D}_{1n})=\hat{\mathcal{S}}_n^{(1)}\cap \hat{\mathcal{S}}_n^{(2)}$ denote the  estimate of $\mathcal{S}$ that results after the \emph{two-step} procedure. We write $\hat{\mathcal{S}}_{n}$ for $\hat{\mathcal{S}}_n(\mathcal{D}_{1n})$  and $\hat{q}_n$ for $\hat{q}_n(\mathcal{D}_{1n})$. As a next step,  use $\hat{\mathcal{S}}_n$ and  $\mathcal{D}_{2n}$ for inference and clustering the covariates. We note here that, conditioned on $\hat{\mathcal{S}}_n$, the model is well-specified. However, the suggested model from the first stage is a \emph{random model}. Its variability appears in the inferential and clustering part in the form of a random index set. 

{\textbf{Dimension-Matching Technique}}\label{DMT}: Since $\hat{\mathcal{S}}_{n}$ is a random set, it's dimension $\hat{q}_n$ may not equal $q$, the dimension of $\mathcal{S}$. However, to establish theoretical guarantees, one needs to compare $\hat{\mathcal{S}}_{n}$ and $\hat{\mathcal{S}}$. To handle this issue we use the dimension-matching technique: supposing that $j \in \hat{\mathcal{S}}_{n} \bigcap \mathcal{S}^c$, we set $\beta_j =0$; on the other hand, if $j \in \hat{\mathcal{S}}_n \bigcap \mathcal{S}$, then the true value is $\beta_j$. Hence, for every $j \in \hat{\mathcal{S}}_n$, the true value of $\beta_j$ is defined and this yields a new vector $\bm{\beta}_{\hat{\mathcal{S}}_n} = (\beta_1, \beta_2,\dots, \beta_{\hat{q}_n})'$, which is referred to as the \emph{dimension-adjusted vector of regression coefficients}. In general, for any $\bm{x} \in \real^q$, we will denote by $\bm{x}_n$ the dimension-adjusted version of $\bm{x}$. In the theoretical results presented below we compare $\bm{\beta}_{\hat{\mathcal{S}}_n}$ with $\hat{\bm{\beta}}_{\hat{\mathcal{S}}_n}$, which is obtained by minimizing
\begin{align}{\label{post:LS}}
Q_n(\bm{\beta}_{\hat{\mathcal{S}}_n}) = \sum\limits_{i \in \mathcal{D}_{2n}} (y_i - \bm{x}_i\bm{\beta})^2.
\end{align}
Since the post-selection estimator of NWMs are based on $\hat{\bm{\beta}}_{\hat{\mathcal{S}}_n}$, theoretical investigation in the following sections require their asymptotic behavior. Our first theorem addresses this issue, and is similar in spirit to Theorem 1 in \cite{KV18}.
%
\begin{theorem}\label{Thm-Asym-LSBeta}
Assume that the assumptions \ref{A1}-\ref{A3} and \ref{P1}-\ref{P6} hold. Then,\\
{\bf[1]} $\hat{\mathcal{S}}_n$ is a weakly consistent estimator of $\mathcal{S}$; that is,
\[
 \lim_{n \rightarrow \infty} \bm{P}(\hat{\mathcal{S}}_n = \mathcal{S})=1.
\]
{\bf[2]} Let $\bm{x} \in \mathbb{R}^q$ and $\bm{x}_n$ denote the dimension-adjusted version of $\bm{x}$. Then,
\[
   \lim_{n \rightarrow \infty} \bm{P}\left( \sqrt{\frac{n}{2}}(\hat{\bm{\beta}}_{\hat{\mathcal{S}}_n} - \bm{\beta}_{\hat{\mathcal{S}}_n}) \le \bm{x}_n \right) = \bm{P}( \bm{Z} \le \bm{x}),
\]
where $\bm{Z} \sim  N(\bm{0}, \sigma^2 \bm{V}_{\mathcal{S}}^{-1})$  and $\bm{V}_{\mathcal{S}} = E[\bm{X}_{\mathcal{S}}'\bm{X}_{\mathcal{S}}]$.
\end{theorem}
The proof of the Theorem is in Appendix \ref{App:B}. Notice that the key difference between Theorem \ref{Thm-Asym-LSBeta} and Lemma \ref{lem:FL} is the bias term $(\bm{V}_{\mathcal{S}}+\Sigma)^{-1}\bm{b}$. The bias term in Theorem \ref{Thm-Asym-LSBeta} is removed, since we split the data into two independent data sets, use one set to select significant variables in (\ref{Model}), and use the other to estimate them. In addition, the weak consistency of $\hat{\mathcal{S}}_n$ facilitates the theoretical result without the bias. Theoretical results regarding the limit distributions of NWMs and consistency of clustering will be based on Theorem \ref{Thm-Asym-LSBeta}.

\subsection{\textbf{Post-selection estimators of NWM and their asymptotic properties}}\label{PSNWM}
In this section, we describe post-selection estimators of the NWM described in Section  \ref{sec:Sec2}. We begin with weighted degree centrality.

\noindent {\bf{Weighted degree centrality:}} We recall from Section \ref{sec:Sec3} that for 
$i \in \hat{\mathcal{S}}_n$ the weighted degree
centrality $D_i$ is given by (\ref{D}),
where $\mathcal{V}(i) = \{j | w_{ij}>0\} \subset \hat{\mathcal{S}}_n$.
By the properties of $f$, it follows that 
$f(\hat{\beta}_i, \hat{\beta}_j) >0$ for each $(i, j) \in \hat{\mathcal{S}}_n \times \hat{\mathcal{S}}_n$.  Furthermore, since the network is fully-connected and $f(\beta_i,\beta_j)=0$ for $i=j$, we can estimate $\mathcal{V}(i)$ by $\hat{\mathcal{S}}_n$. We notice here that the estimate of $\mathcal{V}(i)$ is independent of $i$ due to the use of the post-selection approach. This choice reduces the computational burden in high-dimensional problems by $\hat{q}_n^2$ computations, which is the required number of computations to compute $\mathcal{V}(i)$. Hence, the post-selection estimator of the weighted degree centrality is given by
\[
\hat{D}_{n,i} = \sum\limits_{j \in \hat{\mathcal{S}}_n} \hat{w}_{ij} = \sum\limits_{j \in \hat{\mathcal{S}}_n} f(\hat{\beta}_i,\hat{\beta}_j),
\]
where $\hat{\beta}_i$ and $\hat{\beta}_j$ are the minimizers of (\ref{post:LS}). Let
$\hat{\bm{D}}_n = (\hat{D}_{n,1}, \hat{D}_{n,2},\dots, \hat{D}_{n,\hat{q}_n})$  denote the estimated vector of degree centralities of the vertices in $\hat{\mathcal{S}}_n$, derived using data from $\mathcal{D}_{2n}$. Also, let
${\bm{D}}_n = ({D}_{n,1}, {D}_{n,2},\dots, {D}_{n,\hat{q}_n})$, where
\[
{D}_{n,i} = \sum\limits_{j \in \hat{\mathcal{S}}_n} w_{ij} = \sum\limits_{j \in \hat{\mathcal{S}}_n} f(\beta_i,\beta_j).
\]
Our next result is concerned with the joint limiting distribution of the vector of weighted degree centralities.

\begin{theorem}\label{Thm-Degree-Vec}
Assume that the assumptions \ref{A1}-\ref{A3} and \ref{P1}-\ref{P6} hold. For any $\bm{x} \in \real^q$, let ${\bm{x}}_n$ denote its dimension-adjusted version. Then,
\[
\lim_{n \rightarrow \infty} \bm{P}\left( \sqrt{\frac{n}{2}}(\hat{\bm{D}}_n - \bm{D}_n)\le {\bm{x}}_n \right) =\bm{P}\left( \bm{Z} \le {\bm{x}}\right),
\]
where $\bm{Z} \sim N(0,\bm{\Sigma}_D)$ and
$\bm{\Sigma}_D= \sigma^2 \bm{L}_{D}^{\prime}(\mathcal{S}, \bm{\beta}) \bm{V}_{\mathcal{S}}^{-1}\bm{L}_{D}(\mathcal{S}, \bm{\beta})$ is a positive-definite matrix, and $\bm{L}_D(\mathcal{S}, \bm{\beta})$ is a $q \times q$ matrix whose $(j, k)^{\text{th}}$ element is given by
\[
l_D(j,k; \mathcal{S}, \bm{\beta}) = \sum_{r \in {\mathcal{S}}} \frac{\partial f(\beta_j, \beta_r)}{\partial \beta_k}= 
\begin{cases}
\sum\limits_{r \in \mathcal{S}} \frac{\partial f(\beta_j, \beta_r)}{\partial \beta_j} & j=k \\
\frac{\partial f(\beta_j,\beta_k)}{\partial \beta_k} & j \neq k.
\end{cases}
\]
\end{theorem}

\noindent {\bf{Partial Correlation Weights}:} When the weights on the edges of the network are taken as partial correlations, $f$ depends not just on $\beta_i$ and $\beta_j$ but on the entire vector $\bm{\beta}$. Analyses of these post-selection estimates of partial correlation coefficients then require investigation of their limiting distribution, which is of independent interest. Our next theorem is concerned  with this matter. 
Let ${\bm{\rho}}=(\rho_1, \rho_2,  \rho_3 , \dots, \rho_q)^{\prime}$ denote the vector of population transformed (i.e., $\rho_i= (1+\rho_{i|-i})/2$) partial correlation coefficients, and let ${\bm{\rho}}_{\hat{\mathcal{S}}_n}$ be the corresponding dimension-adjusted vector.  Let $\hat{\bm{\rho}}_{\hat{\mathcal{S}}_n}$ denote the post-selection estimated vector of partial correlation coefficients. Recall that $h(\cdot)$ yields a vector of partial correlation coefficients and is defined on the space $\mathfrak{M}_n$ of all precision matrices with dimension $(\hat{q}_n+1)\times (\hat{q}_n+1)$.  Our goal is to work with the vectorized version of the precision matrix; however, since the precision matrix is symmetric, vectorization will lead to duplicates in the vector, leading to additional collinearity challenges in downstream analyses. For this reason, following \cite{NW90}, we introduce the \emph{elimination matrix} denoted by $\bm{K}_n$. To define $\bm{K}_n$, let $E_{ij}$ be a $(\hat{q}_n+1) \times (\hat{q}_n+1$) matrix with one in its $(i, j)^{th}$ position and zeros elsewhere. Also let $u_{ij}$ be the unit vector of length $(\hat{q}_n+1)(\hat{q}_n+2)/2$ with unity in its $[(j-1)(\hat{q}_n+1) + i - j(j-1)/2]^{th}$ element. Then, the $ (\hat{q}_n+1)(\hat{q}_n+2)/2 \times (\hat{q}_n+1)^2$ elimination matrix is defined to be
\[
\bm{K}_n= \sum\limits_{i \ge j} u_{ij} \,
(\mbox{vec} E_{ij})',
\]
where $\mbox{vec} (E_{ij})$ is $(\hat{q}_n+1)^2 \times 1$ vector obtained by vectorizing $E_{ij}$. Then, for a matrix $\bm{M}_n \in \mathfrak{M}_n$, we define
\[
v(\bm{M}_n)=\bm{K}_n \, \mbox{vec} (\bm{M}_n),
\]
and hence $v(\bm{M}_n)$ is a vector of dimension $(\hat{q}_n+1)(\hat{q}_n+2)/2 \times 1$ representing the lower triangular sub-matrix of $\bm{M}$. Hence, by an abuse of notation, the mapping $h(\cdot)$ that yields the partial correlation coefficients can be  modified to be a mapping from $v(\mathfrak{M}_n)=\{v(\bm{M}): \bm{M} \in \mathfrak{M}_n\}$.
With such a modification the vector of partial derivatives of $h(\cdot)$ with respect to the elements of the vector is well-defined and is denoted by $\nabla h_n$, which is a $\hat{q}_n \times  (\hat{q}_n+1)(\hat{q}_n+2)/2$ matrix. We first describe the joint limiting distribution of $\hat{\bm{\rho}}_{\hat{\mathcal{S}}_n}$, which is a key step required in the proof of our second main result. 

\begin{theorem}{\label{conv:PSPC}}
Assume that the assumptions \ref{A1}-\ref{A3} and \ref{P1}-\ref{P6} hold. For any ${\bm{x}} \in \real^q$, let ${\bm{x}}_n$ denote its dimension-adjusted version. Then the following holds:
\[
\lim_{n \ra \ff} {\bm{P}}\left(\sqrt{\frac{n}{2}}\left({\hat{\bm{\rho}}}_{\hat{\mathcal{S}}_n}-\bm{\rho}_{{\hat{\mathcal{S}}}_n}\right) \le {\bm{x}}_n\right) = {\bm{P}}\left( \bm{Z} \le \bm{x}\right),
\]
where $\bm{Z} \sim N({\bm{0}}, \bm{\Delta}_{\mathcal{S}}(\bm{\rho}))$ and 
$\bm{\Delta}_{\mathcal{S}}(\bm{\rho}) = \nabla h(v(\Gamma_{\mathcal{S}}))\bm{K}\bm{L}_{\mathcal{S}}'\bm{\Delta}_{\mathcal{S}} \bm{L}_{\mathcal{S}} \bm{K}' \nabla h(v(\Gamma_{\mathcal{S}}))'$.  Here $\bm{L}_{\mathcal{S}}= - \Gamma_{\mathcal{S}} \otimes \Gamma_{\mathcal{S}}$, $\bm{\Delta}_{\mathcal{S}} = E[(\mathcal{Y}_{\mathcal{S}}'\mathcal{Y}_{\mathcal{S}}) \otimes (\mathcal{Y}_{\mathcal{S}}'\mathcal{Y}_{\mathcal{S}})] - \mbox{vec} (\Sigma) \mbox{vec} (\Sigma)'$,
  $\bm{K}$ is the elimination matrix, and $\nabla h(v(\Gamma_{\mathcal{S}}))$ is the partial derivatives of $h(v(\Gamma_{\mathcal{S}}))$ with respect to the elements of $v(\Gamma_{\mathcal{S}})$ defined on the space $\mathfrak{M}$ of precision matrices of dimension $(q+1) \times (q+1)$. 
\end{theorem}
Our next main result concerns the joint limiting distribution of degree centralities when the weights are a  function of the partial correlation coefficients.  Specifically, let $\hat{\mathcal{D}}_n(\bm{\rho})$   denote the vector of degree centralities obtained using partial correlation weights, and let $\mathcal{D}_n(\bm{\rho})$ denote the corresponding dimension-adjusted population vector of degree centralities. 

\begin{theorem}{\label{DA_DCPC}}
Assume that the assumptions \ref{A1}-\ref{A3} and \ref{P1}-\ref{P6} hold. For any $\bm{x} \in \real^q$, let ${\bm{x}}_n$ denote its dimension-adjusted version. Then,
\[
\lim_{n \rightarrow \infty} {\bm{P}}\left(\sqrt{\frac{n}{2}}\left(\hat{\bm{D}}_n(\bm{\rho}) - \bm{D}_n({\bm{\rho}})\right) \le {\bm{x}}_n \right)= {\bm{P}}\left( \bm{Z} \le {\bm{x}}\right),
\]
where $\bm{Z} \sim N(0,\Sigma_D({\bm{\rho}}))$,
$\Sigma_D(\bm{\rho})=  \bm{L}_D(\mathcal{S},\bm{\rho})^{\prime} {\bm{\Delta}}_{\mathcal{S}}(\bm{\rho}) \bm{L}_{D}(\mathcal{S},\bm{\rho})$,   ${\bm{\Delta}}_{\mathcal{S}}(\bm{\rho})$ is the limiting covariance matrix in Theorem \ref{conv:PSPC}, and $\bm{L}_D(\mathcal{S},\bm{\rho})$ is a $q \times q$ matrix whose $(j, k)^{\text{th}}$ element is given by
\[
l_D(j,k;\mathcal{S}, {\bm{\rho}}) = \sum_{r \in {\mathcal{S}}} \frac{\partial f(\rho_j, \rho_r)}{\partial \rho_k} = 
\begin{cases}
\sum\limits_{r \in \mathcal{S}} \frac{\partial f(\rho_j, \rho_r)}{\partial \rho_j} & j=k \\
\frac{\partial f(\rho_j,\rho_k)}{\partial \rho_k} & j \neq k.
\end{cases}
\]
\end{theorem}

We now   investigate the properties of the clustering coefficient. As described previously, the post-selection estimator of the weighted clustering coefficient of $i \in \hat{\mathcal{S}}_n$ is given by
\[
\hat{C}_{n,i} = \frac{1}{(\hat{q}_n-1)(\hat{q}_n-2)}\sum\limits_{ (j,k) \in \hat{\mathcal{N}}_n(i)} f(\hat{\beta}_j,\hat{\beta}_k),
\]
where $\hat{\mathcal{N}}_n(i)=\{(j, k)\in \hat{\mathcal{S}}_n \times \hat{\mathcal{S}}_n| j \ne k \ne i\}$ and $\hat{q}_n=|\hat{\mathcal{S}}_n|$. The population version is given by
\[
  C_{n,i} = \frac{1}{(\hat{q}-1)(\hat{q}-2)}\sum\limits_{ (j,k) \in \hat{\mathcal{N}}_n(i)} f(\beta_j,\beta_k).
\]
Let $\hat{\bm{C}}_n = (\hat{C}_{n,1}, \hat{C}_{n,2},\dots, \hat{C}_{n,\hat{q}_n})'$ denote the vector of estimated clustering coefficients, and let $\bm{C}_n = (C_{n,1}, C_{n,2},\dots, C_{n,\hat{q}_n})'$ denote the dimension-adjusted vector of population-clustering coefficients. Our next result is concerned with the joint asymptotic behavior of a vector of clustering coefficients.

\begin{theorem}\label{Thm-Clustercoef-Vec}
Assume that the assumptions \ref{A1}-\ref{A3} and \ref{P1}-\ref{P6} hold. For any $\bm{x} \in \real^q$, let ${\bm{x}}_n$ denote its dimension-adjusted version. Then,
\[
\lim_{n \rightarrow \infty} \bm{P}\left( \sqrt{\frac{n}{2}}(\hat{\bm{C}}_n - \bm{C}_n)\le {\bm{x}}_n \right) =\bm{P}\left( \bf{Z} \le {\bm{x}}\right),
\]
where $\bf{Z} \sim N(0,\bm{\Sigma}_C)$,
$\bm{\Sigma}_C = \sigma^2 \bm{L}_C(\mathcal{S},\bm{\beta})'\bm{V}_{\mathcal{S}}^{-1}\bm{L}_C(\mathcal{S},\bm{\beta})$ is a positive definite matrix, and $\bm{L}_C$ is a $q \times q$ matrix whose $(j,k)^{th}$ element is given by
\begin{align*}
    l_C(j,k; \mathcal{S}, \bm{\beta}) = \frac{1}{(q-1)(q-2)} \sum\limits_{(u,v) \in \mathcal{N}(j)} \frac{\partial f(\beta_u,\beta_v)}{\partial \beta_k}= \frac{1}{(q-1)(q-2)}
    \begin{cases}
    0 & j=k \\
    \sum\limits_{(k,u) \in \mathcal{N}(j)} \frac{\partial f(\beta_k,\beta_u)}{\partial \beta_k} & j \neq k,
    \end{cases}
\end{align*}
where $\mathcal{N}(i) = \{ (j,k) \in \mathcal{S} \times \mathcal{S} | j \neq k \neq i\}$.
\end{theorem}

Our next main result concerns the joint limiting distribution of clustering coefficients when the weights are a function of the transformed partial correlation coefficients $\rho_i$.  Specifically, let $\hat{\mathcal{C}}_n(\bm{\rho})$   denote the vector of clustering coefficients obtained using partial correlation weights, and let $\mathcal{C}_n(\bm{\rho})$ denote the corresponding dimension-adjusted population vector of clustering coefficients. 

\begin{theorem}{\label{CA_CCPC}}
Assume that the assumptions \ref{A1}-\ref{A3} and \ref{P1}-\ref{P6} hold. For any $\bm{x} \in \real^q$, let ${\bm{x}}_n$ denote its dimension-adjusted version. Then,
\[
{\bm{P}}\left(\sqrt{\frac{n}{2}}\left(\hat{\bm{C}}_n(\bm{\rho}) - \bm{C}_n({\bm{\rho}})\right) \le {\bm{x}}_n \right)= {\bm{P}}\left( \bf{Z} \le {\bm{x}}\right),
\]
where $\bf{Z} \sim N(0,\bm{\Sigma}_C({\bm{\rho}}))$,
$\bm{\Sigma}_C(\bm{\rho})=  L_C(\mathcal{S},\bm{\rho})^{\prime} {\bm{\Delta}}_{\mathcal{S}}(\bm{\rho}) L_{C}(\mathcal{S},\bm{\rho})$, with ${\bm{\Delta}}_{\mathcal{S}}(\bm{\rho})$ being the limiting covariance matrix in Theorem \ref{conv:PSPC}, and $L_C(\bm{\rho})$ is a $q \times q$ matrix whose $(j, k)^{\text{th}}$ element is given by
\begin{align*}
l_C(j,k; \mathcal{S},\bm{\rho}) &= \frac{1}{(q-1)(q-2)} \sum\limits_{(u,v) \in \mathcal{N}(j)} \frac{\partial f(\rho_u,\rho_v)}{\partial \rho_k}
= \frac{1}{(q-1)(q-2)}
    \begin{cases}
    0 & j=k \\
    \sum\limits_{(k,u) \in \mathcal{N}(j)} \frac{\partial f(\rho_k,\rho_u)}{\partial \rho_k} & j \neq k,
    \end{cases}
\end{align*}
where $\mathcal{N}(i) = \{ (j,k) \in \mathcal{S} \times \mathcal{S} | j \neq k \neq i\}$.
\end{theorem}

\subsection{\textbf{Examples}}
In this section, we consider two   functions $f $ and describe the asymptotic results of the previous subsection. Specifically, working with the transformed partial correlation, consider $f_1:[0,1]\times [0,1] \ra \real^+$ defined by
(\ref{Choice:f}). 
A simple calculation shows that $f_1(\cdot, \cdot)$ is a semi-metric. For this choice of $f_1(\cdot, \cdot)$, the limiting covariance matrix of the degree centrality is given by $\bm{L}_D(\mathcal{S},\bm{\rho})'\bm{\Delta}_{\mathcal{S}}(\bm{\rho})\bm{L}_D(\mathcal{S},\bm{\rho})$, where the $(i,j)^{th}$ element of $\bm{L}_D(\mathcal{S},\bm{\rho})$, denoted by $l_D(j,k;\bm{\rho})$, is given by
\[
l_D(j,k;\mathcal{S},\bm{\rho}) = \frac{\partial D_i}{\partial \beta_j}=
\begin{cases}
-\frac{\sqrt{2}\rho_j}{\sqrt{1-\rho_j^2}} & \text{ if } i \neq j \\
-\frac{\sqrt{2}(q-1)\rho_i}{\sqrt{1-\rho_i^2}} & \text{ if } i = j.
\end{cases}
\]
The limiting covariance matrix of the cluster coefficient is given by $\bm{L}_C(\mathcal{S},\bm{\rho})'\bm{\Delta}_{\mathcal{S}}(\bm{\rho})\bm{L}_C(\mathcal{S},\bm{\rho})$, where the $(i,j)^{th}$ element of $\bm{L}_C(\mathcal{S},\bm{\rho})$, denoted by $l_C(i,j;\mathcal{S},\bm{\rho})$, is given by
\[
l_C(i,j;\mathcal{S},\bm{\rho}) = \frac{\partial C_i}{\partial \beta_j}=
\begin{cases}
-\frac{\sqrt{2}\rho_j}{(q-1)\sqrt{1-\rho_j^2}} & \text{ if } i \neq j \\
0 & \text{ if } i = j.
\end{cases}
\]
Another interesting choice of $f(\cdot, \cdot)$ is given by  $f_2(x, y)= \sqrt{x^2+y^2}$.
Simple algebra shows that
\[
f_1^2(x,y)=4\left(1 + \sqrt{1 - f_2^2(x,y) - \frac{1}{2}(f_2^2(x^2,y^2)-f_2^4(x,y))}\right)-2f_2(x,y).
\]
However, this expression does not help in deriving the covariance matrix of NWM from that given by the function $f_1(x,y)$. It can be seen that the limiting covariance of the degree centrality is $\bm{L}_D(\mathcal{S},\bm{\rho})'\bm{\Delta}_{\mathcal{S}}(\bm{\rho})\bm{L}_D(\mathcal{S},\bm{\rho})$, where the $(i,j)^{th}$ element of $\bm{L}_D(\mathcal{S},\bm{\rho})$, denoted by $l_D(j,k;\mathcal{S},\bm{\rho})$, is given by
\[
l_D(j,k;\mathcal{S},\bm{\rho}) = \frac{\partial D_i}{\partial \beta_j}=
\begin{cases}
\frac{\rho_j}{\sqrt{\rho_i^2 +\rho_j^2}} & \text{ if } i \neq j \\
\sum\limits_{k \in \mathcal{S}(i)}\frac{\rho_i}{\sqrt{\rho_k^2+\rho_i^2}}  & \text{ if } i = j.
\end{cases}
\]
The limiting covariance matrix of the cluster coefficient is given by $\bm{L}_C(\bm{\rho})'\bm{\Delta}_{\mathcal{S}}(\bm{\rho})\bm{L}_C(\bm{\rho})$, where the $(i,j)^{th}$ element of $\bm{L}_C(\bm{\rho})$, denoted by $l_C(i,j;\bm{\rho})$, is given by
\[
l_C(i,j;\bm{\rho}) = \frac{\partial C_i}{\partial \beta_j}=\frac{1}{(q-1)(q-2)}
\begin{cases}
\sum\limits_{\substack{u \in \mathcal{S}(i) \\ u \neq j}}\frac{\rho_j}{\sqrt{\rho_j^2+\rho_u^2}} & \text{ if } i \neq j \\
0 & \text{ if } i = j.
\end{cases}
\]

\subsection{\textbf{Estimating the Limiting Covariance Matrix of NWM}}
We now turn to estimating the limiting covariance matrix in Theorems \ref{Thm-Degree-Vec} through   \ref{Thm-Clustercoef-Vec}. Below, we show consistency of $\hat{\bm{\Delta}}_{\mathcal{S}}(\bm{\rho})$, which is obtained by replacing the parameter values in the formula for ${\bm{\Delta}}_{\mathcal{S}}(\bm{\rho})$ by the post-selection estimators. 

\begin{theorem}{\label{conv:DA_pcov}}
Assume that the assumptions \ref{A1}-\ref{A3} and \ref{P1}-\ref{P6} hold. Then the following hold:\\
{\bf[1]} $\hat{\bm{\Delta}}_{\mathcal{S}}(\hat{\bm{\rho}} )$ converges to $\bm{\Delta}_{\mathcal{S}}(\bm{\rho})$ in probability as $n \rightarrow \infty$.\\
{\bf[2]} Let $\bm{\Delta}_{\hat{\mathcal{S}}_n}(\bm{\rho})$ denote the dimension-adjusted matrix of $\bm{\Delta}_{\mathcal{S}}(\bm{\rho})$. Then, $ \hat{\bm{\Delta}}_{\hat{\mathcal{S}}_n}(\hat{\bm{\rho}})$ converges to $\bm{\Delta}_{\hat{\mathcal{S}}_n}(\bm{\rho})$ in probability as $n \rightarrow \infty$.
\end{theorem}

\section{\textbf{Detection of the Clusters}}\label{Clustering}

In this section, we introduce a new method to detect clusters of covariates that have similar association (to be made precise below) with the response variable. The proposed method uses the estimates of NWMs of the graph and their joint asymptotic behavior.

\subsection{\textbf{Shortcomings of Unsupervised grouping Methods}}

Before introducing our grouping algorithm, we discuss the limitations of unsupervised clustering methods. Recall that our objective is to detect clusters of covariates with respect to the response variable. 
Hence, it is important to account for the response variable when grouping variables. However, it is well-known that the difference between supervised learning and  unsupervised learning (as in supervised and unsupervised clustering) is the availability of the output, which is a response variable in a regression setting.  Unsupervised clustering methods -- such as K-mean clustering or spectral clustering, adapted to our context -- use distances and/or similarity measures between covariates, and detects clusters of covariates based on them. However, unlike clustering observations, clustering covariates raises two issues: (i) how the distance between two covariates is defined, and (ii) how this distance measure incorporates the response variable. Even if we can find a metric to define the distance between two covariates, it is still difficult to account for the response variable. Hence, an unsupervised clustering method cannot give clusters that take into account the association between the response variable and covariates. On the other hand, the weighted implicit network uses the regression coefficient to define the weights; that is, the network already incorporates the association between each covariate and the response variable into the weight between two covariates. Hence, if an appropriate clustering algorithm is established, the weighted implicit network can address this challenge. The next subsection introduces our clustering (grouping) algorithm, which detects the clusters of covariates with respect to the response variable.

\subsection{\textbf{Definitions of Clusters}}
We begin by introducing the definition of clusters of covariates in $\mathcal{S}$ in terms of degree centrality. Clusters based on the clustering coefficients can be defined in a similar manner. Let $K$ denote the number of clusters and $\tau$ denote the tuning parameter that determines the size of clusters. There is no statistically rigorous method to choose an appropriate number of clusters, but one common way is to use Intra-Cluster Correlation Coefficient (ICC), which measures the similarity of clustered data by accounting for the within-cluster and between-cluster variances. In Section \ref{Numerical1}, we provide a numerical study to identify the number of groups using the intra-cluster correlation coefficient. Additionally, let $D_1^* = \max\limits_{i \in \mathcal{S}} D_i$. Let $v_1^*$ denote an index for which the maximum is achieved.   Then, the first cluster, which is denoted by $\mathcal{C}_{\tau}^{(1)}$, is given by
\[
    \mathcal{C}_{\tau}^{(1)} = \{ j | |D_1^* - D_j| \le \tau \text{ for } j \in \mathcal{S} \}.
\]
Now let $D_2^* = \max\limits_{i \in \mathcal{S} - \mathcal{C}_{\tau}^{(1)}} D_i$, and let $v_2^*$ denote an index for which the maximum is achieved. The second cluster, which is denoted by $\mathcal{C}_{\tau}^{(2)}$, is given by
\[
    \mathcal{C}_{\tau}^{(2)} = \{ j | |D_2^* - D_j| \le \tau \text{ for } j \in \mathcal{S} - \mathcal{C}_{\tau}^{(1)} \}.
\]
In the same manner, the remaining $K-2$ clusters can be defined with the corresponding maxima 
$D_i^*$ and maximizers $v_i^*$. However, the underlying issue is that we do not know $\mathcal{S}$, the true active predictor set. Hence, as described in Section \ref{Data-Split}, we split the data randomly into two parts and obtain $\hat{\mathcal{S}}_n$ from the first part of the data by using a consistent regularization method. We estimate the network-wide metrics for all covariates in $\hat{\mathcal{S}}_n$ from $\mathcal{D}_{2n}$. Based on $\hat{\mathcal{S}}_n$, we now define the clusters in the population. 
Now we let $D_{n,1}^* = \max\limits_{i \in \hat{\mathcal{S}}_n} D_i$; also let $v_{n,1}^*$ denote an index where this maximum is achieved. Then, the first cluster in the population, which is denoted by $\mathcal{C}_{n,\tau}^{(1)}$, is given by
\[
    \mathcal{C}_{n,\tau}^{(1)} = \{ j | |D_{n,1}^* - D_j| \le \tau \text{ for } j \in \mathcal{S} \}.
\]
Now let $D_{n,2}^* = \max\limits_{i \in \hat{\mathcal{S}}_n - \mathcal{C}_{n,\tau}^{(1)}} D_i$. The second cluster, which is denoted by $\mathcal{C}_{n,\tau}^{(2)}$, is given by
\[
    \mathcal{C}_{n,\tau}^{(2)} = \{ j | |D_{n,2}^* - D_j| \le \tau \text{ for } j \in \hat{\mathcal{S}}_n - \mathcal{C}_{n,\tau}^{(1)} \}.
\]
In the same manner, the rest of $K-2$ clusters based on $\hat{\mathcal{S}}_n$ can be defined. We denote the corresponding maxima by $D_{n, i}^*$ and the corresponding maximizers by  $v_{n, i}^*$. 

\subsection{\textbf{Sequential testing-based clustering}}\label{seqclustering}
The basic idea of the method is as follows. First we set the number of clusters in the feature set by thresholding the NWM. As an example based on weighted degree centrality, given a vertex $i$ and $\tau >0$  we set  all covariates $j$  whose degree centralities satisfy $|D_i-D_j| < \tau$ to belong to the same cluster. This defines a cluster in the population. Thus, the number of clusters in the population is determined by the thresholding parameter $\tau$. Now, to identify the clusters in the data, we perform several hypothesis tests 
$H_{ij}: |D_i - D_j| \le \tau$ for $1 \le j \le q$ using a student-type statistic. Specifically, set
\[
t_{ij,n} = \frac{|\hat{D}_i - \hat{D}_j|-\tau}{\hat{\sigma}_P},
\]
where $\hat{\sigma}_P$ is an estimate of the standard deviation obtained from Theorem \ref{Thm-Degree-Vec}. Hence, it follows from Theorem \ref{Thm-Degree-Vec} that as $n \ra \ff$, $t_{ij,n}$ converges in distribution to a standard normal random variable. Based on the asymptotic distribution, if we reject $H_{ij}$ we assign $i$ and $j$  to different clusters; otherwise, under
non-rejection, we assign them to the same cluster. 

Operationally, we start with a vertex with the largest (smallest) network-wide metric. We seek to find all those vertices that are within $\tau$ units of the maximum (minimum). It is a calibrator, and it controls the number of vertices in the cluster. Instead of $\tau$, we can alternatively use a desired type I error of the hypothesis test, which is denoted by $\alpha$, to control the size of the cluster.
The sequential testing based clustering algorithm can be described as follows:

\begin{algorithm}[H]
\caption{Sequential testing-based clustering using degree centrality}
\label{Alg1}
\begin{algorithmic}[1]
\item[]
\noindent {\textbf{Step 1}}: Choose the desired number of clusters $K$, and set $k=0$, $\hat{\mathcal{S}}_n^{(0)}=\emptyset$.
\newline
\noindent {\textbf{Step 2}}: Set ${\mathcal{C}}^{(k)}_{n,\tau,\alpha}=\hat{\mathcal{S}}_n-\hat{\mathcal{S}}_n^{(k)}$.
\newline
\noindent {\textbf{Step 3}}: Let $v^*= \mbox{argmax}_{i \in {\mathcal{C}}^{(k)}_{n,\tau,\alpha}} {D}_{i}$. 
\newline
\noindent {\textbf{Step 4}}: Test the hypothesis $H_{v^*j}: |D_{v^*}-D_j| \le \tau$ at level $\alpha$ using the statistic $t_{v^*j,n}$ for $ j \in {\mathcal{C}}^{(k)}_{n,\tau,\alpha}$. 
\newline
\noindent {\textbf{Step 5}}: Increment $k$ by 1, i.e., $ k\ra k+1$. Define the $k^{\mbox{th}}$ cluster as follows:
$$\hat{\mathcal{C}}^{(k+1)}_{n,\tau,\alpha} = \{ j \in \hat{\mathcal{C}}^{(k)}_{n,\tau,\alpha} | H_{v^*j} ~ \mbox{is not rejected at level} ~ \alpha \}.$$
\noindent {\textbf{Step 6}}: If $k < K$ then set $\hat{\mathcal{S}}_n^{(k)}=\hat{\mathcal{C}}^{(k+1)}_{n,\tau,\alpha}$ and go to Step 2.  Else stop.
\end{algorithmic}
\end{algorithm}
In the above algorithm, one can replace  degree centrality by clustering coefficient. Our next result establishes that for all these choices of NWM, the sequential testing method consistently estimates all the true clusters. Before we state the main result, we introduce one more notation. Let $\hat{\mathcal{C}}_{n,\tau,\alpha}^{(k)}$ denote the $k^{th}$ cluster estimated from the above algorithm using the estimated active predictor set ${\hat{\mathcal{S}}}_n$. We also denote by 
$\hat{\mathcal{C}}_{\tau, \alpha}^{(k)}$ the $k^{th}$ cluster estimated using the above algorithm by the oracle; that is, $\hat{\mathcal{C}}_{\tau, \alpha}^{(k)}$ is estimated using the above algorithm, replacing ${\hat{\mathcal{S}}}_n$ by ${{\mathcal{S}}}$. Note that $\hat{\mathcal{C}}_{\tau, \alpha}^{(k)}$ is a random set, and consists of those estimated regression coefficients whose NWM are ``close'' based on the threshold $\tau$.  Finally, let $\mathcal{C}_{\tau}^{(k)}$ denote the $k^{th}$ true cluster in the population.

\begin{theorem}\label{Thm-Consistency-Seq} (\textbf{Consistency of clustering under the sequential testing method}) Assume that the assumptions \ref{A1}-\ref{A3} and \ref{P1}-\ref{P6} hold. Let $\tau >0$ be fixed. Let $K=K(\tau)$ be the number of clusters and $\mathcal{C}_{\tau}^{(k)}$ and $\hat{\mathcal{C}}_{n,\tau,\alpha_n}^{(k)}$ denote the $k^{th}$ true cluster and $k^{th}$ estimated cluster, where $k= 1,\dots, K$. Assume that 
$\alpha_n \ra 0$.  Then, 
\[
\bm{P}(\hat{\mathcal{C}}_{n,\tau,\alpha_n}^{(k)}= \mathcal{C}_{\tau}^{(k)}) \rightarrow 1 \mbox{ \ \ \ as $n \rightarrow \infty$},
\]
for $k = 1,\dots,K$. That is, the probability of detecting all the $K$ clusters converges to one.
\end{theorem}

\section{\textbf{Numerical Studies}}\label{NS}
In this section, we describe our simulations and provide the results. We   show how our sequential clustering algorithm outperforms unsupervised clustering algorithms; we also show  how well our proposed clustering algorithm can identify clusters in high-dimensional settings. Besides these numerical studies, we present additional simulations studies of computational time to estimate $\bm{\Delta}_{\mathcal{S}}(\bm{\rho})$ in Theorem \ref{conv:PSPC}, and regarding the asymptotic behavior of NWMs. Before we present the numerical study results, we begin with a discussion about the computational complexity in the covariance formula in Theorem \ref{conv:PSPC}.

\noindent {\bf{Computational Complexity}:} Using the formula in Theorem \ref{conv:PSPC}, the above estimate of $\bm{\Delta}_{\mathcal{S}}(\bm{\rho})$ involves the computation of precision matrices and Kronecker products along with derivatives of these functions. Each Kronecker product involves at least $\hat{q}_n^4$ computations; estimation of $\bm{\Delta}_{\mathcal{S}}(\bm{\rho})$ requires a product of two Kronecker products of matrices, as well as a couple of products of $\hat{q}_n \times \hat{q}_n$ matrices. The process involves at least $\hat{q}_n^{10}$ computations. For large $\hat{q}_n$, this can be computationally expensive. To facilitate  computation  we propose to use the following \emph{Model-Estimated Post-Selection Resampling} (MEPSRS) to estimate the limiting covariance matrices of the centralities.
MEPSRS first  splits the data into two parts, ${\mathcal{D}}_{1n}$ and ${\mathcal{D}}_{2n}$.
We use ${\mathcal{D}}_{1n}$ to estimate the active predictor set, viz. $\hat{\mathcal{S}}_n$, and resample ${\mathcal{Y}}({\mathcal{D}}_{2n})$ $m$ times. Using the resampled data, we fit the model to the feature set $\hat{\mathcal{S}}_{n}$ and compute the weighted degree centralities based on bootstrap estimates of the transformed partial correlation coefficients. Let ${\bm{\rho}}_{\hat{\mathcal{S}}_{n}}^{(j)}$
denote an estimated vector of transformed partial correlation coefficients of  dimension 
$\hat{q}_n$ using the resampled data, and let ${\bm{D}}_{n, j}^*({\bm{\rho}}_{\hat{\mathcal{S}}_{n}}^{(j)})$ denote the corresponding vector of degree centralities. Then the bootstrap estimate of the covariance matrix of the vector of degree centralities, denoted by $\hat{\bm{\Sigma}}_D^B$, is given by
\[
\hat{\bm{\Sigma}}_D^B = \frac{1}{B} \sum\limits_{j=1}^B ({\bm{D}}_{n, j}^*({\bm{\rho}}_{\hat{\mathcal{S}}_{n}}^{(j)}) - \bm{\mu}_D^B)({\bm{D}}_{n, j}^*({\bm{\rho}}_{\hat{\mathcal{S}}_{n}}^{(j)}) - \bm{\mu}_D^B)', 
\]
where $\bm{\mu}_D^B =  \frac{1}{B} \sum\limits_{j=1}^B {\bm{D}}_{n, j}^*({\bm{\rho}}_{\hat{\mathcal{S}}_{n}}^{(j)})$.
The above algorithm is described succinctly  as follows:
\begin{algorithm}[H]
\caption{Bootstrap procedure}
\label{Alg2}
\begin{algorithmic}[1]
\item[]
\noindent {\textbf{Step 1}}: With $\hat{q}_n$ selected covariates from variable selection, let $(y_i,\bm{x}_i^{\hat{\mathcal{S}}_n})$ denote the row vector of the response variable and selected covariates for $i = 1,\dots,n/2$.
\newline
\noindent {\textbf{Step 2}}: Resample $(y_i,\bm{x}_i^{\hat{\mathcal{S}}_n})$ of size $n/2$ with replacement.
\newline
\noindent {\textbf{Step 3}}: Using these resampled samples, compute the NWM for each covariate.
\newline
\noindent {\textbf{Step 4}}: Repeat Steps 1-3 $B$ times.
\end{algorithmic}
\end{algorithm}

\subsection{\textbf{Multiple splits of the data}}
In real data analysis,  when the number of covariates $p$ is too large, the variable selection method and data-splitting method may select more variables than the true model. 
In this case, different data splits may yield different sets of active covariates, which will distort our clustering results. Hence, instead of doing one split we perform multiple splits of the data. This is an extension of the idea of \cite{MB10} to post-selection  clustering problems. Then, we combine results from multiple splits to obtain a cluster. One possible way to combine multiple results is to count the number of appearances of each covariate in multiple splits and assign each covariate to a cluster based on a pre-defined threshold. For example, if the number of multiple splits is 20 and the threshold is 60\%, we assign each covariate to a cluster where it appears more than 12 times out of 20 splits. If $X_1$ appears 14 times in cluster 1 and 6 times in cluster 2, then it is assigned to cluster 1. On the other hand, if $X_2$ appears 11 times in cluster 2 and 9 times in cluster 3, it cannot be assigned to any cluster because of the threshold. Using multiple splits and thresholding, we can reduce the probability of selecting wrong variables in our model and clusters. We adopt this idea in the simulation study and data analysis discussed below.

\begin{algorithm}[H]
\caption{Multiple-splitting algorithm}
\label{Alg3}
\begin{algorithmic}[1]
\item[]
\noindent {\textbf{Step 1}}: Choose the number of multiple splits. Let $M$ denote the desired number of clusters.
\noindent {\textbf{Step 2}}: Split the data into two part. Let $\mathcal{D}_{1n}^{(m)}$ and $\mathcal{D}_{2n}^{(m)}$ denote the two parts of data set in $m^{th}$ split for $m=1,\dots,M$.
\newline
\noindent {\textbf{Step 3}}: Obtain $K^{(m)}$, predefined number of clusters by using Algorithm \ref{Alg1}. Let $k^{(m)}$ denote the $k^{th}$ cluster from $m^{th}$ split for $k = 1,\dots,K^{(m)}$.
\newline
\noindent {\textbf{Step 4}}: Repeat Steps 2-3   $M$ times. 
\newline
\noindent {\textbf{Step 5}}: Assign each covariate to a cluster based on the number of appearances out of $M$ splits. 
\end{algorithmic}
\end{algorithm}

\subsection{\textbf{Numerical Experiments on Supervised Clustering}}\label{Numerical1}

\noindent{\textbf{Comparison of unsupervised and supervised clustering methods}}: 

In this subsection we compare our supervised clustering algorithm with two unsupervised clustering methods, K-means clustering and   spectral clustering. Note that in the perspective of detecting groups of variables, the classical unsupervised clustering methods do not exist. Hence, in this context we modify 
the classical algorithms of K-means clustering and spectral clustering to detect clusters of variables. Both unsupervised clustering algorithms require a pre-specified number of clusters, which is denoted by $k$. Below is the modified K-means clustering that we will use in numerical studies:
\begin{algorithm}[H]
\caption{Modified K-means Clustering}
\label{Alg4}
\begin{algorithmic}[1]
\item[]
\noindent {\textbf{Step 1}}: Given a $n \times p$ data matrix, where each row represents an observation of a $1 \times p$ vector, transpose the data matrix. This $p \times n$ matrix contains $p$ observations of a $1 \times n$ vector.
\newline
\noindent {\textbf{Step 2}}: Compute the distance, such as Euclidean distance, between all pairs of $p$ observations.
\newline
\noindent {\textbf{Step 3}}: Proceed with the classical K-means clustering algorithm. 
\end{algorithmic}
\end{algorithm}
The modified spectral clustering algorithm is presented below.
\begin{algorithm}[H]
\caption{Modified Spectral Clustering}
\label{Alg5}
\begin{algorithmic}[1]
\item[]
\noindent {\textbf{Step 1}}: Given a $n \times p$ data matrix, where each row represents an observation of a $1 \times p$ vector, transpose the data matrix. This $p \times n$ matrix contains $p$ observations of a $1 \times n$ vector.
\newline
\noindent {\textbf{Step 2}}: Compute the $p \times p$ similarity matrix, denoted by $\bm{M}$, whose element is the similarity measure between two covariates.
\newline
\noindent {\textbf{Step 3}}: Compute the $p \times p$ degree matrix, denoted by $\bm{G}$, which  is a diagonal matrix whose elements are degrees of covariates.
\newline
\noindent {\textbf{Step 4}}: Derive the Laplacian $\bm{L}$, which is defined as $\bm{M}-\bm{G}$, and compute the first $k$ eigenvectors $\bm{v}_1,\dots,\bm{v}_k$ of $\bm{L}$.
\newline
\noindent {\textbf{Step 5}}: Let $\bm{V} \in \mathbb{R}^{p \times k}$ be the matrix with columns  $\bm{v}_1,\dots,\bm{v}_k$.
\newline
\noindent {\textbf{Step 6}}: Apply K-means clustering to $\bm{V}$ to identify clusters among $p$ covariates.
\end{algorithmic}
\end{algorithm}

\begin{figure}[H]
    \centering
    \includegraphics[scale=.75]{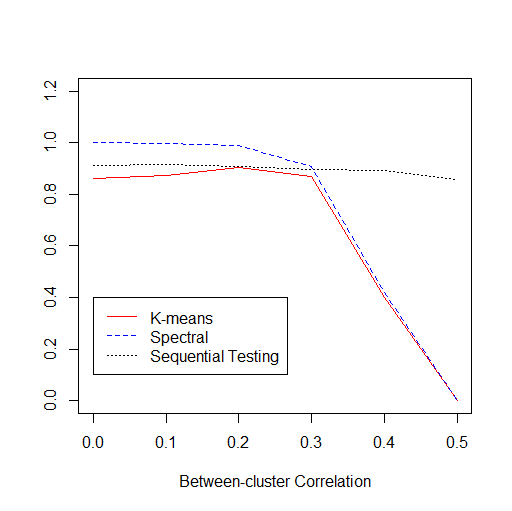}
    \caption{Comparisons among clustering algorithms}
    \label{Fig:UnsVsSup}
\end{figure}

The setting for numerical studies is as follows: the vector of covariates $\bm{X} = (X_1,X_2,\dots,X_p)'$ is generated from a 
$p-$dimensional multivariate normal distribution with mean zero and identity covariance matrix,
where $p=9$. To add ``similarity'' amongst the covariates we make three groups described below. 
We consider two types of correlations: within-cluster correlation and between-cluster correlation. In this numerical study the within-cluster correlation, denoted by $r_w$, is set to .5 for all clusters, and the between-cluster correlation, denoted by $r_b$, is set to six different values.
Given $\bm{x}_i$, the response variable $Y_i$ is generated from $\bm{x}_i\bm{\beta} + \epsilon_i$ where $\epsilon_i \sim N(0,\sigma^2)$ and $\sigma^2 = 1$, and $\bm{\beta}' = (1,-1,3,1,1,2,-1,2,-1)$. Based on this setting, there are 9 covariates and 3 true clusters with respect to $\bm{Y}$, which are $(X_1, X_4, X_5)$, $(X_2, X_7,X_9)$, and $(X_3, X_6, X_8)$. For convenience of notation, we denote the group $(X_3, X_6, X_8)$ as Cluster 1, group $(X_2,X_7, X_9)$ as Cluster 2, and group $(X_1, X_4, X_5)$ as Cluster 3. We compare three clustering algorithms, and Table  \ref{Sim:UnsVsSup}
presents the results based on 5000 simulations. The numbers in the table represent the percentage of times the true clusters are detected correctly. 
\begin{table}[H]
\centering
\begin{tabular}{|c|c|c|c|c|c|}
\hline
$r_b$ & Modified & Modified & Sequential \\
& K-means& Spectral & Clustering \\ \hline
0 & .8624 & 1 & .9104\\
.1 & .8734 & .9984 & .9148 \\
.2 & .9036 & .9908 & .9082\\
.3 & .869  & .9086  & .897 \\
.4 & .4 & .4176 & .8932\\
.5 & .0012  & .0004 & .8574 \\
\hline
\end{tabular}
\caption{Unsupervised clustering methods vs. Sequential testing method}
\label{Sim:UnsVsSup}
\end{table}
Both modified unsupervised clustering algorithms perform well when there is very low between-cluster correlation; the spectral clustering algorithm nearly identifies the correct clusters when clusters are independent or have very low correlation. However, as the between-cluster correlation increases, the performance of the modified unsupervised clustering methods
suffer correspondingly. Once there is very little difference between within-cluster correlation and between-cluster correlation, neither the modified K-means algorithm nor the modified spectral clustering algorithm can identify true clusters. On the other hand,  the sequential testing method maintains its good performance, no matter the degree of between-cluster correlation. It performs similarly when between-cluster correlation is low, and outperforms both modified unsupervised clustering methods as between-cluster correlation gets larger. Figure \ref{Fig:UnsVsSup} 
illustrates the simulation results.

Note that in the simulation study above, we assumed that the true number of clusters is known. If the number of clusters is unknown, the modified K-means clustering and spectral clustering will never identify correct clusters. On the other hand, given that the desired number of clusters is wrong, the sequential clustering algorithm can identify correct clusters. That is, if the desired number of clusters is 2, then the sequential testing clustering algorithm will identify the top two clusters that are associated with the response variable. In our example, it will give us Cluster 1 and Cluster 2.
Unsupervised clustering will be unlikely to detect the correct clusters, since that method attempts  to create two clusters using all the covariates. We performed a numerical study with the same simulation setting, but we set the desired number of clusters to 2 and compared the performance of three clustering algorithms. The results are provided in Appendix \ref{APPC}.

We also performed an extensive study of the proposed methods with agglomerative clustering studied in \cite{BRG13}. Based on the analysis (data not included), we noticed that the methods proposed here perform better than those based on canonical correlations. Additionally, the clustering obtained via the use of NWM using implicit network with correlation weights seems to improve the identification of correct clusters over the use of correlation weights alone, even in the context of unsupervised problems.

\noindent{\textbf{Choosing the number of clusters based on the Intra-Cluster Correlation Coefficient (ICC)}}: In this subsection, we demonstrate the performance of the ICC-based algorithm to choose the number of clusters. The vector of covariates $\bm{X} = (X_1,X_2,\dots,X_p)'$ is generated from a $p-$dimensional multivariate normal distribution  with mean zero and identity covariance matrix.
We consider a multiple linear regression model without an intercept, setting $p=9$ and sample size $n=100$. Given $\bm{x}_i$, the response variable $Y_i$ is generated from $\bm{x}_i\bm{\beta} + \epsilon_i$ where $\epsilon_i \sim N(0,1)$, and $\bm{\beta}' = (1, 1, 1, -1, -1,-1, 2,2,2)$. Based on this setting, there are 3 true clusters: $(X_1, X_2, X_3)$, $(X_4, X_5,X_6)$, and $(X_7, X_8, X_9)$. We implement the algorithm described in Section \ref{seqclustering} to identify the number of clusters. Table \ref{Num_Clust}  presents the results based on 5000 simulations. The numbers in the table record how many times the algorithm identified the corresponding number of clusters.

\begin{table}[H]
\centering
\begin{tabular}{|c|c|c|c|c|c|c|c|c|}
\hline
Number of Clusters & 2 &3 & 4 & 5 & 6 & 7 & 8 & 9\\ \hline
Frequency (Percentage \%) & 175 (3.5) & 4825 (96.5) & 0 (0) & 0 (0) & 0 (0)& 0 (0) & 0 (0) & 0 (0)\\ 
 \hline
\end{tabular}
\caption{ICC Algorithm choosing the number of clusters}
\label{Num_Clust}
\end{table}

The ICC-based algorithm provides promising results, identifying the correct number of clusters more than 95\% of the time.  Table \ref{Num_Clust-case1}
provides the results in the case that 
$\bm{\beta}'$ = (.4,.4,.4, -.2,-.2,-.2,.8,.8,.8).

\begin{table}[H]
\centering
\begin{tabular}{|c|c|c|c|c|c|c|c|c|}
\hline
Number of Clusters & 2 &3 & 4 & 5 & 6 & 7 & 8 & 9\\ \hline
Frequency (Percentage \%) & 329 & 4208 & 456 & 7 (.14) & 0 (0) & 0(0) & 0 (0) & 0 (0) \\ 
& (6.58) & (84.16) & (9.12) & & & & &\\
 \hline
\end{tabular}
\caption{ICC Algorithm choosing the number of clusters}
\label{Num_Clust-case1}
\end{table}

\noindent{\textbf{Clustering based on Supervised Methods}}: In this subsection, we demonstrate the behavior of our supervised clustering algorithm when $p < n$. The vector of covariates $\bm{X} = (X_1,X_2,\dots,X_p)'$ is generated from a $p-$dimensional multivariate normal distribution  with mean zero and identity covariance matrix.
We consider multiple linear regression model without an intercept, where the covariate dimension is $p=20$ and $p=50$, and sample size $n=200$. Given $\bm{x}_i$, the response variable $Y_i$ is generated from $\bm{x}_i\bm{\beta} + \epsilon_i$ where $\epsilon_i \sim N(0,1)$, and $\bm{\beta}' = (1, 1, 1, -1, -1,-1, 2,2,2,0,\dots,0)$, which contains $q=9$ nonzero regression coefficients. Based on this setting, there are 9 truly non-zero covariates and 3 true clusters, which are $(X_1, X_2, X_3)$ (Cluster 2), $(X_4, X_5,X_6)$ (Cluster 3), and $(X_7, X_8, X_9)$ (Cluster 1). 
For the variable selection, we use the \textit{two-step} procedure with SCAD introduced by \cite{FL01}
(with tuning parameters chosen by the cross-validation method), and then perform a hypothesis test. Table \ref{Smallp-seq} represents the results based on 5000 simulations. The numbers in the table represent the proportion of times the true clusters are detected correctly. In addition, $\alpha$ in the table represents the   significance level of hypothesis tests. 

\begin{table}[H]
\centering
\begin{tabular}{|ccc|cccc|}
\hline
NWM & $p$ &$\alpha$ & Cluster 1 & Cluster 2 & Cluster 3 & Combined\\ \hline
\multirow{4}{*}{Degree Centrality} & \multirow{2}{*}{20} &.05 & .958 & .946 &.957 & .938\\ 
& &.1 & .959 & .937 & .939 & .929\\ 
& \multirow{2}{*}{50} & .05 &.951 &.939 & .950 & .932\\
& & .1 & .954 & .935 & .938 & .927 \\ \hline
\multirow{4}{*}{Clustering Coefficient} & \multirow{2}{*}{20} &.05  & .944 & .958 & .983 &.944\\
& &  .1 & .931 & .953 & .982 & .931 \\
& \multirow{2}{*}{50} & .05 & .938 &.953 & .976 & .938\\
& & .1 & .954 & .935 & .938 & .931 \\ \hline
\end{tabular}
\caption{Sequential Testing Method}
\label{Smallp-seq}
\end{table}


In Table \ref{Smallp-seq}, `Cluster 1' indicates the proportion of times that true Cluster 1 is correctly detected. `Cluster 2' and   `Cluster 3' are defined similarly. `Combined' indicates that all three true clusters are correctly detected in the same simulation. From the table, we observe that the proposed clustering method performs very well for all values of $p$. In addition, both degree centrality and clustering coefficient provides similar results, so our clustering algorithm works well regardless of the type of the NWM used.

\section{\textbf{Sports Analytics: Major League Baseball Data}}\label{Sec:SA}
The data for the analysis is collected from the official \cite{MLB16} and \cite{BR16} websites. It contains the statistics of $n=291$ MLB hitters from 30 major league teams in the 2016 season. We only choose the hitters who have more than 251 plate appearances. The logic behind the number `251' is that a hitter has to have plate appearances more than 3.1 $\times$ the number of games played by his team to be officially ranked. Since each team plays 162 games in one season, the minimum number of plate appearances is 502. Since we consider that any player with more than half of the required number of plate appearances makes a substantial contribution to the team's performance of the season, we select players with more than 251 plate appearances. 

In this data set, the variable of interest is Wins Above Replacement (WAR), which measures a player's relative performance over replacing players. WAR measures how much better a player is than players who would be available to replace him. WAR is widely used to evaluate a player's performance, and it is often of interest to find the association between WAR and baseball statistics. Our goal is to detect top clusters that are associated with WAR. 
The details of the $p=34$ covariates are described in Table \ref{Data-variables}. 

\begin{table}[H]
\scriptsize
\begin{tabular}{ccc|ccc}
\hhline{======}
ID & Variable & Description & ID & Variable & Description \\ \hline
& WAR & Wins above replacement & 18 & Pitch/PA & Average number of pitches \\
& & of each player & & & per each plate appearance \\
1 & Age & A player's age in 2016 & 19 & GDP & Number of ground outs  \\
& & & & &made into double plays \\
2 & G & Number of games played by a player & 20 & HBP & Number of hit-by-pitch \\
3 & AB & Number of At-bats of a player & 21 & SH & Number of sacrifice hits\\
4 & R & Number of runs made by a player & 22 & SF & Number of sacrifice flies \\
5 & H & Number of hits & 23 & IBB & Intentional bases on balls\\
6 & X2B & Number of doubles & 24 & GO\_AO & Ratio of ground outs vs. Fly outs  \\
7 & X3B & Number of triples & 25 & GS & Number of games started \\
8 & HR & Number of home runs & 26 & INN & Number of innings played in field  \\
9 & RBI & Number of runs batted in & 27 & PO & Number of putouts \\
10 & SB & Number of stolen bases & 28 & A & Number of assists\\
11 & CS & Number of times a player  & 29 & E & Number of errors\\
& & was caught stealing bases & & &\\
12 & BB & Number of bases on balls & 30 & DP & Number of double plays  \\
& & & & & turned by a player \\
13 & SO & Number of strikeouts & 31 & Field \% & Fielding percentage of a player \\
14 & BA & Batting average throughout the season & 32 & RF.G & Range factor per game\\
& & & & & Number of outs related to the player \\
15 & OBP & On base percentage of a player & 33 & BA \_RISP & Batting average with runners\\
& & & & & in scoring position \\
16 & SLG & Slugging percentage of a player & 34 & Salary & Salary of a player (in million \$) \\
17 & OPS+ & OPS = OBP + SLG. OPS+ is adjusted &  & & \\
& & OPS with respect to the ballpark  & && \\
\hline
\end{tabular}
\caption{\small Variables in the data set}
\label{Data-variables}
\end{table}

Even though WAR already takes into account  some of the baseball statistics (especially offense-type statistics) described in Table \ref{Data-variables}, it is still worth detecting clusters of them, since it can provide   insight into groups of baseball statistics that are more highly associated with WAR. In addition, our analysis provides a way to investigate which factor one has to examine to study the details of WAR. Furthermore, by analyzing both offensive and defensive characteristics of baseball players, we take into account  both offensive and defensive performances of players to study the details of WAR. 

We consider a multiple linear regression model with sparsity. For the variable selection, we applied the multiple-splitting scheme with the two-stage variable selection procedure (SCAD and t-test). The number of splits is 20 and each split contains two sub-samples each of size 156 and 155. Based on 20 splits, we assign each variable to a cluster if it appears at least 50\% of the time. The desired number of cluster is equal to 3. 
We set $\tau=0$ for all 3 clusters.
For the hypothesis tests in the sequential testing method, we set $\alpha$, the significance level, equal to .1 with the Bonferroni adjustment.

There are in total 23 covariates that are selected more than 50\% of the time. The set of estimated active predictors, denoted by $\hat{\mathcal{S}}_n$, is given by
%
%
\begin{align*}
\hat{\mathcal{S}}_n & = \{\mbox{R, H, X2B, X3B, HR, RBI, AB, CS, OBP, Age, SO, SF,  } \\
& \qquad \mbox{IBB,SB, BB, OPS+, HBP, A, SLG, GDP, GS, E, DP }\}.
\end{align*} 


From Table \ref{Cluster-MLB}, we observe that only 16 out of 23 covariates are in the clusters using sequential testing. Variables that are selected more than 50\% of the time from the variable selection procedure -- but are not assigned more than 50\%  of the time in any cluster -- are labeled as `Not in a cluster'. In addition, the initial number of clusters is 3, but we detect 2 clusters. The role of $\tau$ and $\alpha$   make a difference in the results. One possible interpretation of the analysis is as follows: those baseball statistics in the same cluster share a  similar level of association with the WAR. Summarizing the information in these clusters, one may reduce the number of variables that are relevant to WAR and provide a simpler model with few predictors. Or one can choose one variable from each cluster and use them to fit a model with other predictors that do not belong to any cluster.

\begin{table}[H]
\centering
\begin{tabular}{|c|ccc|}
  \hline
 NWM & Cluster 1 & Cluster 2 & Not in a cluster\\ 
  \hline
Degree Centrality & R, H, X2B, X3B, HR, RBI, & AB, CS, OBP, & Age, SO, SF, IBB, \\
& SB, BB, OPS+, HBP, A &SLG, GDP  &GS, E, DP\\
Clustering Coefficient & R, H, X2B, X3B, HR, RBI, & AB, CS, OBP, & Age, SO, SF, IBB, \\
& SB, BB, OPS+ ,HBP, A & SLG, GDP  &GS, E, DP\\
 \hline
\end{tabular}
\caption{Clusters based on Network-wide Metrics}
\label{Cluster-MLB}
\end{table}

\section{\textbf{Breast Cancer Data}}\label{BC}
We next apply our methods to a regression problem involving gene expression data. The data set originates from breast cancer tissue samples  from The Cancer Genome Atlas (TCGA) project. It contains the gene expression levels of 17,814 genes from 536 patients. The response variable in this data set is the gene expression level of the \textit{BRCA1} gene that is identified to increase the risk of early onset of breast cancer. Since the \textit{BRCA1} gene is likely to interact with other genes, our goal is to identify groups of genes that have similar association with \textit{BRCA1}.

As done by \cite{K08}, we first select 3000 genes with the largest variance in their gene expression level. Then, we use the top 500 genes that have the largest absolute correlation with \textit{BRCA1} amongst the 3000 genes. Among 536 observations, we randomly select  half of the observations  for variable selection performed using SCAD and t-test. SCAD has two tuning parameters, which are usually denoted by $\lambda$ and $a$. We fix $a$ at 3.7 and choose $\lambda$ by using  cross-validation. Using the remaining half, we identify clusters. The number of splits is 20 and the threshold percentage of appearances out of 20 splits is 50\%. The number of bootstrap samples is 500 and the desired number of clusters is 3.
We use .05 as a significance level for each hypothesis test in the proposed sequential testing method, and apply Bonferroni's correction to adjust the significance level of each test in multiple hypothesis tests. We also choose $\tau$, the tuning parameter that controls the size of clusters, to be equal to 0 so that we assign genes to the same cluster only if their network-wide metrics are not significantly different. The set of estimated active predictors $\hat{\mathcal{S}}_n$, is given by
\begin{align*}
    \hat{\mathcal{S}}_n & = \{ \mbox{ABCAB, ABHD2,  ACTO4, ACTR10, ACTR3B, ADCK1,}\\
    & \qquad \mbox{ADH1C, ADORA2A, AGA, AKR1CL2, ALG5} \}.
\end{align*}
%
 Table \ref{breastcancer} 
 shows the clusters detected by using our methods
 (the gene expressions are reported as their indices).
\begin{table}[H]
\centering
\begin{tabular}{|c|c|c|}
\hline
& Cluster 1 & Not in cluster \\ \hline
Degree & ADCK1, AKR1CL2,  ALG5 & ABCAB, ABHD2,  ACTO4,  ACTR10, ACTR3B \\
Centrality & & ADH1C, ADORA2A, AGA \\
Clustering & ADCK1, AKR1CL2,  ALG5 & ABCAB, ABHD2,  ACTO4,  ACTR10, ACTR3B \\
Coefficient & & ADH1C, ADORA2A, AGA \\
\hline
\end{tabular}
\caption{Sequential testing based Clustering}
\label{breastcancer}
\end{table}
\newpage
 \begin{center}
    {\Large \bf 
    Appendices:  \\ \vspace{.2in}
    Grouping predictors via network-wide metrics
    }
\end{center}
  \medskip

\date{}

\appendix
\section{\textbf{Regularity Conditions and assumptions}}\label{App:A}
In this section, we describe the assumptions that facilitate the theoretical results. The first set of assumptions concerns the function $f(\cdot,\cdot)$, which provides the weight on the edge of an implicit network. 

\noindent {\bf{Assumptions on $f(\cdot, \cdot)$}}

\begin{enumerate}[label=(A\arabic*)]
\item\label{A1} $f(x,y)=0$ if and only if $x = y$.
\item\label{A2} $\bm{L}_D(\mathcal{S}, \bm{\beta})$ and $\bm{L}_C(\mathcal{S}, \bm{\beta})$ are non-zero matrices.
\item\label{A3} $f(\cdot,\cdot)$ is twice continuously differentiable.
\end{enumerate}
\noindent Now, we describe the assumptions on the penalty function and tuning parameters in (11). Let $p_{\lambda}'(\theta)$ denote the first derivative of $p_{\lambda}(\theta)$. We consider a class of penalties that satisfies the following conditions \citep{KK12}:
\begin{enumerate}[label = (P\arabic*)]
\item\label{P1} $p_{\lambda}'(\cdot)$ is nonnegative, nonincreasing, and continous over $(0,\infty)$.
\item\label{P2} There exists $a>0$ such that $\lim_{\theta \rightarrow 0+} p_{\lambda}'(\theta) = \lambda$, $p_{\lambda}'(\theta) \ge \lambda - \theta/a$ $(0 < \theta < a \lambda)$, and $p_{\lambda}'(\theta) = 0$ $(\theta \ge a \lambda)$.
\end{enumerate}
This class contains the SCAD penalty \citep{FL01} and the MCP \citep{Z10}. Notice that for the MCP penalty \citep{Z10}, the derivative is given by
\begin{align*}
    p_{\lambda}'(\theta) = \frac{(a\lambda - \theta)_+}{a}
\end{align*}
for $a >1$ while, for the SCAD penalty  the derivative is given by
\begin{align*}
    p_{\lambda}'(|\theta|) = \lambda \left\{I(\theta \le \lambda) + \frac{(a\lambda -\theta)_+}{(a-1)\lambda}I(\theta > \lambda) \right\}
\end{align*}
for some $a >2$ and $\theta \ge 0$. In case of using SCAD, the tuning parameter $\lambda$ must be chosen such that the following additional assumptions hold:
\begin{enumerate}[label=(P\arabic*)]
\setcounter{enumi}{2}
\item\label{P3} $p_{\lambda}(\theta)$ has a second order continuous derivative at nonzero components of $\bm{\theta}$ and is nonnegative with $p_{\lambda}(0) =0$.
\item\label{P4} $\lim\inf_{n \rightarrow \infty} \lim\inf_{\theta \rightarrow 0+} p'_{\lambda_n}(\theta)/\lambda_n >0$.
\item\label{P5} $\lambda_n \rightarrow 0$ and $\sqrt{n}\lambda_n \rightarrow \infty$ as $n \rightarrow \infty$ .
\item\label{P6} Let $b_n = \max \{ |p_{\lambda}'' (|\beta_j|)| : \beta_j \neq 0 \}$. Then, $b_n \rightarrow 0$ as $n \rightarrow \infty$.
\end{enumerate}

\section{\textbf{Proofs}}\label{App:B}
In this section, we provide detailed proofs of results presented in the previous sections. We begin with the proof of Proposition \ref{Prop-vector}. 

\subsection*{\textbf{Proof of Proposition \ref{Prop-vector}}}
Using multivariate Taylor expansion \citep{A69}, one can express the vector of degree centralities as follows:
\begin{align}
\sqrt{n}(\hat{\bm{D}}_n - \bm{D}) &= \sqrt{n}(\hat{\bm{\beta}}_{\mathcal{S}} - \bm{\beta}_{\mathcal{S}})'\hat{\bm{L}}_D(\mathcal{S},\hat{\bm{\beta}}) + \sqrt{n}(\hat{\bm{\beta}}_{\mathcal{S}} - \bm{\beta}_{\mathcal{S}})' \hat{\bm{H}}_n^{(D)}(\mathcal{S},\hat{\bm{\beta}})(\hat{\bm{\beta}}_{\mathcal{S}} - \bm{\beta}_{\mathcal{S}})\\
&= J_n(1) + J_n(2) + \bm{B} \nonumber
\end{align}

where
\begin{align}
    J_n(1) &= \sqrt{n}(\hat{\bm{\beta}}_{\mathcal{S}} - \bm{\beta}_{\mathcal{S}}+ \bm{c})' \left[\hat{\bm{L}}_D (\mathcal{S},\hat{\bm{\beta}}) -\hat{\bm{H}}_n^{(D)'}(\mathcal{S},\hat{\bm{\beta}})\bm{c}\right] \\
    J_n(2) &= \sqrt{n}(\hat{\bm{\beta}}_{\mathcal{S}} - \bm{\beta}_{\mathcal{S}} + \bm{c})' \hat{\bm{H}}_n^{(D)}(\mathcal{S},\hat{\bm{\beta}})(\hat{\bm{\beta}}_{\mathcal{S}} - \bm{\beta}_{\mathcal{S}})\\
    \bm{B} &= -\sqrt{n}\bm{c}'\left[\hat{\bm{L}}_D(\mathcal{S},\hat{\bm{\beta}})-\hat{\bm{H}}_n^{(D)'}(\mathcal{S},\hat{\bm{\beta}})\bm{c}\right].
\end{align}
We will first show that $J_n(1)$ converges to a multivariate normal distribution with mean $\bm{0}$ and covariance matrix $\bm{\Sigma}_D$, and $J_n(2)$ converges to 0 in probability. To this end, observe that $\hat{\bm{H}}_n^{(D)}(\mathcal{S},\hat{\bm{\beta}})$ converges in probability to $\bm{H}^{(D)}(\mathcal{S},\bm{\beta})$ by the continuity of $f(\cdot,\cdot)$ and using the continuous mapping theorem. Also, by Lemma 1\ref{lem:FL}, $\sqrt{n}(\hat{\bm{\beta}}_{\mathcal{S}}-\bm{\beta}_{\mathcal{S}}+\bm{c})$ converges to a multivariate normal distribution with mean vector $\bm{0}$ and covariance matrix $\sigma^2(\bm{V}_{\mathcal{S}}+\Sigma)^{-1} \bm{V}_{\mathcal{S}}(\bm{V}_{\mathcal{S}}+\Sigma)^{-1}$. In addition, $\hat{\bm{L}}_D(\mathcal{S},\hat{\bm{\beta}})$ converges to $\bm{L}_D(\mathcal{S},\bm{\beta})$ by the continuity of $f(\cdot,\cdot)$ and using the continuous mapping theorem. Hence, by multivariate Slutsky's Theorem \citep{V00} 
it follows that as $ n \ra \ff$ $J_n(1) \xrightarrow{d} N({\bm{0}}, \bm{\Sigma}_D)$, where $\bm{\Sigma}_D = \sigma^2 [\bm{L}_D(\mathcal{S},\bm{\beta}) + \bm{H}^{(D)'}(\mathcal{S}, \bm{\beta})\bm{c}]' (\bm{V}_{\mathcal{S}} + \Sigma)^{-1} \bm{V}_{\mathcal{S}} (\bm{V}_{\mathcal{S}} + \Sigma)^{-1}[\bm{L}_D(\mathcal{S},\bm{\beta}) + \bm{H}^{(D)'}(\mathcal{S}, \bm{\beta})\bm{c}].$
Next turning to $J_n(2)$: by (ii) in Lemma 1 and the continuous mapping theorem using the continuity of $f(\cdot,\cdot)$, consistency of $\hat{\bm{\beta}}_{\mathcal{S}}$ and convergence of $\hat{\bm{H}}_n^{(D)}(\mathcal{S},\hat{\bm{\beta}})$ to $\bm{H}^{(D)}(\mathcal{S},\bm{\beta})$ in probability, it follows that $J_n(2)$ converges to 0 in probability.  $\blacksquare$

\subsection*{\textbf{Proof of Proposition \ref{Prop-vector2} }}
The proof of this proposition proceeds in the same manner as that of Proposition 1\ref{Prop-vector}. We only provide it here to make the paper self-contained. By multivariate Taylor expansion \citep{A69}, one can express the vector of clustering coefficients as follows:
\begin{align}
\sqrt{n}(\hat{\bm{C}}_n - \bm{C}) &= \sqrt{n}(\hat{\bm{\beta}}_{\mathcal{S}} - \bm{\beta}_{\mathcal{S}})'\hat{\bm{L}}_C(\mathcal{S},\bm{\beta})+ \sqrt{n}(\hat{\bm{\beta}}_{\mathcal{S}} - \bm{\beta}_{\mathcal{S}})' \hat{\bm{H}}_n^{(C)}(\mathcal{S},\bm{\beta})(\hat{\bm{\beta}}_{\mathcal{S}} - \bm{\beta}_{\mathcal{S}})\\
&= J_n(1) + J_n(2) + \bm{B} \nonumber
\end{align}
where
\begin{align}
    J_n(1) &= \sqrt{n}(\hat{\bm{\beta}}_{\mathcal{S}} - \bm{\beta}_{\mathcal{S}}+ \bm{c})'\left[\hat{\bm{L}}_{C,\mathcal{S}}(\hat{\bm{\beta}})-\hat{\bm{H}}_n^{(C)'}(\mathcal{S},\hat{\bm{\beta}})\bm{c}\right]\\
    J_n(2) &= \sqrt{n}(\hat{\bm{\beta}}_{\mathcal{S}} - \bm{\beta}_{\mathcal{S}} + \bm{c})' \hat{\bm{H}}_n^{(C)}(\mathcal{S},\hat{\bm{\beta}})(\hat{\bm{\beta}}_{\mathcal{S}} - \bm{\beta}_{\mathcal{S}})\\
    \bm{B} &= -\sqrt{n}\bm{c}'\left[\hat{\bm{L}}_C(\mathcal{S},\bm{\beta})-\hat{\bm{H}}_n^{(C)'}(\mathcal{S},\hat{\bm{\beta}})\bm{c}\right].
\end{align}
We will first show that $J_n(1)$ converges to a multivariate normal distribution with mean $\bm{0}$ and covariance matrix $\bm{\Sigma}_C$, and $J_n(2)$ converges to 0 in probability. To this end, observe that $\hat{\bm{H}}_n^{(C)}(\mathcal{S},\hat{\bm{\beta}})$ converges in probability to $\bm{H}^{(C)}(\mathcal{S},\bm{\beta})$ by the continuity of $f(\cdot,\cdot)$ and using the continuous mapping theorem. Also, by Lemma 1\ref{lem:FL}, $\sqrt{n}(\hat{\bm{\beta}}_{\mathcal{S}}-\bm{\beta}_{\mathcal{S}}+\bm{c})$ converges to a multivariate normal distribution with mean vector $\bm{0}$ and covariance matrix $\sigma^2(\bm{V}_{\mathcal{S}}+\Sigma)^{-1} \bm{V}_{\mathcal{S}}(\bm{V}_{\mathcal{S}}+\Sigma)^{-1}$. In addition, $\hat{\bm{L}}_C(\mathcal{S},\hat{\bm{\beta}})$ converges to $\bm{L}_C(\mathcal{S},\bm{\beta})$ by the continuous mapping theorem using the continuity of $f(\cdot,\cdot)$. Hence, by multivariate Slutsky's Theorem \citep{V00} it follows that as $ n \ra \ff$ $J_n(1) \xrightarrow{d} N({\bm{0}}, \bm{\Sigma}_C)$, where $\bm{\Sigma}_C = \sigma^2 [\bm{L}_C(\mathcal{S},\bm{\beta}) + \bm{H}^{(C)'}(\mathcal{S}, \bm{\beta})\bm{c}]' (\bm{V}_{\mathcal{S}} + \Sigma)^{-1} \bm{V}_{\mathcal{S}} (\bm{V}_{\mathcal{S}} + \Sigma)^{-1}[\bm{L}_C(\mathcal{S},\bm{\beta}) + \bm{H}^{(C)'}(\mathcal{S}, \bm{\beta})\bm{c}].$
Next turning to $J_n(2)$: by (ii) in Lemma 1\ref{lem:FL} and by the continuous mapping theorem using the continuity of $f(\cdot,\cdot)$, consistency of $\hat{\bm{\beta}}_{\mathcal{S}}$ and convergence of $\hat{\bm{H}}_n^{(C)}(\mathcal{S},\hat{\bm{\beta}})$ to $\bm{H}^{(C)}(\mathcal{S},\bm{\beta})$ in probability, it follows that $J_n(2)$ converges to 0 in probability.  $\blacksquare$

\subsection*{\textbf{Proof of Theorem \ref{Thm-Asym-LSBeta}}}
\textbf{Proof of} {\bf[1]} : Recall that we use a \emph{two-step} procedure for the variable selection; one is the regularization method and the other is the t-statistic for testing $H: \beta_j =0$ vs. $K: \beta_j \neq 0$ for  $j \in \hat{\mathcal{S}}_n^{(1)}$, where $\hat{\mathcal{S}}_n^{(1)}$ denotes the estimated active predictor set obtained using regularization. Also, $\lim_{n \rightarrow \infty} \bm{P}(\hat{\mathcal{S}}_n^{(1)} = \mathcal{S}) =1.$ And let $\hat{\mathcal{S}}_n^{(2)}$ denote the estimated active predictor set obtained from the hypothesis test using $\mathcal{D}_{1n}$; that is
\begin{align*}
    \hat{\mathcal{S}}_n^{(2)} = \{ j \in \hat{\mathcal{S}}_n^{(1)} | |t_{n,j} > c_{\alpha_n} \},
\end{align*}
where $t_{n,j}$ is the t-statistic for testing $H : \beta_j =0$ vs. $K : \beta_j \neq 0$. Also if $\alpha_n \rightarrow 0$ as $n \rightarrow \infty$, $\lim_{n \rightarrow \infty} \bm{P}(\hat{\mathcal{S}}_n^{(2)} = \mathcal{S}) =1.$ To see this, note that
\begin{align*}
    \bm{P}(\hat{\mathcal{S}}_n^{(2)} \ne \mathcal{S}) \le \bm{P}(\hat{\mathcal{S}}_n^{(2)} \subsetneq \mathcal{S}) + \bm{P}(\mathcal{S}\subsetneq  \hat{\mathcal{S}}_n^{(1)}).
\end{align*}
We first notice that 
\begin{align*}
\bm{P}(\hat{\mathcal{S}}_n^{(2)} \subsetneq \mathcal{S}) \le \bm{P}(\hat{\mathcal{S}}_n^{(1)} \subsetneq \mathcal{S}) + \bm{P}([\hat{\mathcal{S}}_n^{(1)}=\mathcal{S}] \cap[\hat{\mathcal{S}}_n^{(2)} \neq \mathcal{S}]).
\end{align*}
By the consistency of the regularization method, the first term on the RHS of the above equation converges to 0. As for the second term, notice that
\begin{align*}
\bm{P}(\hat{\mathcal{S}}_n^{(1)}\cap\hat{\mathcal{S}}_n^{(2)} \neq \mathcal{S}) \le \bm{P}(t_{n,j}
>c_{\alpha_n}\quad \mbox{for some $j \in \mathcal{S}$}),
\end{align*}
and the RHS converges to 0 as $n \rightarrow \infty$ by the consistency of the $t-$test.
Finally,
\begin{align*}
    \bm{P}(\hat{\mathcal{S}}_n = \mathcal{S}) &= \bm{P}(\hat{\mathcal{S}}_n^{(1)} \bigcap \hat{\mathcal{S}}_n^{(2)} = \mathcal{S}) \\
    &= \bm{P}(\hat{\mathcal{S}}_n^{(1)} = \mathcal{S} \text{ and } \hat{\mathcal{S}}_n^{(2)} = \mathcal{S})\\
    &= 1 - \bm{P}(\hat{\mathcal{S}}_n^{(1)} \neq \mathcal{S} \text{ or } \hat{\mathcal{S}}_n^{(2)} \neq \mathcal{S})\\
    &\ge 1 - \bm{P}(\hat{\mathcal{S}}_n^{(1)} \neq \mathcal{S}) - \bm{P}(\hat{\mathcal{S}}_n^{(2)} \neq \mathcal{S})\\
    &\rightarrow 1 \quad \text{as $n \rightarrow \infty$}.
\end{align*}
\textbf{Proof of} {\bf[2]} : Notice that $\hat{\mathcal{S}}_n$ is estimated from $\mathcal{D}_{1n}$ and $\hat{\bm{\beta}}_{\hat{\mathcal{S}}_n}$ is estimated from $\mathcal{D}_{2n}$. We  will use the independence of $\mathcal{D}_{1n}$ and $\mathcal{D}_{2n}$ in the proof. First notice that,
\begin{align*}
\bm{P} \left( \sqrt{\frac{n}{2}}(\hat{\bm{\beta}}_{\hat{\mathcal{S}}_n} - \bm{\beta}_{\hat{\mathcal{S}}_n}) \le \bm{x}_n \right) &= \bm{P} \left( \sqrt{\frac{n}{2}}(\hat{\bm{\beta}}_{\hat{\mathcal{S}}_n} - \bm{\beta}_{\hat{\mathcal{S}}_n}) \le \bm{x}_n , \hat{\mathcal{S}}_n = \mathcal{S} \right)\\
&+ \bm{P} \left( \sqrt{\frac{n}{2}}(\hat{\bm{\beta}}_{\hat{\mathcal{S}}_n} - \bm{\beta}_{\hat{\mathcal{S}}_n}) \le \bm{x}_n, \hat{\mathcal{S}}_n \neq \mathcal{S} \right)\\
&= J_n(1) + J_n(2).
\end{align*}
We will show that $J_n(1) \rightarrow \bm{P}( N(\bm{0}, \sigma^2 \bm{V}_{\mathcal{S}}^{-1}) \le \bm{x}_n)$ and $J_n(2) \rightarrow 0$ as $n \rightarrow \infty$. To this end, note that
\begin{align*}
J_n(1) &= \bm{P} \left( \sqrt{\frac{n}{2}}(\hat{\bm{\beta}}_{\hat{\mathcal{S}}_n} - \bm{\beta}_{\hat{\mathcal{S}}_n}) \le \bm{x}_n \bigg| \hat{\mathcal{S}}_n = \mathcal{S} \right) \bm{P}\left(\hat{\mathcal{S}}_n = \mathcal{S} \right)\\
&= \bm{P} \left( \sqrt{\frac{n}{2}}(\hat{\bm{\beta}}_{\mathcal{S}} - \bm{\beta}_{\mathcal{S}}) \le \bm{x}\right) \bm{P}\left(\hat{\mathcal{S}}_n = \mathcal{S} \right) \\
&\rightarrow \bm{P}\left(N(\bm{0},\sigma^2\bm{V}_{\mathcal{S}}^{-1}) \le \bm{x}\right), \mbox{ \ as $n \rightarrow \infty$,}
\end{align*}
where the penultimate line follows from the independence of $\mathcal{D}_{1n}$ and $\mathcal{D}_{2n}$, and the last convergence comes from the standard theory of ordinary least squares estimator \citep{A85}. Next turning to  $J_n(2)$, 
\begin{align*}
J_n(2) &= \bm{P} \left( \sqrt{\frac{n}{2}}(\hat{\bm{\beta}}_{\hat{\mathcal{S}}_n} - \bm{\beta}_{\hat{\mathcal{S}}_n}) \le \bm{x}_n \bigg| \hat{\mathcal{S}}_n \neq \mathcal{S} \right) \bm{P}\left(\hat{\mathcal{S}}_n \neq \mathcal{S} \right)\\
&\le \bm{P}\left(\hat{\mathcal{S}}_n \neq \mathcal{S} \right) \rightarrow 0,
\end{align*}
by estimation consistency. This completes the proof. $\blacksquare$

\subsection*{\textbf{Proof of Theorem \ref{Thm-Degree-Vec}}} 
By multivariate Taylor expansion \citep{A69}, one can express the vector of degree centralities as follows:
\begin{align}
\sqrt{\frac{n}{2}}(\hat{\bm{D}}_n -\bm{D}) &= \sqrt{\frac{n}{2}}\hat{\bm{L}}_D(\hat{\mathcal{S}}_n,\hat{\bm{\beta}})(\hat{\bm{\beta}}_{\hat{\mathcal{S}}_n} - \bm{\beta}_{\hat{\mathcal{S}}_n}) + \sqrt{\frac{n}{2}}(\hat{\bm{\beta}}_{\hat{\mathcal{S}}_n} - \bm{\beta}_{\hat{\mathcal{S}}_n})' \hat{\bm{H}}_n^{(D)}(\hat{\mathcal{S}}_n,\hat{\bm{\beta}})(\hat{\bm{\beta}}_{\hat{\mathcal{S}}_n} - \bm{\beta}_{\hat{\mathcal{S}}_n})\\
&= J_n(1) + J_n(2) \nonumber
\end{align}
where $\hat{\bm{L}}_D(\hat{\mathcal{S}}_n,\hat{\bm{\beta}})$ is a $\hat{q}_n \times \hat{q}_n$ matrix whose $(j,k)^{th}$ element is given by
\begin{align}\label{LHat:D:Thm}
\hat{l}_D(j,k;\hat{\mathcal{S}}_n,\hat{\bm{\beta}}) =  \sum\limits_{r \in \hat{\mathcal{S}}_n} \frac{\partial f(\hat{\beta}_j, \hat{\beta}_r)}{\partial \beta_k} =
\begin{cases}
\sum\limits_{r \in \hat{\mathcal{S}}_n} \frac{\partial f(\hat{\beta}_j, \hat{\beta}_r)}{\partial \beta_j} & j=k \\
\frac{\partial f(\hat{\beta}_j,\hat{\beta}_k)}{\partial \beta_k} & j \neq k
\end{cases},
\end{align}
where
\begin{align*}
    \frac{\partial f(\hat{\beta}_i, \hat{\beta}_j)} {\partial \beta_j} = \frac{\partial f(\beta_i, \beta_j)} {\partial \beta_j} \bigg|_{\beta_j^* \in (\min(\hat{\beta}_j, \beta_j), \max(\hat{\beta}_j, \beta_j))}.    
\end{align*}
Let $\hat{\bm{H}}_n^{(D)}(\hat{\mathcal{S}}_n,\hat{\bm{\beta}})$ is a $\hat{q}_n \times \hat{q}_n \times \hat{q}_n$ Hessian matrix associated with $\bm{D}$ defined by $\bm{\beta}$ in $\hat{\mathcal{S}}_n$. And $\bm{H}^{(D)}(\mathcal{S},\bm{\beta})$ is a $q \times q \times q$ Hessian matrix associated with $\bm{D}$ defined by $\bm{\beta}$ in $\mathcal{S}$. The $(i,jk)^{th}$ element of $\bm{H}^{(D)}(\mathcal{S},\bm{\beta})$ is given by

\begin{align*}
    H_{i,jk}^{(D)} (\mathcal{S}, \bm{\beta}) =
    \begin{cases}
    \sum\limits_{r \in \mathcal{S}} \frac{\partial^2 f(\beta_i, \beta_r)}{\partial \beta_i^2} & j=k=i \\
    \frac{\partial^2 f(\beta_i,\beta_k)}{\partial \beta_i \partial \beta_k} \text{ or } \frac{\partial^2 f(\beta_i,\beta_j)}{\partial \beta_j \partial \beta_i} & j=i \text{ and } k \neq i \text{, or } j \neq i \text{ and } k=i\\
    \frac{\partial^2 f(\beta_i,\beta_j)}{\partial \beta_j^2} \text{ or } \frac{\partial^2 f(\beta_i,\beta_k)}{\partial \beta_k^2} & j=k\neq i \\
    0 & otherwise
\end{cases},
\end{align*}
Similarly, The $(i,jk)^{th}$ element of $\hat{H}_{i,jk}^{(D)}(\hat{\mathcal{S}}_n,\hat{\bm{\beta}})$ is given by
\begin{align}\label{HHat:D:Thm}
    \hat{H}^{(D)}_{i,jk} (\hat{\mathcal{S}}_n,\hat{\bm{\beta}}) = \frac{\partial^2 \hat{D}_{n,i}}{\partial \beta_j \partial \beta_k} =
\begin{cases}
\sum\limits_{r \in \hat{\mathcal{S}}_n} \frac{\partial^2 f(\hat{\beta}_i, \hat{\beta}_r)}{\partial \beta_i^2} & j=k=i \\
\frac{\partial^2 f(\hat{\beta}_i,\hat{\beta}_k)}{\partial \beta_i \partial \beta_k} \text{ or } \frac{\partial^2 f(\hat{\beta}_i,\hat{\beta}_j)}{\partial \beta_j \partial \beta_i} & j=i \text{ and } k \neq i \text{, or } j \neq i \text{ and } k=i\\
\frac{\partial^2 f(\hat{\beta}_i,\hat{\beta}_j)}{\partial \beta_j^2} \text{ or } \frac{\partial^2 f(\hat{\beta}_i,\hat{\beta}_k)}{\partial \beta_k^2} & j=k\neq i \\
0 & otherwise
\end{cases}.
\end{align}
And $J_n(1) = \sqrt{\frac{n}{2}}\hat{\bm{L}}_D(\hat{\mathcal{S}}_n,\hat{\bm{\beta}})(\hat{\bm{\beta}}_{\hat{\mathcal{S}}_n} - \bm{\beta}_{\hat{\mathcal{S}}_n})'$ and
$J_n(2) = \sqrt{\frac{n}{2}}(\hat{\bm{\beta}}_{\hat{\mathcal{S}}_n} -\bm{\beta}_{\hat{\mathcal{S}}_n})' \hat{\bm{H}}_n^{(D)} (\hat{\mathcal{S}}_n,\hat{\bm{\beta}})(\hat{\bm{\beta}}_{\hat{\mathcal{S}}_n} - \bm{\beta}_{\hat{\mathcal{S}}_n})$. We will first show that $J_n(1)$ converges to a multivariate normal distribution with mean $\bm{0}$ and covariance matrix $\bm{\Sigma}_D$, and $J_n(2)$ converges to 0 in probability. To this end, observe that $\hat{\bm{H}}_n^{(D)}(\hat{\mathcal{S}}_n,\hat{\bm{\beta}})$ converges in probability to $\bm{H}^{(D)}(\mathcal{S},\bm{\beta})$ by the continuous mapping theorem using the continuity of $f(\cdot,\cdot)$ and $\hat{\mathcal{S}}_n \xrightarrow{p} \mathcal{S}$ by {\bf[1]} in Theorem 1. Also, by (ii) in Theorem 1\ref{Thm-Asym-LSBeta}, $\sqrt{\frac{n}{2}}(\hat{\bm{\beta}}_{\hat{\mathcal{S}}_n}-\bm{\beta}_{\hat{\mathcal{S}}_n})$ converges to a multivariate normal distribution with mean vector $\bm{0}$ and covariance matrix $\sigma^2\bm{V}_{\mathcal{S}}^{-1}$. In addition, $\hat{\bm{L}}_D(\hat{\mathcal{S}}_n,\hat{\bm{\beta}})$ converges to $\bm{L}_D(\mathcal{S},\bm{\beta})$ in probability by the continuous mapping theorem using the continuity of $f(\cdot,\cdot)$. Hence, by multivariate Slutsky's Theorem \citep{V00} 
it follows that as $ n \ra \ff$ $J_n(1) \xrightarrow{d} N({\bm{0}}, \bm{\Sigma}_D)$, where $\bm{\Sigma}_D = \sigma^2 \bm{L}_D(\mathcal{S},\bm{\beta})'\bm{V}_{\mathcal{S}}^{-1}\bm{L}_D(\mathcal{S},\bm{\beta})$. Next turning to $J_n(2)$, recall that $J_n(2) = \sqrt{\frac{n}{2}}(\hat{\bm{\beta}}_{\hat{\mathcal{S}}_n} -\bm{\beta}_{\hat{\mathcal{S}}_n})' \bm{H}_n^{(D)}(\hat{\mathcal{S}}_n,\hat{\bm{\beta}})(\hat{\bm{\beta}}_{\hat{\mathcal{S}}_n} - \bm{\beta}_{\hat{\mathcal{S}}_n}) .$  By Theorem 1\ref{Thm-Asym-LSBeta} and the continuous mapping theorem using the continuity of $f(\cdot,\cdot)$, consistency of $\hat{\bm{\beta}}_{\hat{\mathcal{S}}_n}$ and convergence of $\hat{\bm{H}}_n^{(D)}(\hat{\mathcal{S}}_n,\hat{\bm{\beta}})$ to $\bm{H}^{(D)}(\mathcal{S},\bm{\beta})$ in probability, it follows that $J_n(2)$ converges to 0 in probability.  $\blacksquare$

\subsection*{\textbf{Proof of Theorem \ref{conv:PSPC}}}
The proof of the Theorem involves several steps. The first step is to verify that the limiting distribution of sample covariance of $\mathcal{Y}$ is Gaussian with mean vector ${\bm{0}}$ and covariance matrix $\bm{K}\bm{\Delta}_{\mathcal{S}} \bm{K}'$. This is achieved in Lemma \ref{Lemma-Asym-Covhat} below. The second step is concerned with the limiting distribution of the post-selection estimate of the precision matrix. This is dealt with in Lemma \ref{Lemma-Asym-Precision}. Finally, we use these results to derive the joint limit distribution of the sample partial correlation coefficients. 

We begin with the limiting distribution of $\hat{\Sigma}_{\mathcal{S}}$, which is the limiting covariance of the sample covariance of $\mathcal{Y}$, denoted by $\Sigma_{\mathcal{S}}$, as follows \citep{NW90}:
\begin{align}\label{S:Asym-Cov}
\lim_{n \rightarrow \infty}  \bm{P}\left( \sqrt{\frac{n}{2}}v(\hat{\Sigma}_{\mathcal{S}} - \Sigma_{\mathcal{S}}) \le \bm{x} \right) = \bm{P}( N(\bm{0}, \bm{K}\bm{\Delta}_{\mathcal{S}} \bm{K}') \le \bm{x} ),
\end{align}
where $\bm{\Delta}_{\mathcal{S}} = E[\mathcal{Y}_{\mathcal{S}}'\mathcal{Y}_{\mathcal{S}} \otimes \mathcal{Y}_{\mathcal{S}}'\mathcal{Y}_{\mathcal{S}}] - vec(\Sigma_{\mathcal{S}})vec(\Sigma_{\mathcal{S}})'$ and $\bm{K}$ is defined in the statement of Theorem 3. Then, we can establish the following Lemma.
\begin{lem}\label{Lemma-Asym-Covhat}
For any $\bm{x} \in \real^q$, let ${\bm{x}}_n$ denote its dimension adjusted version. Then,
\begin{align}
\bm{P}\left( \sqrt{\frac{n}{2}}v(\hat{\Sigma}_{\hat{\mathcal{S}}_n} - \Sigma_{\hat{\mathcal{S}}_n}) \le \bm{x}_n \right) = \bm{P}( N(\bm{0}, \bm{K}\bm{\Delta}_{\mathcal{S}} \bm{K}') \le \bm{x} ) \mbox{ \ \ \ as $n \rightarrow \infty$}.
\end{align}
\end{lem}
\begin{proof}
Notice that $\hat{\mathcal{S}}_n$ is estimated from $\mathcal{D}_{1n}$ and $\hat{\bm{\beta}}_{\hat{\mathcal{S}}_n}$ is estimated from $\mathcal{D}_{2n}$. We use the independence of $\mathcal{D}_{1n}$ and $\mathcal{D}_{2n}$ for the proof.
\begin{align*}
\bm{P} \left( \sqrt{\frac{n}{2}}v(\hat{\Sigma}_{\hat{\mathcal{S}}_n} - \Sigma_{\hat{\mathcal{S}}_n}) \le \bm{x}_n \right) &= \bm{P} \left( \sqrt{\frac{n}{2}}v(\hat{\Sigma}_{\hat{\mathcal{S}}_n} - \Sigma_{\hat{\mathcal{S}}_n}) \le \bm{x}_n , \hat{\mathcal{S}}_n = \mathcal{S} \right)\\
&+ \bm{P} \left( \sqrt{\frac{n}{2}}v(\hat{\Sigma}_{\hat{\mathcal{S}}_n} - \Sigma_{\hat{\mathcal{S}}_n}) \le \bm{x}_n, \hat{\mathcal{S}}_n \neq \mathcal{S} \right)\\
&= J_n(1) + J_n(2).
\end{align*}
We will show that $J_n(1) \rightarrow \bm{P}(N(\bm{0}, \bm{K}\bm{\Delta}_{\mathcal{S}}\bm{K}') \le \bm{x}_n)$ and $J_n(2) \rightarrow 0$ as $n \rightarrow \infty$. $J_n(1)$ can be represented as follows:
\begin{align*}
J_n(1) &= \bm{P} \left( \sqrt{\frac{n}{2}}v(\hat{\Sigma}_{\hat{\mathcal{S}}_n} - \Sigma_{\hat{\mathcal{S}}_n}) \le \bm{x}_n \bigg| \hat{\mathcal{S}}_n = \mathcal{S} \right) \bm{P}\left(\hat{\mathcal{S}}_n = \mathcal{S} \right)\\
&= \bm{P} \left( \sqrt{\frac{n}{2}}v(\hat{\Sigma}_{\mathcal{S}} - \Sigma_{\mathcal{S}}) \le \bm{x}\right)  \\
&\rightarrow \bm{P}\left(N(\bm{0}, \bm{K}\bm{\Delta}_{\mathcal{S}} \bm{K}') \le \bm{x}\right) \mbox{ \ as $n \rightarrow \infty$.}
\end{align*}
where the last convergence follows from  (\ref{S:Asym-Cov}). We will next show $J_n(2)$ converges to 0. To this end,
\begin{align*}
J_n(2) &= \bm{P} \left( \sqrt{\frac{n}{2}}v(\hat{\Sigma}_{\hat{\mathcal{S}}_n} - \Sigma_{\hat{\mathcal{S}}_n}) \le \bm{x}_n \bigg| \hat{\mathcal{S}}_n \neq \mathcal{S} \right) \bm{P}\left(\hat{\mathcal{S}}_n \neq \mathcal{S} \right)\\
&\le \bm{P}\left(\hat{\mathcal{S}}_n \neq \mathcal{S} \right) \rightarrow 0.
\end{align*}
\end{proof}
Before we prove Theorem 3\ref{conv:PSPC}, we first establish some properties of functions defined on the space of positive definite symmetric  matrices ${\mathcal{A}}$ endowed with Frobenius norm. The following lemma describes the derivative of an inverse of a invertible matrix \citep{PP08}.
\begin{lem}\label{Mat: pro}
Let $z: \mathcal{A} \ra \mathcal{A}$ such that $z(A)=A^{-1}$. Then $z(\cdot)$ is continuous. Furthermore,
\begin{eqnarray}
\frac{\partial v(z(A))}{v(A)}= -(z(A) \otimes z(A)) \equiv \bm{L}(A).
\end{eqnarray}
\end{lem}
\noindent We now state an extended version of the delta method which plays a critical role in several proofs.
\begin{prop}\label{mat:delta}
Let $\{A_n: n \ge 1\}$ be a collection of matrices in $\mathcal{A}$ such that $\sqrt{n}(v(A_n)-v(A)) \xrightarrow{d} N(\bm{0}, \mathfrak{D})$ as $n \ra \ff$. Then 
\begin{eqnarray}
{\bm{P}}\left(\sqrt{n}(v(z(A_n))-v(z(A))) \le {\bm{x}}\right) = {\bm{P}}\left(N({\bm{0}}, \mathfrak{D}^*_z) \le \bm{x}\right)
\end{eqnarray}
where $\mathfrak{D}^*_z=\bm{L}(A)^{\prime} \mathfrak{D} \bm{L}(A)$.
\end{prop}
We are now ready to prove Lemma \ref{Lemma-Asym-Precision}  concerning the limit distribution of the post-selection estimator of the precision matrix.

\begin{lem}\label{Lemma-Asym-Precision}
For any $\bm{x} \in \real^q$, let ${\bm{x}}_n$ denote its dimension adjusted version. Then, the following will hold:\\
{\bf[1]}
\begin{align}
\lim_{n \ra \ff} \bm{P}\left( \sqrt{\frac{n}{2}}\left(v(\hat{\Gamma}_{\mathcal{S}}) - v( \Gamma_{\mathcal{S}})\right) \le \bm{x} \right) = \bm{P}( N(\bm{0},\bm{\Delta}^*_{\mathcal{S}} ) \le \bm{x} ),
\end{align}
where $\bm{\Delta}_{\mathcal{S}}^* = \bm{K}\bm{L}(\Sigma_{\mathcal{S}})'  \bm{\Delta}_{\mathcal{S}} \bm{L}(\Sigma_{\mathcal{S}}) \bm{K}^{\prime}$ and $\bm{L}(\bm{\Sigma}_{\mathcal{S}}) = -(\Gamma_{\mathcal{S}} \otimes \Gamma_{\mathcal{S}})$.\\
{\bf[2]} 
\begin{align}
\lim_{n \ra \ff} \bm{P}\left( \sqrt{\frac{n}{2}}\left(v(\hat{\Gamma}_{\hat{\mathcal{S}}_n}) - v( \Gamma_{\hat{\mathcal{S}}_n})\right) \le \bm{x}_n \right) = \bm{P}( N(\bm{0}, \bm{\Delta}^*_{\mathcal{S}}) \le \bm{x} ).
\end{align}
\end{lem}
\begin{proof}
Let $z: \mathcal{A} \ra \mathcal{A}$ such that $z(A)= A^{-1}$. Then by continuity of $z(\cdot)$, it follows that $z(\hat{\bm{\Sigma}}_S)$ converges in probability to $z({\bm{\Sigma}}_S)$. Now, we apply Proposition \ref{mat:delta} to (\ref{S:Asym-Cov}) we get, as $n \ra \ff$, that
\begin{eqnarray} 
\sqrt{\frac{n}{2}}\left(v(z(\hat{\bm{\Sigma}}_S))-v(z({\bm{\Sigma}}_S))\right) \xrightarrow{d} N (\bm{0}, \bm{\Delta}^*_{\mathcal{S}}).
\end{eqnarray}
Now by another application of Lemma \ref{Mat: pro} and (\ref{S:Asym-Cov}), it follows that $\bm{\Delta}^*_{\mathcal{S}}= \bm{K} L'(\bm{\Sigma}_{\mathcal{S}}) \bm{\Delta}_{\mathcal{S}}  L(\bm{\Sigma}_{\mathcal{S}})\bm{K}^{\prime}$. The proof for {\bf[2]} follows by the conditioning argument as before.
\end{proof}

Now we introduce the limiting distribution of $\hat{\bm{\rho}}_{\mathcal{S}}$. The $(i,j)^{th}$ element of $\nabla h(v(\Gamma_{\mathcal{S}}))$, denoted by $h_{ij}(v(\Gamma_{\mathcal{S}}))$, is given by
\begin{align*}
    h_{ij}(v(\Gamma_{\mathcal{S}})) = \frac{\partial h(v(\Gamma_{\mathcal{S}}))}{\partial \gamma_{v,j}}=
    \begin{cases}
    -\frac{1}{2}\frac{\sqrt{\gamma_{v,i+1}}}{ \gamma_{v,1}^{3/2} \sqrt{\gamma_{v,k}}} & i=1,\dots, q \text{ and } j=1 \\
    \frac{1}{2}\frac{1}{\sqrt{\gamma_{v,i+1}}\sqrt{\gamma_{v,1}}\sqrt{\gamma_{v,k}}} &i=1,\dots, q \text{ and } j = i+1 \\
    -\frac{1}{2}\frac{\sqrt{\gamma_{v,i+1}}}{ \sqrt{\gamma_{v,1}} \gamma_{v,k}^{3/2}} & i=1,\dots, q \text{ and } j=k, 
    \end{cases}
\end{align*}
where $\gamma_{v,i}$ is the $i^{th}$ element of $v(\Gamma_{\mathcal{S}})$ and $k = \sum\limits_{l=0}^i (q-l)+1$. Then, by applying the multivariate delta method, we have the following result:
\begin{align}\label{Asym-Parcor}
\lim_{n \rightarrow \infty} \bm{P}\left(\sqrt{\frac{n}{2}}(\hat{\bm{\rho}}_{\mathcal{S}} -\bm{\rho}_{\mathcal{S}})\le \bm{x} \right)= \bm{P}( N(\bm{0}, \bm{\Delta}_{\mathcal{S}}(\bm{\rho})) \le \bm{x}),
\end{align}
where $\bm{\Delta}_{\mathcal{S}}(\bm{\rho}) = \nabla h(v(\Gamma_{\mathcal{S}}))\bm{K}\bm{L}(\Sigma_{\mathcal{S}})'\bm{\Delta}_{\mathcal{S}} \bm{L}(\Sigma_{\mathcal{S}}) \bm{K}' \nabla h(v(\Gamma_{\mathcal{S}}))'$. By using the conditioning argument as before, the proof of Theorem 3\ref{conv:PSPC} is completed. $\blacksquare$

\subsection*{\textbf{Proof of Theorem \ref{DA_DCPC}}} 
By multivariate Taylor expansion \citep{A69}, one can express the vector of clustering coefficients as follows:
\begin{align}
\sqrt{\frac{n}{2}}(\hat{\bm{D}}_n(\bm{\rho}) - \bm{D}(\bm{\rho})) &= \sqrt{\frac{n}{2}}\hat{\bm{L}}_D(\hat{\mathcal{S}}_n,\hat{\bm{\rho}})(\hat{\bm{\rho}}_{\hat{\mathcal{S}}_n} - \bm{\rho}_{\hat{\mathcal{S}}_n}) + \sqrt{\frac{n}{2}}(\hat{\bm{\rho}}_{\hat{\mathcal{S}}_n} - \bm{\rho}_{\hat{\mathcal{S}}_n})' \hat{\bm{H}}_n^{(D)}(\hat{\mathcal{S}}_n,\hat{\bm{\rho}})(\hat{\bm{\rho}}_{\hat{\mathcal{S}}_n} - \bm{\rho}_{\hat{\mathcal{S}}_n})\\
&= J_n(1) + J_n(2) \nonumber
\end{align}
where $\hat{\bm{L}}_D(\hat{\mathcal{S}}_n,\hat{\bm{\rho}})$ is a $\hat{q}_n \times \hat{q}_n$ matrix whose $(j,k)^{th}$ element is given by
\begin{align}\label{LHat:D:Thm2}
\hat{l}_D(j,k;\hat{\mathcal{S}}_n,\hat{\bm{\rho}})  = \sum_{r \in \hat{\mathcal{S}}_n} \frac{\partial f(\hat{\rho}_j, \hat{\rho}_r)}{\partial \rho_k} = 
\begin{cases}
\sum\limits_{r \in \hat{\mathcal{S}}_n} \frac{\partial f(\hat{\rho}_j, \hat{\rho}_r)}{\partial \rho_j} & j=k \\
\frac{\partial f(\hat{\rho}_j,\hat{\rho}_k)}{\partial \rho_k} & j \neq k.
\end{cases},
\end{align}
where
\begin{align*}
    \frac{\partial f(\hat{\rho}_i, \hat{\rho}_j)} {\partial \rho_j} = \frac{\partial f(\hat{\rho}_i, \hat{\rho}_j)} {\partial \rho_j} \bigg|_{\rho_j^* \in (\min(\hat{\rho}_j, \rho_j), \max(\hat{\rho}_j, \rho_j))}.    
\end{align*}
Let $\hat{\bm{H}}_n^{(D)}(\hat{\mathcal{S}}_n,\hat{\bm{\rho}})$ is a $\hat{q}_n \times \hat{q}_n \times \hat{q}_n$ Hessian matrix associated with $\bm{D}$ defined by $\bm{\rho}$ in $\hat{\mathcal{S}}_n$. And $\bm{H}^{(D)}(\mathcal{S},\bm{\rho})$ is a $q \times q \times q$ Hessian matrix associated with $\bm{D}$ defined by $\bm{\rho}$ in $\mathcal{S}$. The $(i,jk)^{th}$ element of $\bm{H}^{(D)}(\mathcal{S},\bm{\rho})$ is given by

\begin{align*}
    H_{i,jk}^{(D)}(\mathcal{S}, \bm{\rho}) = \frac{\partial^2 D_i(\bm{\rho})}{\partial \rho_j \partial \rho_k} =
    \begin{cases}
    \sum\limits_{r \in \mathcal{S}} \frac{\partial^2 f(\rho_i, \rho_r)}{\partial \rho_i^2} & j=k=i \\
    \frac{\partial^2 f(\rho_i,\rho_k)}{\partial \rho_i \partial \rho_k} \text{ or } \frac{\partial^2 f(\rho_i,\rho_j)}{\partial \rho_j \partial \rho_i} & j=i \text{ and } k \neq i \text{, or } j \neq i \text{ and } k=i\\
    \frac{\partial^2 f(\rho_i,\rho_j)}{\partial \rho_j^2} \text{ or } \frac{\partial^2 f(\rho_i,\rho_k)}{\partial \rho_k^2} & j=k\neq i \\
    0 & otherwise
\end{cases},
\end{align*}
Similarly, The $(i,jk)^{th}$ element of $\hat{H}_{i,jk}^{(D)}(\hat{\mathcal{S}}_n,\hat{\bm{\rho}})$ is given by
\begin{align}\label{HHat:D:Thm2}
    \hat{H}^{(D)}_{i,jk} (\hat{\mathcal{S}}_n,\hat{\bm{\rho}}) = \frac{\partial^2 \hat{D}_{n,i}(\bm{\rho})}{\partial \rho_j \partial \rho_k} =
\begin{cases}
\sum\limits_{r \in \hat{\mathcal{S}}_n} \frac{\partial^2 f(\hat{\rho}_i, \hat{\rho}_r)}{\partial \rho_i^2} & j=k=i \\
\frac{\partial^2 f(\hat{\rho}_i,\hat{\rho}_k)}{\partial \rho_i \partial \rho_k} \text{ or } \frac{\partial^2 f(\hat{\rho}_i,\hat{\rho}_j)}{\partial \rho_j \partial \rho_i} & j=i \text{ and } k \neq i \text{, or } j \neq i \text{ and } k=i\\
\frac{\partial^2 f(\hat{\rho}_i,\hat{\rho}_j)}{\partial \rho_j^2} \text{ or } \frac{\partial^2 f(\hat{\rho}_i,\hat{\rho}_k)}{\partial \rho_k^2} & j=k\neq i \\
0 & otherwise
\end{cases}.
\end{align}
We will first show that $J_n(1)$ converges to a multivariate normal distribution with mean $\bm{0}$ and covariance matrix $\bm{\Sigma}_D(\bm{\rho})$, and $J_n(2)$ converges to 0 in probability. To this end, observe that $\hat{\bm{H}}_n^{(D)}(\hat{\mathcal{S}}_n,\hat{\bm{\rho}})$ converges in probability to $\bm{H}^{(D)}(\mathcal{S},\bm{\rho})$ by the continuous mapping theorem, using the continuity of $f(\cdot,\cdot)$. Also, by Theorem 3\ref{conv:PSPC}, $\sqrt{\frac{n}{2}}(\hat{\bm{\rho}}_{\hat{\mathcal{S}}_n}-\bm{\rho}_{\hat{\mathcal{S}}_n})$ converges to a multivariate normal distribution with mean vector $\bm{0}$ and covariance matrix $\bm{\Delta}_{\mathcal{S}}(\bm{\rho})$. In addition, $\hat{\bm{L}}_D(\hat{\mathcal{S}}_n,\hat{\bm{\rho}})$ converges to $\bm{L}_D(\mathcal{S}, \bm{\rho})$ by the continuous mapping theorem, using the continuity of $f(\cdot,\cdot)$. Hence, by multivariate Slutsky's Theorem \citep{V00} 
it follows that as $ n \ra \ff$ $J_n(1) \xrightarrow{d} N({\bm{0}}, \bm{\Sigma}_D(\bm{\rho}))$, where $\bm{\Sigma}_D(\bm{\rho}) =\bm{L}_D(\mathcal{S},\bm{\rho})' \bm{\Delta}_{\mathcal{S}}(\bm{\rho})\bm{L}_D(\mathcal{S},\bm{\rho})$. Next turning to $J_n(2)$, recall that $J_n(2) = \sqrt{\frac{n}{2}}(\hat{\bm{\rho}}_{\hat{\mathcal{S}}_n} -\bm{\rho}_{\hat{\mathcal{S}}_n})' \bm{H}_n^{(D)}(\hat{\mathcal{S}}_n,\hat{\bm{\rho}})(\hat{\bm{\rho}}_{\hat{\mathcal{S}}_n} - \bm{\rho}_{\hat{\mathcal{S}}_n})$. By Theorem 3\ref{conv:PSPC} and the continuous mapping theorem, using the continuity of $f(\cdot,\cdot)$, consistency of $\hat{\bm{\rho}}_{\hat{\mathcal{S}}_n}$ and convergence of $\hat{\bm{H}}_n^{(D)}(\hat{\mathcal{S}}_n,\hat{\bm{\rho}})$ to $\bm{H}^{(D)}(\mathcal{S},\bm{\rho})$ in probability, it follows that $J_n(2)$ converges to 0 in probability.  $\blacksquare$

\subsection*{\textbf{Proof of Theorem \ref{Thm-Clustercoef-Vec}}} The proof of this theorem proceeds in the same manner as that of Theorem 2\ref{Thm-Degree-Vec}. We only provide it here to make the paper self-contained. By multivariate Taylor expansion \citep{A69}, one can express the vector of clustering coefficients as follows:
\begin{align}
\sqrt{\frac{n}{2}}(\hat{\bm{C}}_n - \bm{C}) &= \sqrt{\frac{n}{2}}(\hat{\bm{\beta}}_{\hat{\mathcal{S}}_n} - \bm{\beta}_{\hat{\mathcal{S}}_n})'\hat{\bm{L}}_C(\hat{\mathcal{S}}_n,\hat{\bm{\beta}}) + \sqrt{\frac{n}{2}}(\hat{\bm{\beta}}_{\hat{\mathcal{S}}_n} - \bm{\beta}_{\hat{\mathcal{S}}_n})' \hat{\bm{H}}_n^{(C)}(\hat{\mathcal{S}}_n,\hat{\bm{\beta}})(\hat{\bm{\beta}}_{\hat{\mathcal{S}}_n} - \bm{\beta}_{\hat{\mathcal{S}}_n})\\
&= J_n(1) + J_n(2) \nonumber
\end{align}
where $\hat{\bm{L}}_C(\hat{\mathcal{S}}_n,\hat{\bm{\beta}})$ and $\hat{\bm{H}}_n^{(C)}(\hat{\mathcal{S}}_n,\hat{\bm{\beta}})$ are obtained by replacing the function of degree centrality in (\ref{LHat:D:Thm}) and ($\ref{HHat:D:Thm}$) with the function of clustering coefficient. Let  $\bm{H}^{(C)}(\mathcal{S}_n,\bm{\beta})$ is a $q \times q \times q$ Hessian matrix associated with $\bm{C}$ defined by $\bm{\beta}$ in $\mathcal{S}$. The $(i,jk)^{th}$ element of $\bm{H}^{(C)}(\mathcal{S},\bm{\beta})$ is given by
\begin{align*}
    H_{i,jk}^{(C)} (\mathcal{S}, \bm{\beta}) = \frac{\partial^2 C_i}{\partial \beta_j \partial \beta_k} = \frac{1}{(q-1)(q-2)}
    \begin{cases}
    \frac{\partial^2 f(\beta_j,\beta_k)}{\partial \beta_j \partial \beta_k} & j\neq k \neq i \\
    \sum\limits_{(k,u) \in \mathcal{N}(i)} \frac{\partial^2 f(\beta_k,\beta_u)}{\partial \beta_k^2} & j=k \text{ and } k \neq i \\
    0 & otherwise
\end{cases},
\end{align*}
We will first show that $J_n(1)$ converges to a multivariate normal distribution with mean $\bm{0}$ and covariance matrix $\bm{\Sigma}_C$, and $J_n(2)$ converges to 0 in probability. To this end, observe that $\hat{\bm{H}}_n^{(C)}(\hat{\mathcal{S}}_n,\hat{\bm{\beta}})$ converges in probability to $\bm{H}^{(C)}(\mathcal{S},\bm{\beta})$ by the continuous mapping theorem, using the continuity of $f(\cdot,\cdot)$. Also, by Theorem 1\ref{Thm-Asym-LSBeta}, $\sqrt{\frac{n}{2}}(\hat{\bm{\beta}}_{\hat{\mathcal{S}}_n}-\bm{\beta}_{\hat{\mathcal{S}}_n})$ converges to a multivariate normal distribution with mean vector $\bm{0}$ and covariance matrix $\sigma^2\bm{V}_{\mathcal{S}}^{-1}$. In addition $\hat{\bm{L}}_C(\hat{\mathcal{S}}_n,\hat{\bm{\beta}})$ converges to $\bm{L}_C(\mathcal{S},\bm{\beta})$ by the continuous mapping theorem, using the continuity of $f(\cdot,\cdot)$. Hence, by multivariate Slutsky's Theorem \citep{V00} 
it follows that as $ n \ra \ff$ $J_n(1) \xrightarrow{d} N({\bm{0}}, \bm{\Sigma}_C)$, where $\bm{\Sigma}_C = \sigma^2 \bm{L}_C(\mathcal{S},\bm{\beta})' \bm{V}_{\mathcal{S}}^{-1} \bm{L}_C(\mathcal{S},\bm{\beta})$. Next turning to $J_n(2)$,  recall that
$J_n(2) = \sqrt{n}(\hat{\bm{\beta}}_{\hat{\mathcal{S}}_n} -\bm{\beta}_{\hat{\mathcal{S}}_n})' \hat{\bm{H}}_n^{(C)}(\hat{\mathcal{S}}_n,\hat{\bm{\beta}})(\hat{\bm{\beta}}_{\hat{\mathcal{S}}_n} - \bm{\beta}_{\hat{\mathcal{S}}_n}) .$  By Theorem 1\ref{Thm-Asym-LSBeta} and the continuous mapping theorem, using the continuity of $f(\cdot,\cdot)$, consistency of $\hat{\bm{\beta}}_{\hat{\mathcal{S}}_n}$ and convergence of $\hat{\bm{H}}_n^{(C)}(\hat{\mathcal{S}}_n,\hat{\bm{\beta}})$ to $\bm{H}^{(C)}(\mathcal{S},\bm{\beta})$ in probability, it follows that $J_n(2)$ converges to 0 in probability.  $\blacksquare$

\subsection*{\textbf{Proof of Theorem \ref{CA_CCPC}}} 
The proof of this theorem proceeds in the same manner as that of Theorem 4\ref{DA_DCPC}. We only provide it here to make the paper self-contained. By multivariate Taylor expansion \citep{A69}, one can express the vector of clustering coefficients as follows:
\begin{align}
\sqrt{\frac{n}{2}}(\hat{\bm{C}}_n(\bm{\rho}) - \bm{C}(\bm{\rho})) &= \sqrt{\frac{n}{2}}\hat{\bm{L}}_C(\hat{\mathcal{S}}_n,\hat{\bm{\rho}})(\hat{\bm{\rho}}_{\hat{\mathcal{S}}_n} - \bm{\rho}_{\hat{\mathcal{S}}_n}) + \sqrt{\frac{n}{2}}(\hat{\bm{\rho}}_{\hat{\mathcal{S}}_n} - \bm{\rho}_{\hat{\mathcal{S}}_n})' \hat{\bm{H}}_n^{(C)}(\hat{\mathcal{S}}_n,\hat{\bm{\rho}})(\hat{\bm{\rho}}_{\hat{\mathcal{S}}_n} - \bm{\rho}_{\hat{\mathcal{S}}_n})\\
&= J_n(1) + J_n(2) \nonumber
\end{align}
where $\hat{\bm{L}}_C(\hat{\mathcal{S}}_n,\hat{\bm{\rho}})$ and $\hat{\bm{H}}_n^{(C)}(\hat{\mathcal{S}}_n,\hat{\bm{\rho}})$ are obtained by replacing the function of degree centrality in (\ref{LHat:D:Thm2}) and ($\ref{HHat:D:Thm2}$) with the function of clustering coefficient. Let $\bm{H}^{(C)}(\mathcal{S},\bm{\rho})$ is a $q \times q \times q$ Hessian matrix associated with $\bm{C}$ defined by $\bm{\rho}$ in $\mathcal{S}$. The $(i,jk)^{th}$ element of $\bm{H}^{(C)}(\mathcal{S},\bm{\rho})$ is given by

\begin{align*}
    H_{i,jk}^{(D)}(\mathcal{S}, \bm{\rho}) = \frac{\partial^2 C_i(\bm{\rho})}{\partial \rho_j \partial \rho_k} = \frac{1}{(q-1)(q-2)}
    \begin{cases}
    \frac{\partial^2 f(\rho_j,\rho_k)}{\partial \rho_j \partial \rho_k} & j\neq k \neq i \\
    \sum\limits_{(k,u) \in \mathcal{N}(i)} \frac{\partial^2 f(\rho_k,\rho_u)}{\partial \rho_k^2} & j=k \text{ and } k \neq i \\
    0 & otherwise
\end{cases},
\end{align*}
We will first show that $J_n(1)$ converges to a multivariate normal distribution with mean $\bm{0}$ and covariance matrix $\bm{\Sigma}_C(\bm{\beta})$, and $J_n(2)$ converges to 0 in probability. To this end, observe that $\hat{\bm{H}}_n^{(C)}(\hat{\mathcal{S}}_n,\hat{\bm{\rho}})$ converges in probability to $\bm{H}^{(C)}(\mathcal{S},\bm{\rho})$ by the continuous mapping theorem, using the continuity of $f(\cdot,\cdot)$. Also, by Theorem 3\ref{conv:PSPC}, $\sqrt{\frac{n}{2}}(\hat{\bm{\rho}}_{\hat{\mathcal{S}}_n}-\bm{\rho}_{\hat{\mathcal{S}}_n})$ converges to a multivariate normal distribution with mean vector $\bm{0}$ and covariance matrix $\bm{\Delta}_{\mathcal{S}}(\bm{\rho})$. In addition, $\hat{\bm{L}}_C(\hat{\mathcal{S}}_n,\hat{\bm{\rho}})$ converges to $\bm{L}_C(\mathcal{S}, \bm{\rho})$ in probability by the continuous mapping theorem, using the continuity of $f(\cdot,\cdot)$. Hence, by multivariate Slutsky's Theorem \citep{V00} 
it follows that as $ n \ra \ff$ $J_n(1) \xrightarrow{d} N({\bm{0}}, \bm{\Sigma}_C(\bm{\rho}))$, where $\bm{\Sigma}_C(\bm{\rho}) =\bm{L}_C(\mathcal{S},\bm{\rho})' \bm{\Delta}_{\mathcal{S}}(\bm{\rho})\bm{L}_C(\mathcal{S},\bm{\rho})$. Next turning to $J_n(2)$, recall that $J_n(2) = \sqrt{\frac{n}{2}}(\hat{\bm{\rho}}_{\hat{\mathcal{S}}_n} -\bm{\rho}_{\hat{\mathcal{S}}_n})' \bm{H}_n^{(C)}(\hat{\mathcal{S}}_n,\hat{\bm{\rho}})(\hat{\bm{\rho}}_{\hat{\mathcal{S}}_n} - \bm{\rho}_{\hat{\mathcal{S}}_n})$.  By Theorem 3\ref{conv:PSPC} and the continuous mapping theorem, using the continuity of $f(\cdot,\cdot)$, consistency of $\hat{\bm{\rho}}_{\hat{\mathcal{S}}_n}$ and convergence of $\hat{\bm{H}}_n^{(C)}(\hat{\mathcal{S}}_n,\hat{\bm{\rho}})$ to $\bm{H}^{(C)}(\mathcal{S},\bm{\rho})$ in probability, it follows that $J_n(2)$ converges to 0 in probability.  $\blacksquare$

\subsection*{\textbf{Proof of Theorem \ref{conv:DA_pcov}}}

Proof of {\bf[1]}.
We can represent $\bm{\Delta}_{\mathcal{S}}(\bm{\rho})$ as a product of matrices as follows:
\begin{align*}
    \bm{\Delta}_{\mathcal{S}}(\bm{\rho}) = \nabla h(v(\Gamma_{\mathcal{S}}))'\bm{L}(\Sigma_{\mathcal{S}})'\bm{\Delta}_{\mathcal{S}} \bm{L}(\Sigma_{\mathcal{S}}) \nabla h(v(\Gamma_{\mathcal{S}})).
\end{align*}
where $\bm{\Delta}_{\mathcal{S}}$ is the covariance matrix of $vec(\bm{\Sigma}_{\mathcal{S}})$, $\bm{L}(\Sigma_{\mathcal{S}}) = - (\bm{\Gamma}_{\mathcal{S}} \otimes \bm{\Gamma}_{\mathcal{S}})$, and $\nabla h(v(\Gamma_{\mathcal{S}}))$ is the partial derivatives of $h(v(\Gamma_{\mathcal{S}}))$. Let the estimate of $\bm{\Delta}_{\mathcal{S}}(\bm{\rho})$, denoted by $\hat{\bm{\Delta}}_{\mathcal{S}}(\bm{\rho})$, be represented as follows:
\begin{align*}
\hat{\bm{\Delta}}_{\mathcal{S}}(\bm{\rho}) =  \nabla h_n(v(\hat{\Gamma}_{\mathcal{S}})) \hat{\bm{L}}(\hat{\Sigma}_{\mathcal{S}})'\hat{\bm{\Delta}}_{\mathcal{S}} \hat{\bm{L}}(\hat{\Sigma}_{\mathcal{S}}) \nabla h_n(v(\hat{\Gamma}_{\mathcal{S}}))'. 
\end{align*}
We will show that the estimate of each term in $\hat{\bm{\Delta}}_{\mathcal{S}}$ converges to the true value and then it will complete the proof by applying the property of convergence in probability \citep{G13}. 
Recall that $\mathcal{Y}_{\mathcal{S}} = (\bm{Y}, \bm{X}_{\mathcal{S}})$ and let $\mathcal{Y}_{\mathcal{S},i} = (Y_i,R_{\mathcal{S},i})$. Then, 
\begin{align*}
    \bm{\Delta}_{\mathcal{S}} = E[\mathcal{Y}_{\mathcal{S},i}'\mathcal{Y}_{\mathcal{S},i} \otimes \mathcal{Y}_{\mathcal{S},i}'\mathcal{Y}_{\mathcal{S},i}] - vec(\bm{\Sigma}_{\mathcal{S}})vec(\bm{\Sigma}_{\mathcal{S}})'
\end{align*} 
Since 
\begin{align*}
    \hat{\bm{\Sigma}}_{\mathcal{S}} = \frac{1}{n} \sum\limits_{i=1}^n \mathcal{Y}_{\mathcal{S},i}'\mathcal{Y}_{\mathcal{S},i} &\xrightarrow{p} \bm{\Sigma}_{\mathcal{S}} \text{ and }\\
    \frac{1}{n} \sum\limits_{i=1}^n (\mathcal{Y}_{\mathcal{S},i}'\mathcal{Y}_{\mathcal{S},i} \otimes \mathcal{Y}_{\mathcal{S},i}'\mathcal{Y}_{\mathcal{S},i}) &\xrightarrow{p} E[\mathcal{Y}_{\mathcal{S},i}'\mathcal{Y}_{\mathcal{S},i} \otimes \mathcal{Y}_{\mathcal{S},i}'\mathcal{Y}_{\mathcal{S},i}]
\end{align*}
as $n \rightarrow \infty$, $\hat{\bm{\Sigma}}_{\mathcal{S}} \xrightarrow{p} \bm{\Sigma}_{\mathcal{S}}$ as $n \rightarrow \infty$. Let $z$ denote a function that is defined in Lemma \ref{Mat: pro}. Then, by continuous mapping theorem and Lemma \ref{Mat: pro}, $\hat{\bm{\Gamma}}_{\mathcal{S}} = z(\hat{\bm{\Sigma}}_{\mathcal{S}}) \xrightarrow{p} z(\bm{\Sigma}_{\mathcal{S}}) = \bm{\Gamma}_{\mathcal{S}}$. Then, by the continuous mapping theorem and the property of convergence in probability, the estimate of $ \bm{L}({\Sigma}_{\mathcal{S}})$, which is denoted by $ \hat{\bm{L}}(\hat{\Sigma}_{\mathcal{S}})$, also converges to $ \bm{L}({\Sigma}_{\mathcal{S}})$; that is
\begin{align*}
    \hat{\bm{L}}(\hat{\Sigma}_{\mathcal{S}}) = - \hat{\bm{\Gamma}}_{\mathcal{S}} \otimes \hat{\bm{\Gamma}}_{\mathcal{S}} \xrightarrow{p} - \bm{\Gamma}_{\mathcal{S}} \otimes \bm{\Gamma}_{\mathcal{S}} =  \bm{L}({\Sigma}_{\mathcal{S}}) \text{ as $n \rightarrow \infty$.}
\end{align*}
By Assumption \ref{A3} and the continuous mapping theorem, $\nabla h_n(v(\hat{\Gamma}_{\mathcal{S}})) \xrightarrow{p}  \nabla h(v(\Gamma_{\mathcal{S}}))$ as $n \rightarrow \infty$. Then, using the property of convergence in probability gives
\begin{align*}
     \hat{\bm{\Delta}}_{\mathcal{S}}(\bm{\rho}) &=  \nabla h_n(v(\hat{\Gamma}_{\mathcal{S}}))'\hat{\bm{L}}(\hat{\Sigma}_{\mathcal{S}})'\hat{\bm{\Delta}}_{\mathcal{S}} \hat{\bm{L}}(\hat{\Sigma}_{\mathcal{S}}) \nabla h(v(\hat{\Gamma}_{\mathcal{S}}))\\ &\xrightarrow{p} \nabla h(v(\Gamma_{\mathcal{S}}))' \bm{L}(\Sigma_{\mathcal{S}})' \bm{\Delta}_{\mathcal{S}} \bm{L}(\Sigma_{\mathcal{S}}) \nabla h(v(\Gamma_{\mathcal{S}}))= \bm{\Delta}_{\mathcal{S}}(\bm{\rho})
\end{align*}
as $n \rightarrow \infty$ by the convergence of each element and the continuity of the product of random variables. The proof is completed. $\blacksquare$

Proof of {\bf[2]}.
Notice that $\hat{\mathcal{S}}_n$ is estimated from $\mathcal{D}_{1n}$ and $\hat{\bm{\beta}}_{\hat{\mathcal{S}}_n}$ is estimated from $\mathcal{D}_{2n}$. We use the independence of $\mathcal{D}_{1n}$ and $\mathcal{D}_{2n}$ for the proof. For $\epsilon >0$,
\begin{align*}
\bm{P} \left( \Vert \hat{\bm{\Delta}}_{\hat{\mathcal{S}}_n} - \bm{\Delta}_{\hat{\mathcal{S}}_n} \Vert > \epsilon \right) &= \bm{P} \left( \Vert \hat{\bm{\Delta}}_{\hat{\mathcal{S}}_n} - \bm{\Delta}_{\hat{\mathcal{S}}_n} \Vert > \epsilon , \hat{\mathcal{S}}_n = \mathcal{S} \right) + \bm{P} \left( \Vert \hat{\bm{\Delta}}_{\hat{\mathcal{S}}_n} - \bm{\Delta}_{\hat{\mathcal{S}}_n} \Vert > \epsilon, \hat{\mathcal{S}}_n \neq \mathcal{S} \right)\\
&= J_n(1) + J_n(2).
\end{align*}
We will show that $J_n(1) \ra 0$ and $J_n(2) \ra 0$ as $n \ra \ff$. $J_n(1)$ can be represented as follows:
\begin{align*}
J_n(1) &= \bm{P} \left(\Vert \hat{\bm{\Delta}}_{\hat{\mathcal{S}}_n} - \bm{\Delta}_{\hat{\mathcal{S}}_n} \Vert > \epsilon \bigg| \hat{\mathcal{S}}_n = \mathcal{S} \right) \bm{P}\left(\hat{\mathcal{S}}_n = \mathcal{S} \right)\\
&= \bm{P} \left(\Vert \hat{\bm{\Delta}}_{\hat{\mathcal{S}}_n} - \bm{\Delta}_{\hat{\mathcal{S}}_n} \Vert > \epsilon \bigg| \hat{\mathcal{S}}_n = \mathcal{S} \right)\\
&= \bm{P} \left(\Vert \hat{\bm{\Delta}}_{\mathcal{S}} - \bm{\Delta}_{\mathcal{S}} \Vert > \epsilon\right) \\
&\ra 0,
\end{align*}
where the penultimate line comes from the independence of $\mathcal{D}_{1n}$ and $\mathcal{D}_{2n}$. Now we will show $J_n(2)$ converges to 0.
\begin{align*}
J_n(2) &= \bm{P} \left( \Vert \hat{\bm{\Delta}}_{\hat{\mathcal{S}}_n} - \bm{\Delta}_{\hat{\mathcal{S}}_n} \Vert > \epsilon \bigg| \hat{\mathcal{S}}_n \neq \mathcal{S} \right) \bm{P}\left(\hat{\mathcal{S}}_n \neq \mathcal{S} \right)\\
&\le \bm{P}\left(\hat{\mathcal{S}}_n \neq \mathcal{S} \right) \ra 0.
\end{align*}
Proof is completed. $\blacksquare$

\subsection*{\textbf{Proof of Theorem \ref{Thm-Consistency-Seq}}}
We  begin by recalling the notations needed for studying the properties of estimated clusters of covariates in $\mathcal{S}$ and $\hat{{\mathcal{S}}}_n$.  Let $\hat{\mathcal{C}}_{\tau,\alpha_n}^{(k)}$ denote the $k^{th}$ cluster of covariates in $\mathcal{S}$, and it can be obtained by replacing $\hat{\mathcal{S}}_n$ with $\mathcal{S}$ in Algorithm 1. Then,
the proof consists of two parts: we will show that (i) $\hat{\mathcal{C}}_{\tau,\alpha_n}^{(k)}$ converges to $\mathcal{C}_{\tau}^{(k)}$ as $n \rightarrow \infty$ in probability and (ii) for each $k=1,2, \cdots K$, the symmetric difference between $\hat{\mathcal{C}}_{n,\tau,\alpha_n}^{(k)}$ and $\hat{\mathcal{C}}_{\tau, \alpha_n}^{(k)}$ converges to $\emptyset$, an empty set, as $n \rightarrow \infty$  in probability. For (i), we will show that the symmetric difference between the true cluster and estimated cluster converges to $\emptyset$ with probability tending to 1. We recall that the symmetric difference between a set $\mathcal{A}$ and a set $\mathcal{B}$ is given by
\begin{align*}
    \mathcal{A} \ \triangle \ \mathcal{B} = (\mathcal{A} - \mathcal{B}) \cup (\mathcal{B} - \mathcal{A}).
\end{align*}
Now, notice that for any $k=1,2, \cdots K$
\begin{align}\label{Thm8-1}
 \bm{P}(\hat{\mathcal{C}}^{(k)}_{\tau,\alpha_n} \ \triangle \ \mathcal{C}^{(k)}_{\tau} = \emptyset) &=  \bm{P}(\hat{\mathcal{C}}^{(k)}_{\tau,\alpha_n}-\mathcal{C}^{(k)}_{\tau}= \emptyset) +  \bm{P}(\mathcal{C}^{(k)}_{\tau}-\hat{\mathcal{C}}^{(k)}_{\tau,\alpha_n}= \emptyset).
\end{align}
We will show that the above quantity converges to one as $n \ra \ff$. The proof relies on 
the consistency of the hypothesis test used in the clustering algorithm. That is, since the type 1 error,  $\alpha_n$, is converging to 0 as $n \rightarrow \infty$, it follows that the power $\beta_n$ of the hypothesis test converges to 1 as $n \rightarrow \infty$. 
Observe that 
\begin{eqnarray*}
\hat{\mathcal{C}}^{(k)}_{\tau,\alpha_n}-\mathcal{C}^{(k)}_{\tau}& =&\{j \in {\mathcal{S}: j \in \mathcal{C}}^{(k)}_{\tau,\alpha_n} ~{and}~ j \notin  \mathcal{C}^{(k)}_{\tau}\}\\
&=& \{j \in \mathcal{S}: t_{j, v_k^*} \le t^*_{\alpha_n}~{and}~  |D_j-D_k^*| > \tau\}
\end{eqnarray*}
By the consistency of the t-test, $\bm{P}( \hat{\mathcal{C}}^{(k)}_{\tau,\alpha_n} \ - \ \mathcal{C}^{(k)}_{\tau} = \emptyset) \ra 1$ as $n \ra \ff$. Next, we note that
\begin{eqnarray*}
\mathcal{C}_{\tau}^{(k)}-\hat{\mathcal{C}}^{(k)}_{\tau,\alpha_n} &=&\{j \in \mathcal{S}: j \in \mathcal{C}_{\tau}^{(k)} ~{and}~ j \notin  \hat{\mathcal{C}}^{(k)}_{\tau,\alpha_n} \}\\
&=& \{j \in \mathcal{S}: t_{j, v_k^*} > t^*_{\alpha_n}~{and}~ |D_j-D_k^*| \le \tau\}
\end{eqnarray*}
Hence, once again by the consistency of the t-test, $\bm{P}( \mathcal{C}^{(k)}_{\tau} \ - \ \hat{\mathcal{C}}^{(k)}_{\tau,\alpha_n} = \emptyset) \ra 0$ as $n \ra \ff$. This completes the proof that the LHS of (\ref{Thm8-1}) converges to 1 as $n \ra \ff$ and hence the proof of (i) is complete. 

Now we turn to the proof of (ii). To this end, observe that 
\begin{align*}
\bm{P} \left( \hat{\mathcal{C}}^{(k)}_{n,\tau,\alpha_n} \ \triangle \ \hat{\mathcal{C}}^{(k)}_{\tau,\alpha_n}=\emptyset \right) &= \bm{P} \left( \hat{\mathcal{C}}^{(k)}_{n,\tau,\alpha_n} \ \triangle \ \hat{\mathcal{C}}^{(k)}_{\tau,\alpha_n}=\emptyset , \hat{\mathcal{S}}_n = \mathcal{S} \right)
+ \bm{P} \left( \hat{\mathcal{C}}^{(k)}_{n,\tau,\alpha} \ \triangle \ \hat{\mathcal{C}}^{(k)}_{\tau,\alpha_n}=\emptyset, \hat{\mathcal{S}}_n \neq \mathcal{S} \right)\\
&= J_n(1) + J_n(2).
\end{align*}
We will show that $J_n(1) \ra 1$ and $J_n(2) \ra 0$ as $n \ra \ff$. Now observe that 
\begin{align*}
J_n(1) &= \bm{P} \left(\hat{\mathcal{C}}^{(k)}_{n,\tau,\alpha_n} \ \triangle \ \hat{\mathcal{C}}^{(k)}_{\tau,\alpha_n}=\emptyset \bigg| \hat{\mathcal{S}}_n = \mathcal{S} \right) \bm{P}\left(\hat{\mathcal{S}}_n = \mathcal{S} \right)\\
&= \bm{P} \left(\hat{\mathcal{C}}^{(k)}_{\tau,\alpha_n} \ \triangle \ \hat{\mathcal{C}}^{(k)}_{\tau,\alpha_n}=\emptyset\right) \bm{P}\left(\hat{\mathcal{S}}_n = \mathcal{S} \right)\\
&= \bm{P}\left(\hat{\mathcal{S}}_n = \mathcal{S} \right) \ra 1~ {\mbox{as}}~ n \ra \ff,
\end{align*}
where the last convergence follows from the consistency of variable selection. Now we will show $J_n(2)$ is equal to 0.
\begin{align*}
J_n(2) &= \bm{P} \left( \hat{\mathcal{C}}^{(k)}_{n,\tau,\alpha_n} \ \triangle \ \hat{\mathcal{C}}^{(k)}_{\tau,\alpha_n}=\emptyset \bigg| \hat{\mathcal{S}}_n \neq \mathcal{S} \right) \bm{P}\left(\hat{\mathcal{S}}_n \neq \mathcal{S} \right)\\
&\le \bm{P}\left(\hat{\mathcal{S}}_n \neq \mathcal{S} \right)  \ra 0 ~ {\mbox{as}}~ n \ra \ff.
\end{align*}

This completes the proof of (ii) and that of the Theorem. $\blacksquare$

\section{\textbf{Additional Numerical Studies}}\label{APPC}

In this section, we provide the additional numerical studies described in Section \ref{NS}.

\textbf{Numerical Experiments on Computational Time}

We provide a small simulation study to compare the bias and the running time of the bootstrap algorithm and the plug-in estimates of the covariance matrix of the estimated precision matrix, which is denoted by $\bm{\Delta}^*$. We perform this study with 10 variables without the regularization. We generate 100 i.i.d observations from a multivariate normal with mean $\bm{0}$ and covariance matrix $\bm{I}$, which is an identity matrix. Because of the computational feasibility, we perform 1000 simulation studies. The average computational time of 1000 simulations of using the exact formula is 6.239 minutes and that of using bootstrap is .116 minutes. And the mean of $\Vert \hat{\bm{\Delta}}_E- \bm{\Delta} \Vert_F$, where $\hat{\bm{\Delta}}_E$ is $\hat{\bm{\Delta}}$ obtained by using the exact formula and $\Vert \cdot \Vert_F$ is the Frobenius norm of a matrix, is 14.450 and the mean of $\Vert \hat{\bm{\Delta}}_B- \bm{\Delta} \Vert_F$, where $\hat{\bm{\Delta}}_B$ is $\hat{\bm{\Delta}}$ obtained by using bootstrap, is 14.688. In addition, the mean of $\Vert \hat{\bm{\Delta}}_E^*- \bm{\Delta}^* \Vert_F$ is 19.067 while the mean of $\Vert \hat{\bm{\Delta}}_B^*- \bm{\Delta}^* \Vert_F$ is 14.540. According to these numerical experiments, the exact formula is 50 times more expensive than the bootstrap approach. In terms of their accuracy in estimating $\bm{\Delta}$, the exact formula and bootstrap based estimate are not very different in estimating $\bm{\Delta}$ and $\bm{\Delta}^*$. Hence we will use the bootstrap approach to estimate the limiting covariance matrix in Theorem \ref{Thm-Degree-Vec} through Theorem \ref{Thm-Clustercoef-Vec}.

\textbf{Numerical Experiments on Centralities}

We now demonstrate the behavior of network-wide metrics of our implicit network. The vector of covariates $\bm{X} = (X_1,X_2,\dots,X_p)'$ is generated from a multivariate normal distribution ($p-$dimensional) with mean vector zero and covariance matrix is a $p-$dimensional identity matrix. We consider multiple linear regression model without an intercept with the covariate dimension $p=10$, and sample size 100 and 300. Given $\bm{x}_i$, the response variable $Y_i$ is generated from $\bm{x}_i\bm{\beta} + \epsilon_i$ where $\epsilon_i \sim N(0,\sigma^2)$ and $\sigma^2 = 1$, and $\bm{\beta}' = (1, 1, 1, 1, 1,1, 1,2,2,2)$. The implicit network was constructed based on the function (ii) in Section \ref{ParCor}. The true value of the degree centrality and the clustering coefficient of our implicit network is 9.748 and .545, respectively. Based on the 1000 simulations, the bias of degree centrality at $n=100$ and $n=300$ are -.008 (.323) and -.014 (.001), respectively. Numbers in parentheses are the standard errors. The bias of clustering coefficient at $n=100$ and $n=300$ are -.014 (.180) and -.001 (.003), respectively.
The following histograms represent the distribution of the degree centrality and clustering coefficient. Among the 10 covariates, we choose $X_1$ to provide histograms.
\begin{figure}[H]
    \centering
    \subfloat[Degree Centrality]{{\includegraphics[scale=.35]{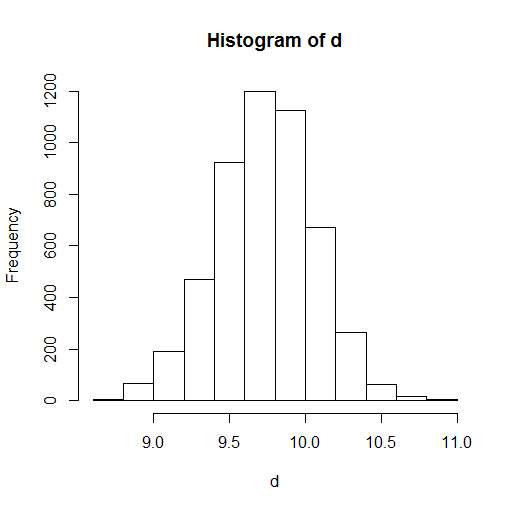} }}
    \label{NWM100-d}
    \qquad
    \subfloat[Clustering Coefficient]{{\includegraphics[scale=.35]{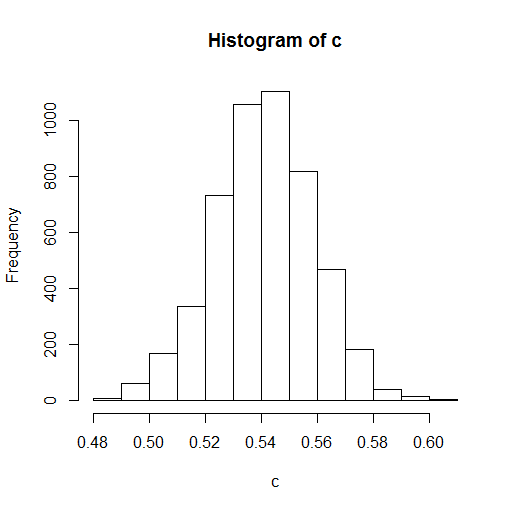} }}
    \caption{Network-wide metrics with $n=100$}
    \label{NWM100-c}
\end{figure}
\begin{figure}[H]
    \centering
    \subfloat[Degree Centrality]{{\includegraphics[scale=.35]{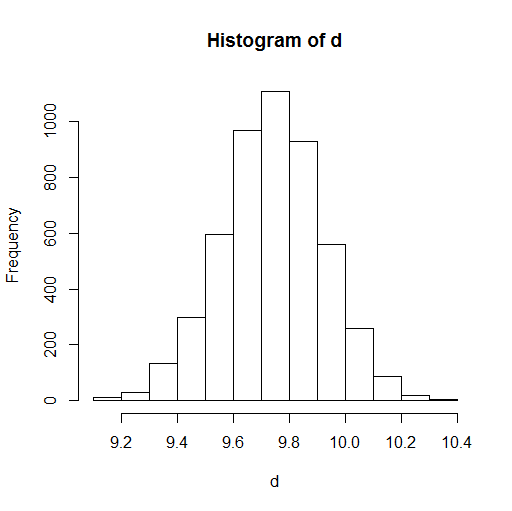} }}
    \label{NWM300-d}
    \qquad
    \subfloat[Clustering Coefficient]{{\includegraphics[scale=.35]{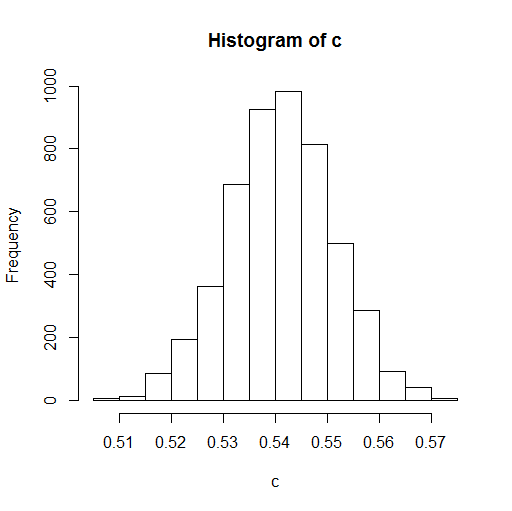} }}
    \caption{Network-wide metrics with $n=300$}
    \label{NWM300-c}
\end{figure}
The histograms above shows that the degree centrality and clustering coefficient both follow a normal distribution, which corresponds with the Theorem \ref{Thm-Degree-Vec} and Theorem \ref{Thm-Clustercoef-Vec} in Section \ref{PSNWM}.

\textbf{Numerical Experiments with a wrong number of clusters}

We repeat the same numerical study described in Section \ref{Numerical1}, but here we set the desired number of clusters to 2 and 4 instead of 3. In this numerical study, we will compare the performance of three clustering algorithms, which are the modified K-means clustering, the spectral clustering, and the sequential testing clustering algorithm, when we give a wrong number of clusters to detect. The simulation setting is same as that in Section \ref{Numerical1}, but the number of simulations is now set to 1000. If the desired number of cluster is 2, then a clustering algorithm must identify the top two clusters that are associated with the response variable. In case of 4, a clustering algorithm must detect all three true clusters and the fourth clusters must be empty. The table in the below represents the proportion of times that each clustering algorithm identifies the correct clusters.

\begin{table}[H]
\centering
\begin{tabular}{|c|c|c|c|c|c|c|}
\hline
$r_b$ & Number of & Modified & Modified & Sequential \\
& Clusters & K-means& Spectral & Clustering \\ \hline
0& 2 & 0 & 0 & .901\\
.1&  & 0 & 0 & .894\\
.2&  & 0 & 0 & .894\\
.3& & 0 & 0 & .872\\
.4&  & 0 & 0 & .875\\
.5&  & 0 & 0 & .867\\
0& 4 & 0 & 0 & .91\\
.1&  & 0 & 0 & .9\\
.2& & 0 & 0 & .9\\
.3&  & 0 & 0 & .878\\
.4&  & 0 & 0 & .887\\
.5&  & 0 & 0 & .858\\
\hline
\end{tabular}
\end{table}

From the table above, we observe that unsupervised clustering methods cannot identify correct clusters when the true number of clusters is unknown. On the other hand, the sequential testing clustering algorithm identifies the true clusters even though the number of clusters is not known in advance. When the pre-specified number of clusters is larger than the true number of clusters, it can identify all correct clusters. If the pre-specified number of clusters (let's say it is $k$) is smaller than the truth, it can identify first $k$ clusters. Hence, the proposed clustering algorithm outperforms unsupervised clustering algorithms when the true number of clusters is unknown.

While eleven genes were selected in the estimated active predictor set, the sequential testing method detected one cluster with 3 variables, using both degree centrality and clustering coefficient. One may now summarize the information contained in the detected cluster and use it to provide a model with other genes that are not in the cluster.

\section{\textbf{Concluding remarks}}\label{Conclusion}

We propose a new method to group predictors that have a similar association with the response variable of interest. Our method uses the properties of the network-wide metrics on the implicit network, and a sequential hypothesis testing procedure to define clusters. Under a population model for groups, conditions are provided that guarantee detection of clusters with probability tending to one as the sample size diverges to infinity. Additionally, the proposed  method takes into account model-selection uncertainty in the detection of clusters of predictors, which seems to be first such result in the literature. A model-assisted bootstrap approach -- with reduced computational burden -- to assess uncertainty in the estimates of network-wide metrics is also provided and illustrated numerically. The numerical experiments also show that adoption of NWM for clustering even in the context of unsupervised problems may yield improved results. The proposed methods are illustrated with examples from sports analytics and the study of breast cancer.

While detailed computations are provided for partial correlation weights, other weight functions such as Pearson's correlation, distance correlation, and mutual information can also be used as weights. Asymptotic theory related to these weights is left for future research.

\vspace{5mm}
\noindent \textit{Disclaimer}: This report is released to inform interested parties of research and to encourage discussion. The views expressed on statistical issues are those of the authors and not necessarily those of the U.S. Census Bureau.

\bibliographystyle{asa}
\bibliography{ref}

\begin{thebibliography}{26}
\newcommand{\enquote}[1]{``#1''}
\expandafter\ifx\csname natexlab\endcsname\relax\def\natexlab#1{#1}\fi

\bibitem[{Altenbuchinger et~al.(2020)Altenbuchinger, Weihs, Quackenbush, J{\"o}rgen, and Zacharias}]{AWQ20}
Altenbuchinger, M., Weihs, A., Quackenbush, J., J{\"o}rgen, G.~H., and Zacharias, H. (2020), \enquote{Gaussian and Mixed Graphical Models as (multi-) omics data analysis tools,} \textit{Biochimica et Biophysica Acta (BBA)-Gene Regulatory Mechanisms}, 1863.

\bibitem[{Amemiya(1985)}]{A85}
Amemiya, T. (1985), \textit{Advanced econometrics}, Harvard university press.

\bibitem[{Antoniou and Tsompa(2008)}]{AT08}
Antoniou, I. and Tsompa, E. (2008), \enquote{Statistical analysis of weighted networks,} \textit{Discrete dynamics in Nature and Society}, 2008.

\bibitem[{Apostol(1969)}]{A69}
Apostol, T.~M. (1969), \enquote{Calculus, Volume II: Multi-variable calculus and linear algebra, with applications to differential equations and probability. Blaisdell Publishing Co., Ginn and Co., Waltham, Mass,} \textit{Toronto, Ont}.

\bibitem[{Baseball-Reference(2016)}]{BR16}
Baseball-Reference (2016), \enquote{{Baseball Reference 2016 Regular Season Stats},} \url{https://www.baseball-reference.com/}.

\bibitem[{B{\"u}hlmann et~al.(2013)B{\"u}hlmann, R{\"u}timann, van~de Geer, and Zhang}]{BRG13}
B{\"u}hlmann, P., R{\"u}timann, P., van~de Geer, S., and Zhang, C.-H. (2013), \enquote{Correlated variables in regression: clustering and sparse estimation,} \textit{Journal of Statistical Planning and Inference}, 143, 1835--1858.

\bibitem[{Crane(2018)}]{C18}
Crane, H. (2018), \textit{Probabilistic foundations of statistical network analysis}, Chapman and Hall/CRC.

\bibitem[{De~la Fuente et~al.(2004)De~la Fuente, Bing, Hoeschele, and Pedro}]{DBH04}
De~la Fuente, A., Bing, N., Hoeschele, I., and Pedro, M. (2004), \enquote{Discovery of meaningful associations in genomic data using partial correlation coefficients,} \textit{Bioinformatics}, 20, 3565--3574.

\bibitem[{Fan and Li(2001)}]{FL01}
Fan, J. and Li, R. (2001), \enquote{Variable Selection via Nonconcave Penalized Likelihood and its Oracle Properties,} \textit{Journal of the American Statistical Association}, 96, 1348.

\bibitem[{Fan and Lv(2011)}]{FL11}
Fan, J. and Lv, J. (2011), \enquote{Nonconcave penalized likelihood with NP-dimensionality,} \textit{IEEE Transactions on Information Theory}, 57, 5467--5484.

\bibitem[{Gut(2013)}]{G13}
Gut, A. (2013), \textit{Probability: a graduate course}, vol.~75, Springer Science \& Business Media.

\bibitem[{Khalili and Vidyashankar(2018)}]{KV18}
Khalili, A. and Vidyashankar, A. (2018), \enquote{Hypothesis Testing in Finite Mixture of Regressions: Sparsity anad Model Selection Uncertainty,} \textit{Canadian Journal of Statistics}, 46, 429--457.

\bibitem[{Kim et~al.(2008)Kim, Choi, and Oh}]{K08}
Kim, Y., Choi, H., and Oh, H.-S. (2008), \enquote{Smoothly clipped absolute deviation on high dimensions,} \textit{Journal of the American Statistical Association}, 103, 1665--1673.

\bibitem[{Kim and Kwon(2012)}]{KK12}
Kim, Y. and Kwon, S. (2012), \enquote{Global optimality of nonconvex penalized estimators,} \textit{Biometrika}, 99, 315--325.

\bibitem[{Lopez-Fernandez et~al.(2004)Lopez-Fernandez, Robles, Gonzalez-Barahona, et~al.}]{LRG04}
Lopez-Fernandez, L., Robles, G., Gonzalez-Barahona, J.~M., et~al. (2004), \enquote{Applying social network analysis to the information in CVS repositories,} in \textit{International workshop on mining software repositories}, IET, pp. 101--105.

\bibitem[{Meinshausen and B{\"u}hlmann(2010)}]{MB10}
Meinshausen, N. and B{\"u}hlmann, P. (2010), \enquote{Stability selection,} \textit{Journal of the Royal Statistical Society: Series B (Statistical Methodology)}, 72, 417--473.

\bibitem[{MLB(2016)}]{MLB16}
MLB (2016), \enquote{{MLB 2016 Regular Season Stats},} \url{https://www.mlb.com/}.

\bibitem[{Muirhead(2009)}]{MRJ09}
Muirhead, R.~J. (2009), \textit{Aspects of multivariate statistical theory}, vol. 197, John Wiley \& Sons.

\bibitem[{Neudecker and Wesselman(1990)}]{NW90}
Neudecker, H. and Wesselman, A.~M. (1990), \enquote{The asymptotic variance matrix of the sample correlation matrix,} \textit{Linear Algebra and its Applications}, 127, 589--599.

\bibitem[{Opsahl et~al.(2010)Opsahl, Agneessens, and Skvoretz}]{OAS10}
Opsahl, T., Agneessens, F., and Skvoretz, J. (2010), \enquote{Node centrality in weighted networks: Generalizing degree and shortest paths,} \textit{Social networks}, 32, 245--251.

\bibitem[{Petersen et~al.(2008)Petersen, Pedersen, et~al.}]{PP08}
Petersen, K.~B., Pedersen, M.~S., et~al. (2008), \enquote{The matrix cookbook,} \textit{Technical University of Denmark}, 7, 510.

\bibitem[{Reverter and Chan(2008)}]{RC08}
Reverter, A. and Chan, E.~K. (2008), \enquote{Combining partial correlation and an information theory approach to the reversed engineering of gene co-expression networks,} \textit{Bioinformatics}, 24, 2491--2497.

\bibitem[{Tibshirani(1996)}]{T96}
Tibshirani, R. (1996), \enquote{Regression shrinkage and selection via the lasso,} \textit{Journal of the Royal Statistical Society. Series B (Methodological)}, 267--288.

\bibitem[{Van~der Vaart(2000)}]{V00}
Van~der Vaart, A.~W. (2000), \textit{Asymptotic statistics}, vol.~3, Cambridge university press.

\bibitem[{Yuan and Lin(2006)}]{YL06}
Yuan, M. and Lin, Y. (2006), \enquote{Model selection and estimation in regression with grouped variables,} \textit{Journal of the Royal Statistical Society: Series B (Statistical Methodology)}, 68, 49--67.

\bibitem[{Zhang et~al.(2010)}]{Z10}
Zhang, C.-H. et~al. (2010), \enquote{Nearly unbiased variable selection under minimax concave penalty,} \textit{The Annals of statistics}, 38, 894--942.

\end{thebibliography}

\end{document}